\documentclass[acmsmall,nonacm,screen]{acmart}

\setcopyright{none}

\acmJournal{PACMPL}
\acmVolume{1}
\acmNumber{PLDI} 
\acmYear{2025}
\acmMonth{6}
\acmDOI{10.1145/nnnnnnn.nnnnnnn}
\startPage{1}


\usepackage{booktabs}
\usepackage{subcaption}
\usepackage{xspace}
\usepackage{xparse}
\usepackage{microtype}
\usepackage{listings}
\usepackage{tikz}
\usepackage{float}
\usepackage{multicol}
\usepackage{amsbsy}
\usepackage{wrapfig}
\usepackage{mathtools}
\usepackage{multicol}
\usepackage{multirow}

\usepackage{colortbl}

\usetikzlibrary{calc}

\usepackage{mathtools}
\usepackage{amsbsy}
\usepackage{float}
\usepackage{microtype}
\usepackage{listings}
\usepackage{wrapfig}
\usepackage[group-separator={,}]{siunitx}

\usepackage{enumitem}
\usepackage{xcolor}
\usepackage{pifont}

\newcommand{\todoinline}[1]{}

\newcommand{\ignore}[1]{}

\usepackage{cleveref}

\newcommand{\figref}[1]{\Cref{#1}}
\newcommand{\secref}[1]{\Cref{#1}}

\definecolor{keywordcolor}{rgb}{0.5,0,0.5}
\definecolor{textgray}{gray}{0.4}
\definecolor{mygray}{rgb}{0.5,0.5,0.5}
\newcommand\code[1]{\lstinline[mathescape=true,basicstyle=\ttfamily\normalsize]|#1|}
\newcommand\codesmall[1]{\lstinline[mathescape=true,basicstyle=\ttfamily\footnotesize]|#1|}

\newcommand{\figscale}{0.38}

\definecolor{DeegenKeywordColor}{rgb}{0.667, 0.051, 0.569} 
\definecolor{CKeywordColor}{rgb}{0.745, 0.616, 0.278} 
\definecolor{BlueBoxColor}{rgb}{0.0, 0.588, 1.0} 

\newcommand\deegenKeyword[1]{\textcolor{DeegenKeywordColor}{\ttfamily\normalsize #1}}

\definecolor{deegenKwdDarkColorDark}{rgb}{0.567, 0.043, 0.484} 
\definecolor{CTypeColor}{rgb}{0.0, 0.456, 0.830} 
\definecolor{LLVMKeyWordColor}{rgb}{0.235, 0.560, 0.0} 
\definecolor{LLVMCommentColor}{rgb}{1, 0.550, 0.0} 

\definecolor{Gray}{rgb}{0.37, 0.37, 0.37} 

\newcommand\deegenKwdDark[1]{\textcolor{deegenKwdDarkColorDark}{\ttfamily\normalsize #1}}
\newcommand\CType[1]{\textcolor{CTypeColor}{\ttfamily\normalsize #1}}
\newcommand\CKeyword[1]{\textcolor{CKeywordColor}{\ttfamily\normalsize #1}}
\newcommand\CComment[1]{\textcolor{Gray}{\ttfamily\normalsize #1}}
\newcommand\ssa[1]{\textcolor{Gray}{\ttfamily\normalsize #1}}
\newcommand\llvmKwd[1]{\textcolor{LLVMKeyWordColor}{\ttfamily\normalsize #1}}

\definecolor{CKeywordColorDark}{rgb}{0.633, 0.524, 0.236} 
\newcommand\CKeywordDark[1]{\textcolor{CKeywordColorDark}{\ttfamily\normalsize #1}}

\setcitestyle{nosort}

\renewcommand\footnotetextcopyrightpermission[1]{}

\makeatletter
\renewcommand*{\@fnsymbol}[1]{\ensuremath{\ifcase#1\or \dagger\or \ddagger\or
   \mathsection\or \mathparagraph\or \|\or **\or \dagger\dagger
   \or \ddagger\ddagger \else\@ctrerr\fi}}
\makeatother

\begin{document}

\title{Deegen: A JIT-Capable VM Generator for Dynamic Languages}
\subtitle{No More Compromise Between Performance and Engineering Cost\vspace{-0.1em}}

\author{Haoran Xu}
\authornote{$^\ddagger$Email addresses: haoranxu@stanford.edu (Haoran Xu), kjolstad@stanford.edu (Fredrik Kjolstad).}
\affiliation{%
    \institution{Stanford University}
    \streetaddress{353 Jane Stanford Way}
    \city{Stanford}
    \state{CA}
    \postcode{94305}
    \country{USA}
}
 \email{haoranxu@stanford.edu}

\author{Fredrik Kjolstad}
\authornotemark[2]
\affiliation{%
    \institution{Stanford University}
    \streetaddress{353 Jane Stanford Way}
    \city{Stanford}
    \state{CA}
    \postcode{94305}
    \country{USA}
}
 \email{kjolstad@stanford.edu}

\renewcommand{\shortauthors}{H. Xu and F. Kjolstad}

\begin{abstract}
Building a high-performance JIT-capable VM for a dynamic language has traditionally required a tremendous amount of time, money, and expertise. 
We present Deegen, a meta-compiler that allows users to generate a high-performance JIT-capable VM for their own language at an engineering cost similar to writing a simple interpreter.
Deegen takes in the execution semantics of the bytecodes implemented as C++ functions, and automatically generates a two-tier VM execution engine with a state-of-the-art interpreter, a state-of-the-art baseline JIT, and the tier-switching logic that connects them into a self-adaptive system. 

We are the first to demonstrate the automatic generation of a JIT compiler, and the automatic generation of an interpreter that outperforms the state of the art.
Our performance comes from a long list of optimizations supported by Deegen, including bytecode specialization and quickening, register pinning, tag register optimization, call inline caching, generic inline caching, JIT polymorphic IC, JIT IC inline slab, type-check removal and strength reduction, type-based slow-path extraction and outlining, JIT hot-cold code splitting, and JIT OSR-entry. These optimizations are either employed automatically, or guided by the language implementer through intuitive APIs. As a result, the disassembly of the Deegen-generated interpreter, baseline JIT, and the generated JIT code rivals the assembly code hand-written by assembly experts in state-of-the-art VMs.


We implement LuaJIT Remake (LJR), a standard-compliant Lua 5.1 VM, using Deegen. Across 44 benchmarks, LJR's interpreter is on average 179\% faster than the official PUC Lua interpreter, and 31\% faster than LuaJIT's interpreter. 
LJR's baseline JIT has negligible startup delay, and its execution performance is on average 360\% faster than PUC Lua and only 33\% slower (but faster on 13/44 benchmarks) than LuaJIT's optimizing JIT.

\end{abstract}


\maketitle

\section{Introduction}
\label{sec:introduction}

Dynamic languages are widely used for their productivity. But they are also known for being slow because they cannot be statically compiled to efficient native code. Just-In-Time (JIT) compilers are used to improve their performance through profile-guided speculative compilation at runtime, but building a good JIT-capable Virtual Machine (VM) is notoriously difficult. For example, every JavaScript VM in modern browsers is built through centuries of engineering time from some of the best compiler engineers in the world, with a huge build cost estimated at \$225M~\cite{chrisCumminsThesis} for V8~\cite{v8jit} and $\sim$\$50M~\cite{boehm2009cocomo} for JSC~\cite{applejsc} and SM~\cite{spidermonkey}.

The high cost has its reason: to yield good performance on both short-running and long-running workloads, the VM needs an optimized interpreter and multiple tiers of JIT compilers to balance startup delay (the time to generate the JIT code) and execution performance~\cite{speculationinjsc, v8turbofan, v8sparkplug, spidermonkeyDesign}.\footnote{This paper assumes that readers are familiar with modern multi-tier VM architectures. For readers seeking an introduction to the topic, we recommend reading the section \href{https://webkit.org/blog/10308/speculation-in-javascriptcore/\#:~:text=The\%20tiers\%20of\%20JavaScriptCore.}{``Overview of JavaScriptCore''} in ~\cite{speculationinjsc}.} For best performance, the interpreter needs to be written in assembly~\cite{v8ignition,jscAsmInterpreter,luajitAsmInterpreter}, as producing a state-of-the-art interpreter from a high-level language like C is beyond the capability of general-purpose compilers~\cite{luajitAsmInterpreterMotivation}. 
In a JIT compiler, the logic in the generated JIT code is hand-crafted by programmers directly in assembly instructions, which differ for each JIT tier, or even each hardware ISA.
For example, V8 once had \textit{nine} baseline JIT implementations for different ISAs~\cite{10.1145/3033019.3033025}. Domain-specific dynamic language optimizations, such as hidden classes~\cite{10.1145/74877.74884}, inline caching~\cite{10.5555/646149.679193}, type speculation~\cite{10.1145/191080.191116}, and watchpoints~\cite{10.1145/2542142.2542143} are required to get good performance, but they require high expertise to get right. Furthermore, the lower VM tiers must tier up to the higher tiers when the code gets hot, and in the case of a speculation failure or watchpoint invalidation, the higher tiers must exit to the lower tiers using on-stack replacement (known as an OSR-exit or deoptimization)~\cite{10.1145/143095.143114}. All of these complications make it notoriously difficult to write a modern dynamic language VM with a multi-tier JIT. 

Such engineering complexity is unaffordable for most VM developer groups. Repurposing a VM from another language is also not an option, as any small difference in the language semantics can become a big change in the assembly implementation, thus a sweeping change to the interpreter and the JIT compilers. As a result, developers have to fall back to simpler designs with worse performance. Many VMs do not have a JIT at all. And for those with a JIT, the design is almost always single-tiered for its relative simplicity,
even though the lack of a baseline JIT can cause significant performance degration on real-world large programs~\cite{hhvmCgo21}. Some (e.g., PyPy~\cite{pypy} and LuaJIT~\cite{luajit}) employ tracing JIT instead of method JIT, which simplifies engineering but has uneven (workload-dependent) and less predictable performance~\cite{mozzilaAbandonTracingJit,convergemetatracing}. And some (e.g., the  PHP 8 JIT~\cite{engebreth2021php}) are primarily optimized for long-running numerical workloads, which are less sensitive to startup delay and easier to optimize. However, such a JIT has little benefit for most real-world uses of dynamic languages~\cite{phpWordpressPerf}. 


To reduce engineering cost, Truffle~\cite{trufflePaper} pioneered the idea of decoupling the JIT compiler from the language: Truffle implements a generic JIT that generates code by partial evaluation~\cite{partialevaluationbook} of a guest language AST interpreter on the input AST program at runtime, also known as the first Futamura projection~\cite{futamuraPaper}. While Truffle achieves remarkably high JIT peak throughput, the approach has fundamental limitations. Since the expensive partial evaluation happens at runtime, the startup delay of the Truffle JIT is significantly higher than conventional JIT compilers~\cite{truffleRubyStartupDelay, truffleSeminarTalk}. The Truffle interpreter is designed to facilitate partial evaluation, but this comes at the cost of performance and memory: \citet{Marr2022GraalWorkshop} showed that Truffle's Python and Ruby interpreters were $\sim$10x slower than CPython and CRuby while consuming more memory as of 2022.\footnote{The Truffle developers have informed us that Truffle's interpreter performance has improved since then.} As a result, for many real-world use cases where the startup delay matters (e.g., client apps, and server apps with continuous deployment), Truffle's performance is unsatisfactory~\cite{truffleRubyStartupDelay}.

%

\begin{figure}
    \centering
    \includegraphics[scale=\figscale]{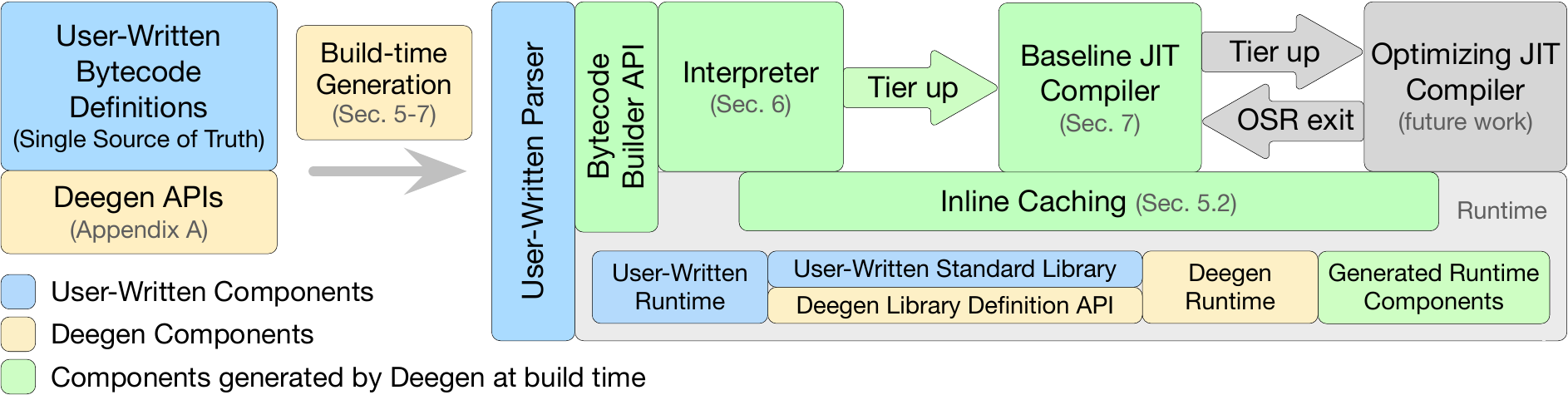}
    \vspace{-0.9em}
    \caption{
        Overview of the Deegen framework that generates a JIT-capable VM automatically. 
        \label{fig:overview}
    }
    \vspace{-1.1em}
\end{figure}

To overcome these limitations, we present Deegen, a fundamentally new approach to building high-performance VMs at low engineering cost. 
Deegen is a compiler generator that automatically \textit{generates} a VM execution engine from C++ bytecode semantics \textit{at build time} (\figref{fig:overview}). In Futamura's taxonomy, Deegen is a second Futamura projector.\footnote{In the sense that Deegen statically generates a compiler. However, our internal design is different from Futamura's proposal.
} Specifically, the user describes the execution semantics of each bytecode in C++ inside our framework. The description is compiled to LLVM IR~\cite{lattner2004} using the unmodified Clang compiler. Deegen then performs analysis, transformation, and optimization at the LLVM IR level to automatically generate the whole execution engine. The generated JIT compiler generates code using Copy-and-Patch~\cite{copyandpatch}, a light-weight and fast code generation technique that works by composing and configuring pre-built \textit{stencils}. Thus, the generated system is self-contained, and does not need LLVM at runtime. 

Deegen currently automatically generates a VM with a state-of-the-art interpreter, a state-of-the-art baseline JIT, and the tier-switching logic that connects the two tiers, which will be the topic of this paper. We already have a concrete design and a partial implementation to generate the third-tier optimizing JIT as well, but we leave it to a future paper.
We are the first to demonstrate:
\begin{enumerate}
    \item A practical compiler-compiler that works for an industrial-strength dynamic language (Lua).
    \item The automatic generation of a two-tier JIT-capable VM and the associated profiling, tier-up and OSR-entry logic that connects the tiers into a self-adaptive system. Furthermore, everything is generated from a single source of truth (C++ bytecode semantics).
    \item The automatic generation of a state-of-the-art interpreter whose code quality can match or surpass existing state-of-the-art interpreters hand-coded in assembly by experts.
    \item The automatic generation of a state-of-the-art baseline JIT compiler that has a negligible startup delay and generates high quality machine code that rivals existing state-of-the-art baseline JITs. This is achieved by an improved Copy-and-Patch technique that transparently supports polymorphic inline caching, hot-cold code splitting, and other optimizations.
        \item The design of a bytecode semantic description framework that facilitates build-time analysis, transformation, and optimization of the bytecode semantics; allows easy expression of common dynamic language optimizations (e.g., logic specialization, inline caching, and type speculation); and features an intuitive, flexible, and user-friendly interface.
\end{enumerate}

To evaluate Deegen, we use it to implement LuaJIT Remake (LJR)\footnote{Deegen and LuaJIT Remake are available under the Apache 2.0 license at \url{https://github.com/luajit-remake/luajit-remake}.}, a standard-compliant Lua 5.1~\cite{lua51standard} VM. To keep our research focused, however, we did not implement garbage collection\footnote{To fairly compare to LuaJIT and PUC Lua, we also turned off their garbage collection in benchmarks.} or the Lua C bindings, and we did not implement all of the Lua standard~library. 
Across 44 benchmarks, LJR's interpreter is on average 179\% faster than the official PUC Lua interpreter~\cite{puclua51}, and 31\% faster than the interpreter in LuaJIT 2.1~\cite{luajit21}. LJR's baseline JIT has negligible startup delay, and its execution performance is on average 360\% faster than PUC Lua, and 33\% slower than LuaJIT's optimizing JIT. Despite that a baseline JIT is \textit{by design} not intended to compete with an optimizing JIT on peak throughput, we are already faster than LuaJIT on 13/44 benchmarks. We believe the performance gap with LuaJIT in peak JIT throughput will be closed once our optimizing JIT lands. 
The goal of this paper is to describe Deegen's architecture and design, and shed light on the key ideas, tricks, and observations that made Deegen possible. 

\section{Illustrative Example of Bytecode Semantics}
\label{sec:bytecode-semantic-input}


\begin{wrapfigure}{r}{0.54\linewidth}
    \centering
    \vspace{-0.9em}
    \includegraphics[width=\linewidth]{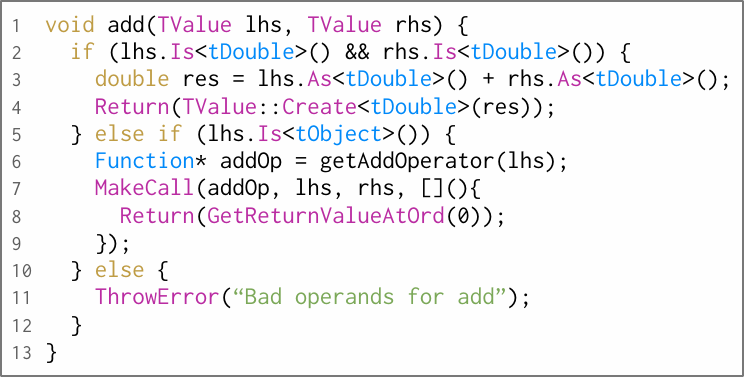}
    \vspace{-2em}
    \caption{
        C++ semantics for a hypothetical \code{add} bytecode.
        \label{fig:bytecode-add-example}
    }
    \vspace{-1.3em}
\end{wrapfigure}

For a quick sense of the Deegen framework, \figref{fig:bytecode-add-example} illustrates an example definition of a hypothetical \code{add} bytecode, with the Deegen APIs colored in \deegenKeyword{\textrm{purple}} (see \secref{appendix:full-deegen-api-reference} for the full Deegen API reference). In this example, the bytecode takes two operands \code{lhs} and \code{rhs}, each being a \textit{boxed value} (i.e., a value and its type). If both operands are of type \texttt{double} (line 2), the addition is performed by simply returning the sum of the two numbers using the \deegenKeyword{Return} API (line 4), which will store the result and pass control to the next bytecode. Otherwise, for the purpose of exposition, we assume the guest language defined some way (e.g., some overloaded operator resolution rule) to obtain a guest language function that shall be called if \code{lhs} is an object (line 6). The \deegenKeyword{MakeCall} API then calls the function with the specified arguments (line 7). The \deegenKeyword{MakeCall} API does not return. Instead, it takes a continuation function, and when the callee returns, the control flow will be transferred to the continuation. The continuation is similar to the main function, except that it can access the list of values returned by the callee. In our example, the continuation simply returns the first value (accessed via the \deegenKeyword{GetReturnValueAtOrd} API) as the result of the \code{add} (line 8). Finally, the \deegenKeyword{ThrowError} API in line 11 throws an exception, which will trigger the stack unwinder to propagate the exception up the call stack until it is eventually caught.  

\begin{wrapfigure}{r}{0.42\linewidth}
    \centering
    \vspace{-0.3em}
    \includegraphics[width=\linewidth]{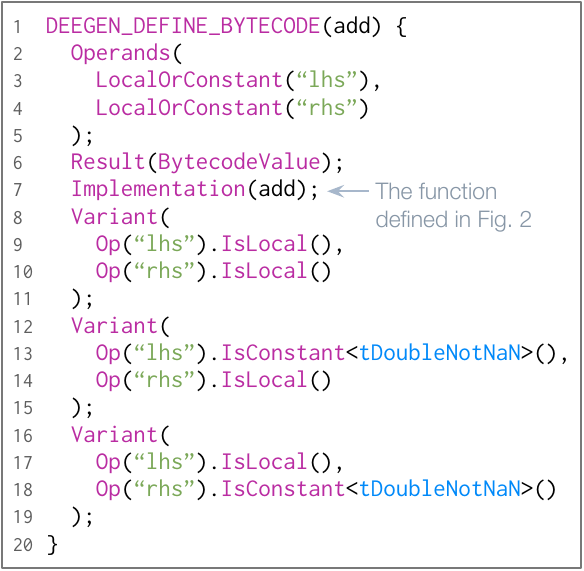}
    \vspace{-2em}
    \caption{
        Bytecode specification for the hypothetical \code{add} bytecode (C++ code).
        \label{fig:specialization-language}
    }
    \vspace{-1.3em}
\end{wrapfigure}

The execution semantic is not everything that defines a bytecode: we also need to know, for example, where the operands come from, and whether the bytecode produces an output value and/or can branch to another bytecode. This is achieved by the Deegen bytecode specification shown in \figref{fig:specialization-language}. Line 2--4 specifies the bytecode operands \code{lhs} and \code{rhs}, each may be either a local in the stack frame or a constant in the constant table. Line 6 specifies that the bytecode produces an output but will not perform a branch. Line 7 specifies the execution semantics: the \code{add} function in ~\figref{fig:bytecode-add-example}. The rest of the lines specifies the \textit{variants} of the bytecode: adding two locals, adding a local with a constant, and adding a constant with a local. In a hand-coded VM, the variants need to be implemented by hand, which is laborious and error-prone. Deegen's \deegenKeyword{Variant} API avoids such logic duplication. Furthermore, users may specify the statically-known type of the constants (line 13) and speculated types of the locals (not shown). Deegen understands the guest language type lattice, and can leverage this information to optimize the execution semantics accordingly (\secref{sec:type-based-optimization}). For example, if \code{rhs} is known to be a \texttt{double}, the \code{rhs.}\deegenKeyword{Is}\code{<tDouble>()} check will be deduced to be trivially true and optimized out.

\begin{wrapfigure}{r}{0.28\linewidth}
    \centering
    \vspace{-0.65em}
    \includegraphics[width=\linewidth]{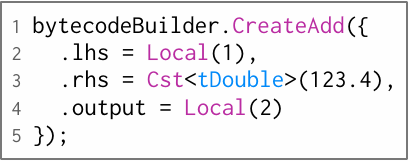}
    \vspace{-2em}
    \caption{
        Emit an add bytecode using the bytecode builder API.
        \label{fig:bytecode-builder-api}
    }
    \vspace{-2em}
\end{wrapfigure}

Deegen is unaware of the guest language syntax. The user is responsible for implementing the parser that builds up the bytecode stream from the program source. For this purpose, Deegen generates a rich set of bytecode-builder APIs from the bytecode semantics as a C++ header file. \figref{fig:bytecode-builder-api} shows the API that appends an \code{add} bytecode to the end of the bytecode stream. As one can see, the API hides all details about the internal bytecode representation and variant selection, and is type-safe and robust by design. For example, it is impossible to get the argument order wrong, to forget to provide an operand, or to supply a value of a bad type.

\section{Generated VM Architecture}
\label{sec:generated-vm-architecture}

Deegen automatically generates a two-tier VM execution engine, with its high-level architecture illustrated in \figref{fig:overview}. At VM runtime, the user-written parser translates each source program function to a bytecode sequence by calling the Deegen-generated \textit{bytecode builder APIs}. The bytecode sequence is then executed by the \textit{interpreter}.
Deegen automatically emits \textit{profiling} logic in the generated interpreter to detect hot functions. These functions are then tiered up to the \textit{baseline JIT} for higher execution throughput, at the cost of a negligible startup delay and the memory to hold the JIT code. With future work, execution can further tier up to a speculative \textit{optimizing JIT}. 
The rest of this section provides an overview of each component of the generated system.

\paragraph{Interpreter} Deegen generates a continuation-passing~\cite{steele1977} interpreter (\figref{fig:interpreter}). Each bytecode is implemented by a native function and, at the end of the function, control is transferred directly to the next bytecode (aka., direct-threading~\cite{10.1145/362248.362270}) via a tail call (a \texttt{jmp} instruction). Deegen supports many optimizations to improve performance, including bytecode specialization, dynamic quickening~\cite{10.1145/1899661.1869633}, register pinning~\cite{ghcRegisterPinning}, tag registers~\cite{jscTagRegisterOptimization}, monomorphic inline caching~\cite{speculationinjsc}, and slow-path outlining~\cite{luajitAsmInterpreterMotivation}. 

\begin{wrapfigure}{r}{0.31\linewidth}
    \centering
    \vspace{-1.2em}
    \begin{minipage}[b]{0.91\linewidth}
    \includegraphics[width=\linewidth]{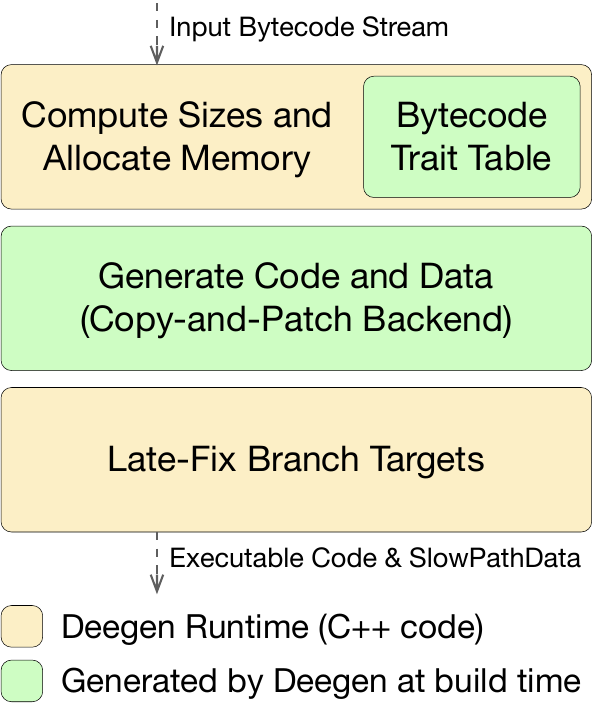}
    \vspace{-1.7em}
    \end{minipage}
    \caption{
        The runtime baseline JIT compilation pipeline. 
        \label{fig:baseline-jit-compilation-pipeline}
    }
    \vspace{-1.3em}
\end{wrapfigure}

\paragraph{Baseline JIT} The primary goal of the baseline JIT is to compile as quickly as possible. Generating good code is certainly desirable, but a secondary priority. To this end, the baseline JIT generated by Deegen lowers a bytecode stream directly to machine code, without going through any analyses or IRs, as illustrated in \figref{fig:baseline-jit-compilation-pipeline}. Since compilation speed is the top concern, inline caching (IC) is the only high-level optimization: the baseline JIT employs polymorphic IC~\cite{10.5555/646149.679193} based on self-modifying code. Nevertheless, many low-level optimizations are applied at build time to improve the machine code, including binding-time analysis~\cite{jones1988automatic} to identify runtime constants that can be burnt into the instructions, IC inline slab~\cite{javascriptcoreInlineSlab} to eliminate branches for monomorphic ICs, hot-cold splitting~\cite{javascriptcoreHotColdSplitting} for better code locality, and jump-to-fallthrough transform to eliminate unnecessary jumps. Through an improved Copy-and-Patch~\cite{copyandpatch} technique, we are able to employ these optimizations automatically and without compromising compilation speed.

\begin{wrapfigure}{r}{0.36\linewidth}
    \centering
    \vspace{-1em}
    \includegraphics[width=\linewidth]{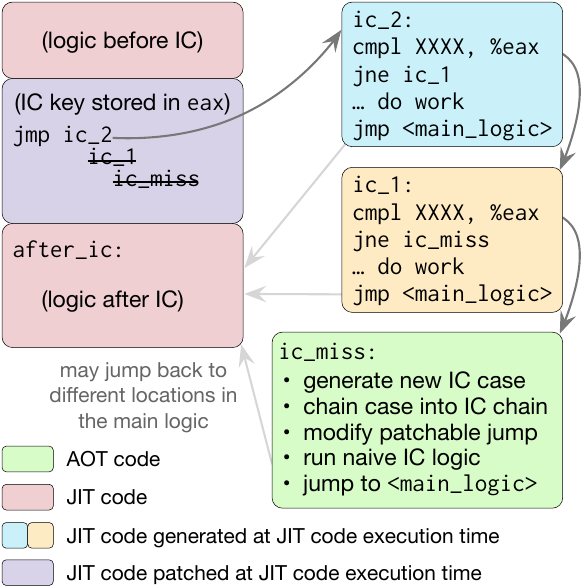}
    \vspace{-2em}
    \caption{
        A simplified illustration of how polymorphic IC works in the JIT.
        \label{fig:polymorphic-ic}
    }
    \vspace{-1em}
\end{wrapfigure}

\paragraph {Inline Caching} Inline caching (IC) is a critical optimization for all VM tiers. Deegen has two~IC mechanisms: \textit{call IC} and \textit{generic IC}. The call IC optimizes guest language function calls: Deegen automatically applies direct/closure dual-mode call IC~\cite{javascriptcoreCallIc}. The generic IC (\secref{sec:generic-inline-caching}) is a set of APIs that allows users to express arbitrary IC semantics (for example, to optimize object access). Deegen lowers the IC semantics to highly optimized implementations that are customized to fit the different needs and constraints of each VM tier. The interpreter uses monomorphic IC with dynamic quickening, since the dispatch overhead in polymorphic IC is high without code generation. The baseline JIT, on the other hand, uses polymorphic IC (\figref{fig:polymorphic-ic}), and leverages the capability of JIT code and self-modifying code for maximum performance. This is another prime example where Deegen lowers a single source of truth (the IC semantics) in different ways to best fit each use case.

\paragraph{Polymorphic IC} Polymorphic IC~\cite{10.5555/646149.679193} is a powerful optimization for the JIT compiler.
The most efficient polymorphic IC employs generation of new JIT code and self-modification of existing JIT code on the fly, as illustrated in \figref{fig:polymorphic-ic}.
In a hand-written VM, this is achieved by carefully planning all the logic pieces (the different IC cases, the support logic, etc.) in assembly, and then hand-craft the IC compiler that emits and patches those logic pieces in machine code. In Deegen, however, all the implementations must be generated automatically from the IC semantics in C++ code.
Clearly, this is a challenging task: at the minimum, we need to lower the C++ semantics to a piece of machine code that can be dynamically patched to append new logic pieces, and LLVM is simply not designed for this. As it turns out, the core trick is to repurpose a little-known LLVM IR node called \code{CallBr} to trick LLVM into doing the heavy lifting, and use \textit{assembly transforms} to finish the remaining bits. \secref{sec:jit-generator} explains our design in more detail.

\paragraph{Profiling}
The VM must detect hot functions so they can be compiled by the JIT compiler. Deegen accomplishes this by automatically emitting profiling logic into the generated interpreter to compute the number of bytecodes executed in each function. Specifically, Deegen generates appropriate accounting logic when a branch is taken and when the function exits (either normally or due to an exception). This gives us accurate profiling information at minimal performance overhead.

\paragraph{Tier-up and OSR-entry} 
The interpreter can enter the baseline JIT at function entries (called a \textit{tier-up}) or at customizable points inside the function body (called an \textit{OSR-entry} or a \textit{hot-loop transfer}). Tier-up is fully automatic. Deegen transparently injects profiling logic to detect hot functions. When a function reaches a hotness threshold, compilation is triggered and future calls to the function will use the optimized version. Tier-up can only elevate future calls, but for long-running functions, we also want to elevate the current invocation: this is supported by OSR-entry. OSR-entry requires simple user input: the user specifies the bytecode kinds that are worth checking for OSR-entry (usually the loop jumps). Deegen will generate OSR-entry checks for these bytecodes. Then, if the current-running function has met the hotness criteria, it will trigger compilation as soon as one of these bytecodes is executed, and OSR-enter the JIT code from that bytecode.

\paragraph{Library Functions in C/C++} Deegen provides a simple set of APIs for built-in library functions in C/C++ to call guest language functions and to throw exceptions (\secref{appendix:cpp-library-definition-apis}). Providing richer APIs with more control, e.g., stack introspection, is merely engineering and left as future work. 

\paragraph{Bytecode Builder API} Deegen is agnostic of the guest language syntax and the user is required to implement the parser that translates the source program to bytecodes. For this purpose, Deegen generates flexible, user-friendly and robust bytecode builder APIs (see \figref{fig:bytecode-builder-api} and \secref{appendix:bytecode-builder-apis}), with advanced features such as inspecting and replacing bytecodes. The APIs are generated as a C++ header file, and fully work for C++ IDE features such as syntax highlighting and autocompletion. 

\paragraph{Limitations} Deegen has ambitious goals, but to keep our research focused and reasonably scoped, several areas are left as future work. First, Deegen is agnostic of garbage collection (GC). Specifically, the inline caching (IC) entities generated by Deegen may reference GC-tracked objects, so to make Deegen production-ready, we would need to provide either an integrated GC or a way for the user-written GC to track the IC entities. Second, we do not consider compatibility of C-bindings with reference implementations: supporting efficient C-binding with backward-compatibility has been an active research area on its own~\cite{hpy,sulong,cinder,trufflerubycapi}. Third, Deegen only targets the x86-64 ISA. While Deegen offloads most heavy lifting of code generation to LLVM at build time, portability still requires non-trivial work due to our lack of access and experience with other ISAs. Finally, Deegen does not support concurrent compilation, so we do not have to think about race conditions between the JIT compiler and the application code.

\section{The Bytecode Semantic Description Framework Design}\label{sec:semantic-description-framework}

The most important design decision about our bytecode semantic description framework is to decide what concepts it should understand. As \citet{ricetheorem} proved, all non-trivial semantic properties of programs are undecidable, so the knowledge we can extract from generic C++ code is fundamentally limited. We must have domain-specific specialized models to support domain-specific optimizations. However, this inevitably hurts generality and penalizes use cases that fall outside the model, so finding the balance is critical. This section discusses our current model and design choices. We recognize that as an initial proof of concept, our model is far from optimal in both generality and performance. Nevertheless, we leave further design space exploration as future work: it requires a lot of engineering and experimentation, but is orthogonal to the core ideas presented in this paper.

\paragraph{Execution Model} We assume a register-based bytecode-machine model where each virtual register (i.e., stack slot) stores an 8-byte boxed value. Each bytecode operand may be a virtual register in the stack frame, a constant boxed value in the constant table, an integer literal value, or a consecutive range of virtual registers in the stack frame. We recognize that stack-based bytecode can be more compact, but we leave supporting it as future work. We made this decision because the bytecode semantics is the single source of truth in Deegen, and supporting multiple formats of bytecode semantics would add significant engineering complexity with limited research novelty. Similarly, we currently do not support 4-byte or 16-byte boxing, so we can focus on the research novelty. 

\paragraph{Boxing Scheme} The boxing scheme~\cite{Gudeman93representingtype} is fundamental to every dynamic language VM. Many schemes with different tradeoffs have been used in practice, such as SMI-boxing~\cite{v8smiboxing}, NaN-boxing~\cite{luajitNaNBoxing}, NuN-boxing~\cite{spidermonkeyNunboxing}, and trivial boxing. Therefore, Deegen does not hardcode a particular scheme, but provides APIs for users to describe it, so that Deegen can reason about the scheme and automatically apply relevant optimizations (\secref{sec:type-based-optimization}).

\begin{wrapfigure}{R}{0.32\linewidth}
    \centering
    \vspace{-1.35em}
    \includegraphics[width=\linewidth]{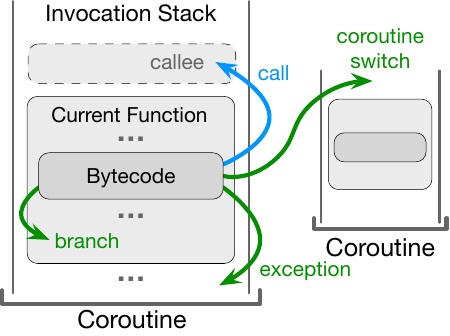}
    \vspace{-2.2em}
    \caption{
        Inter-bytecode control flow.
        \label{fig:inter-bytecode-control-flow}
    }
     \vspace{-3em}
\end{wrapfigure}

\paragraph{Inter-Bytecode Control Flow} The control flow transfer between bytecodes needs to be implemented differently in each VM tier, and the ability to understand the CFG is important for the future optimizing JIT. Thus, users must use Deegen APIs to transfer control between bytecodes (e.g., see \figref{fig:bytecode-add-example}). 
\figref{fig:inter-bytecode-control-flow} illustrates the control flow constructs supported by Deegen: guest language function call, branch, exception, and switching execution to another stackful coroutine (to implement \texttt{yield} and \texttt{resume}). These constructs should be sufficient for a wide range of languages. 

\paragraph{Function Calls} As just explained, calls to guest language functions must be done using Deegen APIs (e.g., \deegenKeyword{MakeCall}, \deegenKeyword{MakeTailCall}, etc.). Additionally, the logic after the call must be written in continuation-passing style (CPS) as a separate function (see \figref{fig:bytecode-add-example}). These constraints grant significant benefits. First, they hide the complicated implementation details of the call (such as variadic arguments handling) from the user, and lets Deegen automatically generate optimized code behind the scenes. Second, CPS makes it easier to support speculative inlining and stackful coroutines. Finally, being able to understand calls allows important high-level optimizations (e.g., call inline caching and speculative inlining) to happen automatically and transparently. Of course, the downside is that if a language requires a call concept not supported by Deegen, it would have to be soft-implemented using existing mechanisms, which hurts performance. To avoid this as much as possible, Deegen provides built-in support for variadic arguments, variadic calls, proper tail calls, in-place calls, and functions returning multiple or variadic results, which we believe is sufficient for a wide range of languages. Deegen can also be extended to support new call concepts as needed.

\paragraph{Lexical Captures} Many languages support first-class functions with lexical capturing. We require users to implement lexical captures using the Deegen-provided \textit{upvalue} mechanism, which is based on Lua's upvalue~\cite{luaUpvalue} implementation.\footnote{While the word ``upvalue'' is Lua jargon, the upvalue mechanism (also called \textit{flat closures}~\cite{10.1145/800055.802037}) is a language-neutral method of implementing first-class lexical-scoped functions that can be used for any language.} We made this requirement as the future optimizing JIT must know which virtual registers are captured to reason about the data flow of a function. Nevertheless, supporting alternate schemes with different performance/memory tradeoffs (e.g., JSC's \textit{scope chain} mechanism) is completely possible and left as future work.

\begin{wrapfigure}{R}{0.34\linewidth}
    \centering
    \vspace{-0.7em}
    \includegraphics[width=\linewidth]{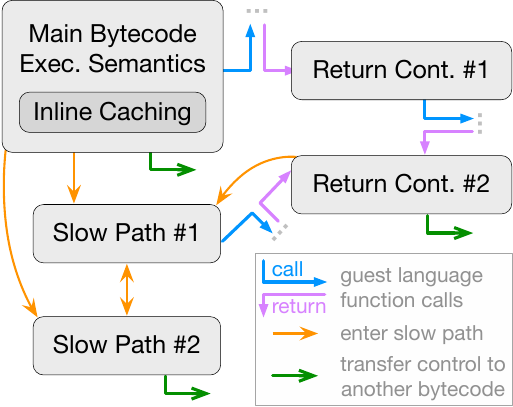}
    \vspace{-2.1em}
    \captionsetup{justification=centering}
    \caption{
        Bytecode components and \\the control flow between them.
        \label{fig:bytecode-components}
    }
     \vspace{-1.5em}
\end{wrapfigure}

\paragraph{Intra-Bytecode Control Flow} The execution semantics of a bytecode consists of multiple components. \figref{fig:bytecode-components} is an example of the different kinds of components in a bytecode, and the control flow between them. We have seen the \textit{main component} and the \textit{return continuation} in \figref{fig:bytecode-add-example}. 
The third kind, the \textit{slow path}, will be covered in the next paragraph. There is no restriction on how control flow transfers between components: for example, return continuations can make further calls and potentially recurse. The main component is special in that it supports inline caching. Call IC (\secref{sec:generated-vm-architecture}) is automatically employed for calls made by the main component. Users may also employ \textit{generic IC}, which we will describe later in the section.

\paragraph{Slow Paths} Slow paths are pervasive in dynamic languages: for example, many languages have complex rules to handle an add of two non-numbers or a call to a non-function. It makes no sense to compile these rarely-used and huge slow paths into JIT code at runtime. Deegen provides two mechanisms to specify such slow paths. First, the user may manually implement the slow~path execution semantics as a C++ function just like the other execution semantics components, except that it can take additional arguments for context. The slow path can then be entered using the \deegenKeyword{EnterSlowPath} API, which is a CPS-style control transfer that never returns. Second, the user may specify type hints for the bytecode operands. Since Deegen understands the boxing scheme, it can analyze the execution semantics and automatically split apart the fast and slow path based on the type hints (\secref{sec:type-based-optimization}). At runtime, the JIT compiler will compile the main component and any return continuations transitively used by it into JIT code. On the other hand, implementations for the slow paths and their associated return continuations are generated as ahead-of-time (AOT)~code at build time. This separation is critical to keep the JIT code size small and the JIT compilation fast.

\begin{wrapfigure}{r}{0.36\linewidth}
    \centering
    \vspace{-1.4em}
    \includegraphics[width=\linewidth]{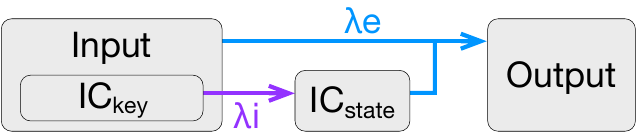}
    \vspace{-2.1em}
    \caption{
        A computation is eligible for IC iff it meets the above characterization.
        \label{fig:ic-abstraction}
    }
    \vspace{-1.4em}
\end{wrapfigure}

\paragraph{Generic Inline Caching} Inline caching can greatly speed~up object accesses, a pervasive operation in dynamic languages. However, it is hard to provide a universal object representation that fits all languages. Thus, Deegen does not understand objects, but instead provides a generic IC mechanism for users to express any IC semantics. Observe that a computation is eligible for IC if and only if it can be decomposed into an \textit{idempotent} computation $\lambda_i:\textrm{IC}_{\textrm{key}}\mapsto\textrm{IC}_{\textrm{state}}$ followed by a cheap computation $\lambda_e:\langle\textrm{Input},\textrm{IC}_{\textrm{state}}\rangle\mapsto\textrm{Output}$, as shown in \figref{fig:ic-abstraction}. In that case, we can cache the $\langle\textrm{IC}_{\textrm{key}}, \textrm{IC}_{\textrm{state}}\rangle$ we have seen, and when we see a cached $\textrm{IC}_{\textrm{key}}$ again, we can skip the idempotent $\lambda_i$ computation, and directly feed the cached $\textrm{IC}_{\textrm{state}}$ to $\lambda_e$. In Deegen's generic IC scheme, users write $\lambda_i$ and $\lambda_e$ as C++ lambdas, and $\textrm{IC}_{\textrm{state}}$ as local variables defined inside $\lambda_i$ that are captured by $\lambda_e$. This provides a clean and flexible API for users to define any IC semantics: see \secref{sec:generic-inline-caching} for detail.

\paragraph{Bytecode Variants} Dynamic languages often involve bytecodes with similar implementations, such as adding two locals vs. adding a local and a constant. Many bytecodes can also be specialized for common cases for better performance, such as a call with exactly one argument. Deegen supports both use cases with the bytecode variant system, which acts on top of the execution semantic descriptions to generate different bytecode variants for better performance and less engineering cost. The \deegenKeyword{Variant} API (e.g., see \figref{fig:specialization-language}) generates a bytecode variant from a specialization description. For example, a boxed operand can be specialized to be a constant or a local and to have a speculated or statically known type mask, and a literal operand can be specialized to have a certain value. When building a bytecode, Deegen will automatically select the most specialized variant, so variant selection is transparent to the user-written parser. Deegen's C++ nature also makes it trivial to implement functionally similar bytecodes (e.g., arithmetics) using C++ templates.  




\section{Semantic-Level Optimizations}
\label{sec:semantic-optimizations}

Deegen compiles bytecode execution semantics to LLVM IR to generate the VM. However, LLVM is designed for static languages. While it has a powerful generic optimizer, it does not support critical dynamic language optimizations such as type-based optimizations and inline caching. This section describes how Deegen implements these domain-specific optimizations on its own.

\subsection{Type-Based Optimizations}\label{sec:type-based-optimization}

To maintain single-source-of-truth, Deegen supports automatically generating optimized versions of the execution semantics based on the known and speculated type information. For example, if an operand is a constant \code{tDouble}, we can eliminate checks for whether it is a double. Similarly, if an operand is speculated to be a double, the code that handles the non-double case should be moved into a separate slow path function, or trigger an OSR exit in the future optimizing JIT. 

Understanding the boxing scheme is a prerequisite for such type-based optimizations. Deegen does not hardcode a boxing scheme, but allows users to describe their scheme by specifying:

\vspace{0.3em}

\begin{minipage}{.76\linewidth}
\begin{itemize}[leftmargin=1em]
    \item The set $\mathbb{T}$ of all \textit{base types} in the language. For example, \figref{fig:type-lattice} shows the type hierarchy in LuaJIT Remake, where the leaf nodes are the \textit{base types}, and each non-leaf node is a set of base types. The base types can be finer-grained than the language-exposed types: for example, \texttt{double} is further divided into \texttt{NaN} and \texttt{NotNaN} for better code specialization. 
    \item A list of \textit{type checkers}, each described by a tuple $\langle S, c, d, e\rangle$, where $S\subsetneq\mathbb{T}$ is a set of types and $c :\mathbb{V}\rightarrow bool$ checks if the type of a boxed value belongs to $S$. Functions $d$ and $e$ are optional: if every unboxed value whose type is in $S$ can be represented by C++ type $\mathbb{C}$, then decoder $d:\mathbb{V}\rightarrow\mathbb{C}$ unboxes a value and encoder $e:\mathbb{C}\rightarrow\mathbb{V}$ does the reverse (e.g., see \secref{appendix:boxing-scheme-description-apis}).
\end{itemize}  
\end{minipage}
\hfill
\begin{minipage}{.23\linewidth}
    \centering
    \vspace{-0.1em}
    \includegraphics[scale=\figscale]{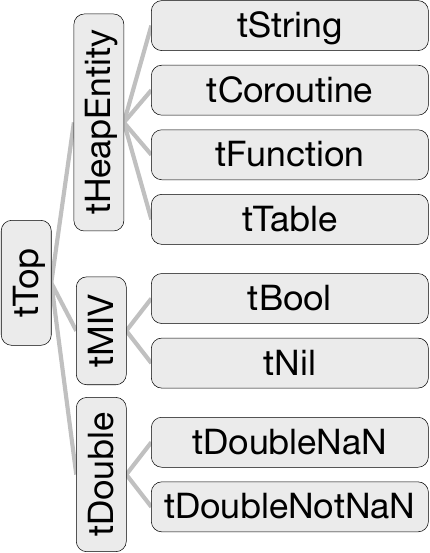}
    \vspace{-1em}
    \captionof{figure}{
        Types in LJR.
        \label{fig:type-lattice}
    }
\end{minipage}

\vspace{-0.31em}

\begin{itemize}[leftmargin=2em]
    \item A list of type-checker \textit{strength reduction rules}, each described by a tuple $\langle P, Q, r\rangle$, where $P\subsetneq\mathbb{T}$ is the precondition set of types that a boxed value $v$ is known to have, $Q\subsetneq P$ is the set of types to check, and $r:\mathbb{V}\rightarrow bool$ is the optimized function to check whether $v$'s type is in $Q$ given the precondition $P$. For example, if $v$ is known to be a \code{tHeapEntity} in \figref{fig:type-lattice} and we want to check if it is a \code{tTable}, then we can skip checking if it represents a pointer value. 
\end{itemize}

Deegen generates the implementations for the \deegenKeyword{TValue}\code{::}\deegenKeyword{Is}\code{<T>}, \deegenKeyword{As}\code{<T>} and \deegenKeyword{Create}\code{<T>} APIs (as used in \figref{fig:bytecode-add-example}), which are just wrappers around the user-specified type checkers $\langle S, c, d, e\rangle$. It also needs to pass all the user description down to the LLVM IR level in a parsable format. We solve both problems with C++ metaprogramming and macro tricks, which we omit due to space.

Another challenge is how Deegen can identify the type check APIs in LLVM IR, as LLVM might inline them. The solution is to compile with \code{-O0}
, then immediately add LLVM's \code{noinline} attribute to these API functions. After Deegen's optimizations are done, we remove the \code{noinline} attribute so the LLVM optimizer can work fully again. This controlled de-abstraction trick is widely used in Deegen so the various Deegen passes can run at the most suitable abstraction level: before any transformation, after inlining \code{AlwaysInline}, after SSA construction, after local simplification, etc.



We are now ready to explain the type-based optimization pass. At its core is an algorithm $\mathcal{A}$ that takes in a C++ function $f(v_1,\cdots,v_n)$ and a type predicate $p:\mathbb{T}^n\rightarrow bool$. $\mathcal{A}$ removes or strength-reduces the type checks in $f$ and produces an optimized function $\mathcal{A}(f, p)$, so that $f$ and $\mathcal{A}(f, p)$ are equivalent for any input $\langle v_1,\cdots,v_n\rangle\in\mathbb{V}^n$ where $p(\textrm{typeof}(v_1),\cdots,\textrm{typeof}(v_n))$ is true.

Note that $\mathcal{A}$ can be used as a black box to generate the fast and slow paths based on the known and speculated types. Again we use \texttt{add} (\figref{fig:bytecode-add-example}) as an example: if \code{lhs} is statically known to be \code{tDouble} and \code{rhs} is speculated (i.e., runtime check is needed) to be \code{tDoubleNotNaN}, we can define:

\vspace{-1.5em}

\begin{align*}
        p_1(t_1, t_2)&:=t_1\in\texttt{tDouble}\wedge t_2\in\texttt{tDoubleNotNaN}  \\[-0.3em]
    p_2(t_1, t_2)&:=t_1\in\texttt{tDouble}\wedge t_2\notin\texttt{tDoubleNotNaN}
\end{align*}

\vspace{-0.2em}

Then, the fast path can be implemented as follows: if  \code{rhs.}\deegenKeyword{Is}\code{<tDoubleNotNaN>()} is true, then call $\mathcal{A}(f, p_1)$, otherwise call the slow path function. And the slow path function is simply $\mathcal{A}(f, p_2)$.

The key idea to algorithm $\mathcal{A}$ is to compute the possible types of each operand at each basic block. Formally, let $b\in\mathbb{B}$ be a basic block in $f$, we compute $M:\langle b, i\rangle\mapsto S$ where $S\subset\mathbb{T}$ is the set of possible types of $v_i$ when execution reaches basic block $b$. The following algorithm computes $M$:
\begin{enumerate}
    \item $\forall b\in\mathbb{B}, \forall i\in[1, n]$, let $M(b, i)\leftarrow \emptyset$.
    \item $\forall\langle t_1,\cdots,t_n\rangle\in\mathbb{T}^n$ such that $p(t_1,\cdots,t_n)$ is true:
    \begin{enumerate}
        \item Replace each type check API call in $f$ to \texttt{true} or \texttt{false} assuming $v_i$ has type $t_i$.
        \item Run Sparse Conditional Constant Propagation~\cite{10.1145/103135.103136} to compute the set of reachable blocks $R\subset\mathbb{B}$ from function entry after SCCP.
        \item $\forall b\in R, \forall i\in[1,n]$, let $M(b,i)\leftarrow M(b,i)\cup \{t_i\}$. Undo all modifications to $f$.
    \end{enumerate}
\end{enumerate}

Then, for a type check $c$ in basic block $b$ that checks if the type of $v_i$ is in $S$, it is trivially true if $M(b, i)\subset S$, and trivially false if $S\cap M(b,i)=\emptyset$. If neither is the case, for each strength reduction rule $\langle P,Q,r\rangle$, $r$ can replace $c$ iff $M(b,i)\subset P$ and $S\cap M(b,i)=Q\cap M(b,i)$, and $\lnot r$ can replace $c$ iff $M(b,i)\subset P$ and $M(b,i)\setminus S=Q\cap M(b,i)$. Note that a type checker $\langle S,c,d,e\rangle$ is also a strength reduction rule $\langle\mathbb{T},S,c\rangle$, thus should also be considered.
If multiple rules qualify, we pick the cheapest based on user-provided cost estimation. We omit an argument of correctness due to space.

Note how this algorithm is tailored to the use cases of Deegen: it runs SCCP $|\mathbb{T}|^n$ times, which can take a fraction of a second. This is an unacceptable cost for a runtime analysis but totally fine at build time, and allows us to get very accurate analysis results to generate highly optimized code. 

\subsection{Generic Inline Caching}
\label{sec:generic-inline-caching}

Deegen supports \textit{generic inline caching} (\secref{sec:semantic-description-framework}), a set of APIs for users to express any IC semantics. \figref{fig:ic-abstraction} illustrated the four core concepts: $\textrm{IC}_{\textrm{key}}, \textrm{IC}_{\textrm{state}}, \lambda_i, \lambda_e$, and the relations between them.

\begin{figure}[b]
    \vspace{-1em}
    \begin{minipage}[b]{0.53\linewidth}
    \centering
    \includegraphics[height=11.6em]{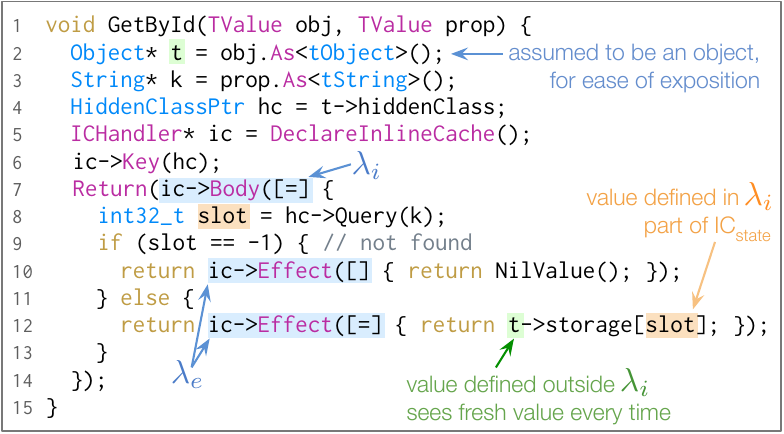}
    \vspace{-2em}
    \caption{Use of Generic IC to optimize object access.}
    \label{fig:generic-ic-api-example}
    \end{minipage}%
    \hfill
    \begin{minipage}[b]{0.4655\linewidth}
    \centering
    \includegraphics[height=11.6em]{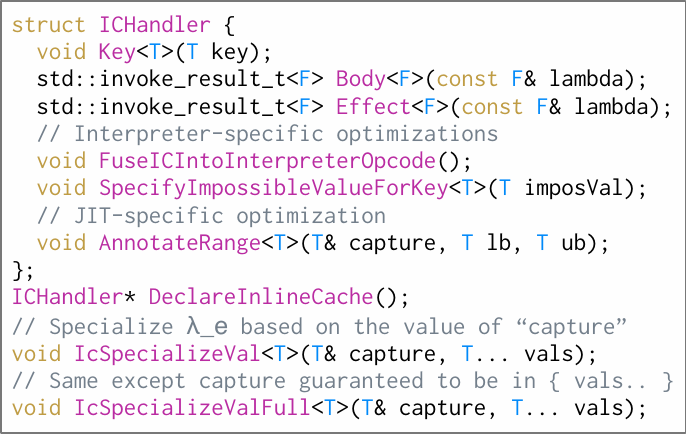}
    \vspace{-2em}
    \caption{Generic IC API definitions.}
    \label{fig:generic-ic-api-list}
    \end{minipage}%
\end{figure}

\figref{fig:generic-ic-api-list} lists the APIs for generic IC. 
\figref{fig:generic-ic-api-example} gives an example that uses these APIs to optimize \texttt{GetById}, a bytecode that returns the value of a \textit{fixed} property of an object (e.g., \texttt{item\textbf{.name}}). 
We use \deegenKeyword{DeclareInlineCache} to create an IC (line 5), and specify the $\textrm{IC}_{\textrm{key}}$ (must be an integer) to be the hidden class of the object (line 6). We then use the \deegenKeyword{Body} API to specify the idempotent computation $\lambda_i$ (line 7). Inside $\lambda_i$, we query the hidden class of \texttt{obj} to find the storage slot for property \texttt{prop} (line 8). 
This operation is indeed idempotent, because \texttt{prop} is a string constant with respect to the bytecode\footnote{Note that every instantiation of the bytecode has its own IC instance (thus ``inline'' caching), so despite that different instantiations of \codesmall{GetById} can have different \texttt{prop} values, \texttt{prop} is always constant with respect to its own IC instance.} and the hidden class query is idempotent by design.
Finally, we use the \deegenKeyword{Effect} API to specify different $\lambda_e$ computations (line 10 and 12) depending on whether the property is found.\footnote{\deegenKeyword{\footnotesize{Effect}} must be used in the form of \texttt{\textcolor{CKeywordColorDark}{return}} \codesmall{ic->}\deegenKeyword{\footnotesize{Effect}}\codesmall{([=]\{..\});} That is, no further $\lambda_i$ computation is allowed after it. We also do not support defining $\lambda_e$ as multiple lambdas or recursive $\lambda_e$, though it can be useful to support it in the future.} $\textrm{IC}_{\textrm{state}}$ is not explicitly specified: since $\lambda_i$ is idempotent, the local variables declared in $\lambda_i$ and captured by a $\lambda_e$ naturally forms the $\textrm{IC}_{\textrm{state}}$ for this $\lambda_e$. For example, the $\textrm{IC}_{\textrm{state}}$ of the $\lambda_e$ on line 10 is empty, and the $\textrm{IC}_{\textrm{state}}$ of the $\lambda_e$ on line 12 consists of variable \code{slot} but not variable \texttt{t}.

An IC entry is created when the \deegenKeyword{Effect} API is executed, which conceptually consists of the $\langle\textrm{IC}_{\textrm{key}}, \lambda_e, \textrm{IC}_{\textrm{state}}\rangle$ during this execution. For example, when a \texttt{GetById} with $\texttt{prop}=$``\texttt{x}'' is executed on object \texttt{o1} with hidden class \texttt{H}, and property \texttt{x} is found in slot 5, we will create an IC entry with $\textrm{IC}_{\textrm{key}}=\texttt{H}$, $\lambda_e=\textrm{(line 12)}$, $\textrm{IC}_{\textrm{state}}=\{\texttt{slot}=5\}$. When this bytecode is executed again on a new input, at the time the \deegenKeyword{Body} API is called, instead of executing $\lambda_i$, we check if there exists an IC entry whose $\textrm{IC}_{\textrm{key}}$ matches the new key. If so, we can skip the idempotent computation $\lambda_i$ and directly produce the output by executing $\lambda_e(\textrm{NewInput}, \textrm{IC}_{\textrm{state}})$. For example, when the \texttt{GetById} bytecode is executed again on a different object \texttt{o2} with the same hidden class \texttt{H}, we will execute the $\lambda_e$ in line 12 directly with variable $\texttt{slot}=5$ but variable \texttt{t} being the new object \texttt{o2}, thus correctly giving the value of \texttt{o2.x} without going through the expensive hidden class query.

In real-world languages, retrieving a property from an object is a lot more complex than in the \figref{fig:generic-ic-api-example} example. Different actions need to be taken depending on, e.g., whether the property is stored in the inlined or outlined storage, whether a metatable/prototype may exist, and whether the array part is homogeneous-typed or free of holes. This results in many \deegenKeyword{Effect} lambdas with only slightly different logic. The \deegenKeyword{IcSpecializeVal} API listed in \figref{fig:generic-ic-api-list} solves this problem. It takes an $\textrm{IC}_{\textrm{state}}$ variable and a list of values, and generates specialized versions of the \deegenKeyword{Effect} lambda for each listed value of the $\textrm{IC}_{\textrm{state}}$ variable. Multiple uses of \deegenKeyword{IcSpecializeVal} in the same \deegenKeyword{Effect} lambda compose as a Cartesian product. This template-like mechanism allows easy creation of many \deegenKeyword{Effect} lambdas without code duplication or loss of performance. \figref{fig:generic-ic-api-list} also lists three optional APIs for interpreter and JIT optimizations, which we describe in \secref{sec:interpreter-generator,sec:jit-generator}.

The C++ lambdas only serve as a user-friendly mechanism to express the semantics. They do not exist in the generated interpreter and JIT: in fact, there is not even a function call if the IC hits.
This is needed for performance, as a $\lambda_e$ can often be reduced to a few instructions, so the overhead of even a single extra instruction matters. We must desugar the nice APIs and remove all overhead.

The Deegen frontend performs the initial desugaring. It analyzes the LLVM IR to recover all the IC information ($\textrm{IC}_{\textrm{key}}, \textrm{IC}_{\textrm{state}}, \lambda_i, \lambda_e$), lowers syntactic sugars (e.g., \deegenKeyword{IcSpecializeVal}), rewrites lambdas to normal functions to remove overheads, and finally creates a magic function call that encodes all the information, so the later passes can easily access what they need for the lowering. This is not trivial, but we omit the details as it is mostly engineering. We explain how we generate highly optimized interpreter and JIT implementations from the IC semantics in \secref{sec:interpreter-generator,sec:jit-generator}.

\subsection{Other Domain-Specific LLVM Optimizations}

We also implemented a few smaller domain-specific LLVM optimizations, which we discuss below.  

\paragraph{Tag Register Optimization} Boxing scheme operations often involve large 64-bit constants, which are costly to materialize and bloat the JIT code size. 
The tag register~\cite{jscTagRegisterOptimization} technique solves this problem by  register-pinning~\cite{ghcRegisterPinning} large constants needed by the boxing scheme. We implemented this optimization by scanning the LLVM IR to look for large constants that can be represented by adding a 8-bit offset value to a tag register, and replace them accordingly.

\paragraph{Call IC Check Hoisting} A common pattern when making a guest language function call is
\begin{center}
    \texttt{\CKeywordDark{if} (callee.\deegenKeyword{Is}<\CType{tFunction}>()) \{  \deegenKeyword{MakeCall}(callee, ...); \} \CKeywordDark{else} \{ ... \}}
\end{center}
If the call IC hits, the callee must be a function (since it has been checked when the IC is created), so the \texttt{if}-check is redundant. Thus, for the above code pattern, we can hoist the call IC check above the \texttt{if}-check to reduce a type-check if the call IC hits. We call this \textit{call IC check hoisting}.

\paragraph{NaN-related Optimization} LuaJIT Remake uses NaN-boxing~\cite{luajitNaNBoxing}, which boxes non-double values by storing them as NaN-payloads~\cite{ieee754nan}. This makes testing whether a double is NaN a common operation. LLVM often fails to generate optimal code, likely because NaN-testing is very rare elsewhere. We observed two missed optimizations: (1) Given two doubles, checking that neither is NaN can be done by one \texttt{fcmp} IR node. (2) If (1) is followed immediately by an ordered comparison, one can reuse the flags from the previous comparison in x86-64 since \texttt{ucomisd} sets the parity, sign and zero flags all at once. LLVM backend will only do this fusion if the IR is in a specific pattern (out of several equivalent patterns). We implemented an IR rewrite to fix both cases.

\section{Interpreter Design and Generation}
\label{sec:interpreter-generator}

This section explains how we generate a highly-optimized interpreter from the bytecode semantics. 

\begin{figure}[b]
    \vspace{-0.5em}
    \begin{minipage}{0.67\linewidth}
    \centering
    \includegraphics[width=\linewidth]{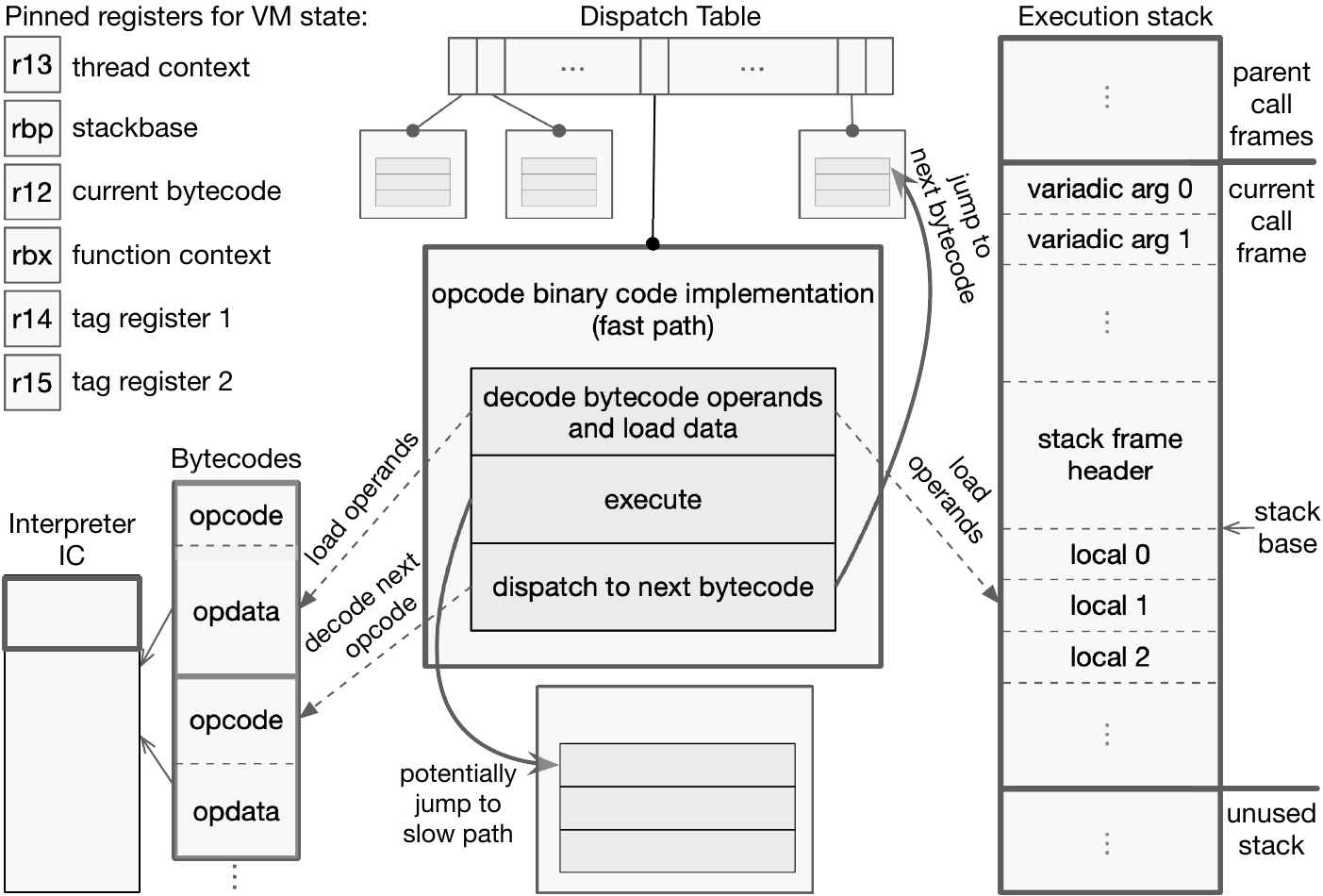}
    \vspace{-1.49em}
    \caption{Architecture of the generated interpreter.}
    \label{fig:interpreter}
    \end{minipage}
    \hfill
    \begin{minipage}{0.31\linewidth}
    \centering
    \includegraphics[width=\linewidth]{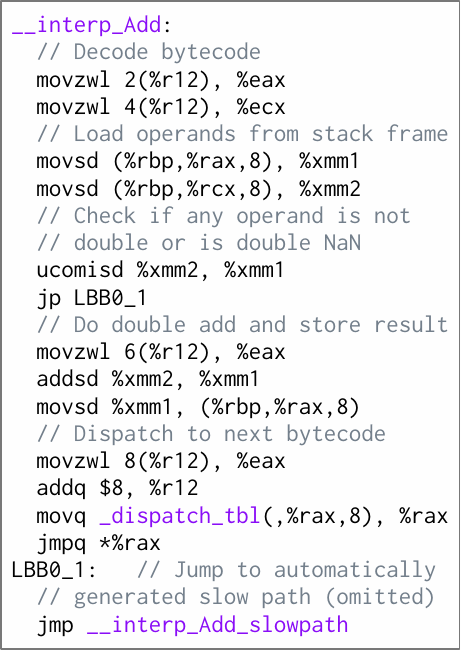}
    \vspace{-1.9em}
    \caption{Real disassembly of the\\ \texttt{Add} bytecode in LJR interpreter.}
    \label{fig:interpreter-add-assembly}
    \end{minipage}%
\end{figure}

\subsection{Interpreter Design}
\label{sec:interpreter-design}

Deegen generates a continuation-passing~\cite{steele1977} interpreter with register-pinning~\cite{ghcRegisterPinning}, as shown in \figref{fig:interpreter}. 
\textit{Continuation-passing} means that each bytecode is implemented by a stand-alone machine function, and control transfer is implemented by tail calls. \textit{Register-pinning} means that important VM states are always passed around the interpreter in fixed CPU registers.

At implementation level, we use LLVM's GHC calling convention (GHCcc)~\cite{ghcConvention} to~achieve both. GHCcc supports guaranteed-tail-call, allowing us to do continuation-passing. It also has no callee-saved registers, which is a perfect match for continuation-passing. In GHCcc, all arguments are passed in registers, so taking a value as argument and passing it to the continuation in the same argument position effectively pins the value to the CPU register corresponding to that argument. 

We currently register-pin six VM states (\figref{fig:interpreter} top-left), though this can be easily changed to be user-configurable in the future. For best performance, they are pinned in the six callee-saved registers of the C calling convention, so the interpreter does not need to save them across C calls.

Some functions need to take additional states: for example, the slow path can take additional user-defined arguments, and the return continuation takes information about the return values. These states are passed using the remaining available arguments in GHCcc, thus also in registers. 

The interpreter uses a custom stack (instead of the C stack) to record invocations, which is required to support stackful coroutines. \figref{fig:interpreter} (right) illustrates the layout of one stack frame, consisting of three parts: the variadic arguments, the frame header, and the locals. The base pointer points at local 0, allowing fast access to both the locals and the information in the frame header.

The execution of a bytecode conceptually has three steps (\figref{fig:interpreter} middle). First, we decode the bytecode, and the operands are retrieved from the locals or the constant table as needed. Second, we execute the execution semantics of the bytecode specified by the user. Finally, control is transferred to another bytecode or another component of the bytecode (e.g., a slow path) by a tail call.

Inline caching (both call IC and generic IC) in the interpreter is implemented by monomorphic IC~\cite{speculationinjsc} with optional dynamic quickening~\cite{10.1145/1899661.1869633}. Since the IC is monomorphic (i.e., it only caches one entry), the maximum space needed is statically known. Thus, when building the bytecode stream, we reserve space in an interpreter IC array if the bytecode uses IC\footnote{The IC is not embedded in the bytecode but aggregated by IC definition into arrays, so GC can work through them easily.}, so no dynamic allocation happens at runtime (\figref{fig:interpreter} bottom-left). 

\subsection{Interpreter Generation}
\label{sec:interpreter-generation}

At a high level, the interpreter is generated from the bytecode semantic descriptions in four steps: 
\begin{enumerate}
    \item The Deegen frontend takes in the raw LLVM IR compiled from the C++ description, parses all the Deegen API calls, and produces a list of \textit{lowering tasks} that each lowers one bytecode component (\figref{fig:bytecode-components}) of one bytecode variant to the interpreter implementation.
    \item For each task, run the semantics optimizer (\secref{sec:semantic-optimizations}) to optimize the execution semantics.
    \item Generate the concrete interpreter implementation from the optimized semantics.
    \item Link all the generated logic back to the original LLVM module, then compile it to final library. This link-back step is needed to correctly handle linkage issues with runtime calls. 
\end{enumerate}

In step (1), we parse the LLVM IR generated by Clang to recover the bytecode specifications (\figref{fig:specialization-language}), and figure out the definition of all the bytecode variants. We then parse all uses of intra-bytecode control flow APIs to determine the bytecode components needed by each bytecode variant. Finally, we perform preliminary desugaring for complex Deegen APIs like generic IC and guest language function calls, so later passes can easily access information about these API calls. 

In step (3), we create an LLVM function with the prototype based on the register pinning scheme described in \secref{sec:interpreter-design}, so the function has access to all VM states. We then emit the function body that decodes the bytecode based on the bytecode variant definition, loads operands from the locals or the constant table, and calls the execution semantics function with the operands. Next, we mark the execution semantics function \texttt{AlwaysInline} and let LLVM inline it. Finally, we lower all Deegen APIs (e.g., \deegenKeyword{Return} and \deegenKeyword{MakeCall}) to concrete implementations (these concrete implementations usually need to access the VM states, which is why we must first apply inlining).

We will explain the Deegen API lowering process using two examples: \deegenKeyword{MakeCall} and generic IC. As explained in \secref{sec:semantic-description-framework}, the \deegenKeyword{MakeCall} family consists of many APIs. The arguments may consist of singleton values, ranges of locals, variadic arguments of the function, and variadic results of the previous bytecode, and the call may be must-tail or in-place. It is not trivial to figure out how the new frame can be built with minimal data movements in all cases. Furthermore, call IC needs to be employed for calls from the main component. It takes a lot of engineering to make all of these work and to make them efficient. Fortunately, all of this happens automatically behind the scenes: Deegen takes the work so the user can take a rest. 

\begin{wrapfigure}{r}{0.39\linewidth}
    \centering
    \vspace{-2em}
    \includegraphics[width=\linewidth]{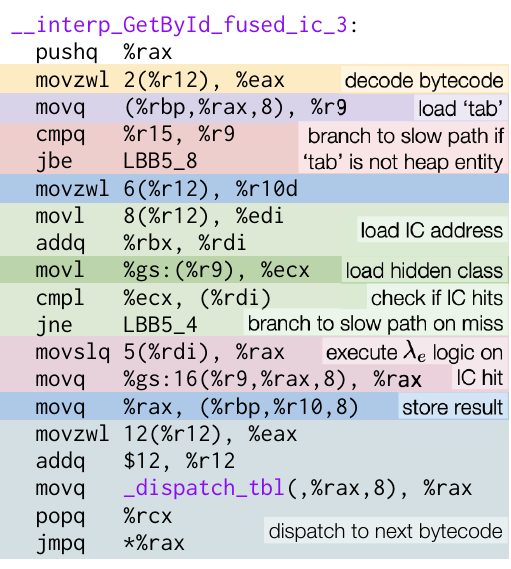}
    \vspace{-2.35em}
    \caption{
        Real disassembly of a quickened \texttt{GetById} bytecode in LJR interpreter.
        \label{fig:interpreter-getbyid-assembly}
    }
    \vspace{-1.5em}
\end{wrapfigure}

The generic IC is another performance-critical API. It is fairly straightforward to lower the IC semantics to a monomorphic interpreter IC implementation: before executing $\lambda_i$, we check if the cached IC entry exists and the $\textrm{IC}_{\textrm{key}}$ matches. If so, we can skip the $\lambda_i$, and use a switch to execute the cached $\lambda_e$ logic. Otherwise, we execute $\lambda_i$, and inside $\lambda_i$, we insert logic before each \deegenKeyword{Effect} API to populate the monomorphic IC entry accordingly. 
The simple implementation above works well, but we can do better if we have more information. \figref{fig:generic-ic-api-list} lists two optional APIs the user may use to further optimize interpreter IC performance.  \deegenKeyword{SpecifyImpossibleValueForKey} can be used to specify a constant that is never a valid $\textrm{IC}_{\textrm{key}}$. In that case, we can initialize the cached $\textrm{IC}_{\textrm{key}}$ to the impossible key, so at bytecode execution time, it is safe to not check whether an IC exists, thus saving a branch. The \deegenKeyword{FuseICIntoInterpreterOpcode} API may be used if the bytecode only employs one generic IC and the IC is not executed in a loop. In that case, for each $\lambda_e$, we generate a specialized bytecode implementation where the IC always executes this $\lambda_e$ on hit. At runtime, whenever the IC is updated, we quicken~\cite{10.1145/1899661.1869633} the bytecode to the specialized version for the cached $\lambda_e$, so we avoid an expensive indirect branch to run the correct $\lambda_e$ when the IC hits.

\figref{fig:interpreter-add-assembly} is the real disassembly for the \texttt{Add} bytecode in LJR, and \figref{fig:interpreter-getbyid-assembly} is the real disassembly for a quickened \texttt{GetById} in LJR where the property is found in the inlined storage and metatable is known to not exist. The meaning of each register in the assembly is listed in \figref{fig:interpreter} (top-left). As one can see, the code generated by Deegen rivals assembly interpreters hand-coded by experts.

\vspace{-0.1em}

\section{Baseline JIT Compiler Design and Generation}
\label{sec:jit-generator}

This section explains how we generate a highly-optimized baseline JIT from the bytecode semantics. 

\vspace{-0.1em}

\subsection{Baseline JIT Compiler Design}
\label{sec:baseline-jit-design}

\begin{figure}[b]
    \begin{minipage}[b]{\textwidth}
    \includegraphics[width=\linewidth]{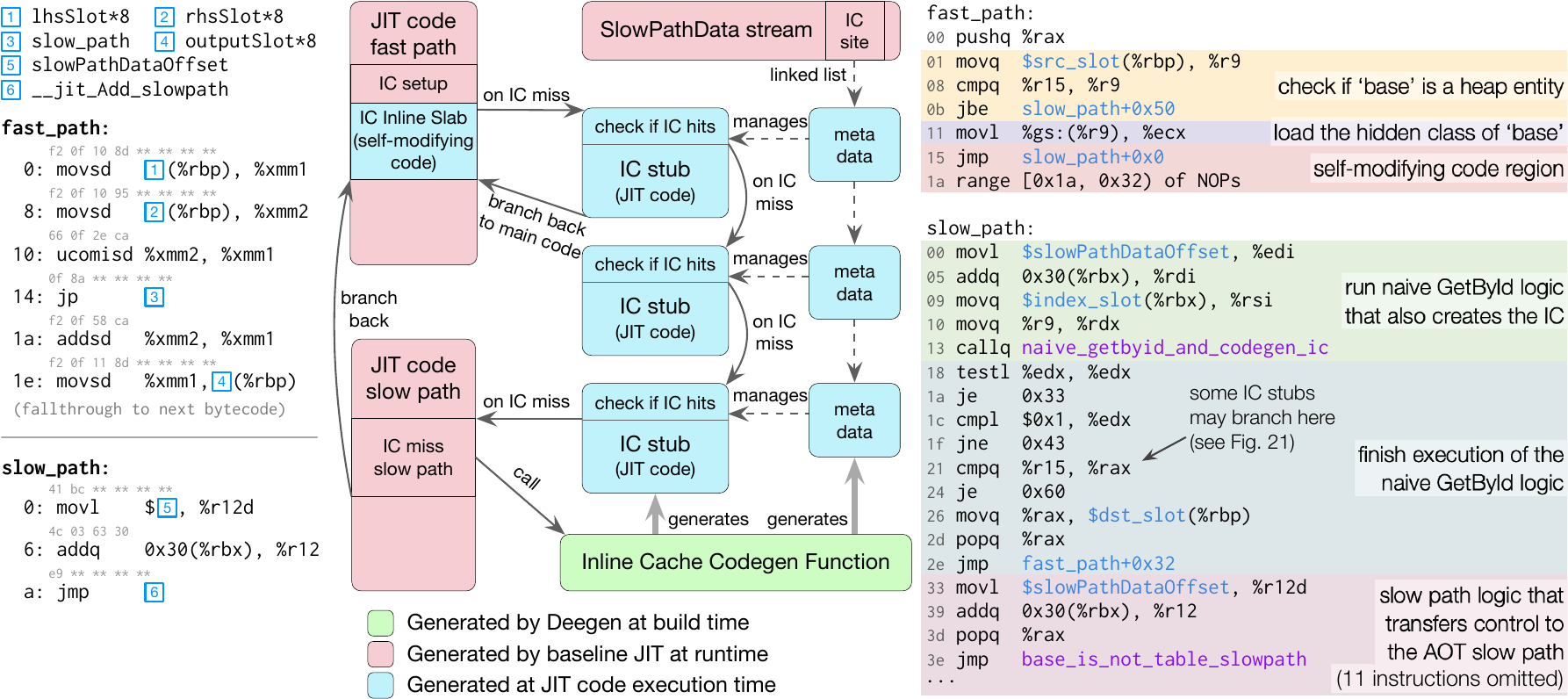}
    \end{minipage}
    \begin{minipage}[b]{\textwidth}
        \begin{minipage}[t]{0.219\linewidth}
        \begin{minipage}[t]{0.97\linewidth}
            \vspace{-4.185em}
            \caption{
                Real disassembly of the JIT code for\\ \texttt{Add} in LJR baseline JIT.\\ Generated by \figref{fig:baseline-codegen}.
                \label{fig:add-baseline-jit-code}
            }
        \end{minipage}
        \end{minipage}
        \begin{minipage}[t]{0.355\linewidth}
        \begin{minipage}[t]{0.02\linewidth}\end{minipage}
        \begin{minipage}[t]{0.97\linewidth}
            \vspace{-2em}
            \caption{
                Diagram of the polymorphic IC design in the baseline JIT.
                \label{fig:polymorphic-ic-design}
            }
        \end{minipage}
        \end{minipage}
        \begin{minipage}[t]{0.41\linewidth}
        \begin{minipage}[t]{0.02\linewidth}\end{minipage}
        \begin{minipage}[t]{0.97\linewidth}
            \vspace{-2em}
            \caption{
                Real disassembly of the \texttt{GetById} JIT code. See \figref{fig:ic-entry-baseline-jit-assembly} for IC stubs disassembly.
                \label{fig:getbyid-baseline-jit-assembly}
            }
        \end{minipage}
        \end{minipage}
    \end{minipage}
    \vspace{-2em}
\end{figure}

A baseline JIT is designed to compile fast at the cost of the quality of the generated JIT code. One can roughly think of the JIT code as the interpreter implementations for the bytecodes concatenated together, but with the optimizations listed below. To demonstrate, we use the disassembly of the~real JIT code generated by the baseline JIT for LJR's \texttt{Add} bytecode as an example (\figref{fig:add-baseline-jit-code}).
\begin{enumerate}
    \item Interpreter dispatches (indirect branches) become direct branches into the corresponding JIT code locations, or eliminated to a fallthrough altogether. As shown in \figref{fig:add-baseline-jit-code}, the interpreter dispatch is gone, and the JIT fast path simply falls through to the next bytecode.
    \item Runtime constants and their derived constant expressions are pre-computed and burnt into the generated JIT code machine instructions as literals, instead of loaded from memory or computed at runtime. Typical examples are bytecode contents and $\textrm{IC}_{\textrm{state}}$ contents in the generated IC stubs. \figref{fig:add-baseline-jit-code} shows these burnt-in constants as \textcolor{BlueBoxColor}{\fboxrule=.1em\fboxsep=.1em\fbox{1}},\ \textcolor{BlueBoxColor}{\fboxrule=.1em\fboxsep=.1em\fbox{2}}, etc. As one can see, the JIT code has no bytecode-decoding logic, and the operands to the \texttt{Add} are directly loaded from the stack frame at pre-computed offsets (e.g., \textcolor{BlueBoxColor}{\fboxrule=.1em\fboxsep=.1em\fbox{1}} is constant expression \texttt{lhsSlot$\times$8}).
    \item The cold logic is separated from the fast path by hot-cold code splitting~\cite{javascriptcoreHotColdSplitting}. As shown in \figref{fig:add-baseline-jit-code}, the rarely-executed logic that sets up the context and branches to the AOT slow path is moved to a separate JIT slow path section. This improves cache locality, and allows the JIT fast path to fall through to the next bytecode, thus removing a branch.
    \item Inline caching (both call IC and generic IC) becomes polymorphic~\cite{10.5555/646149.679193}, and leverages the capability of JIT code and self-modifying code to achieve drastically better performance than the interpreter IC. We will elaborate on this later in the section.
\end{enumerate}

\vspace{-2.5em}

\newcommand{\bigfigheight}{25.5em}
\begin{figure}[b]
    \begin{minipage}[b]{0.245\textwidth}
        \includegraphics[height=\bigfigheight]{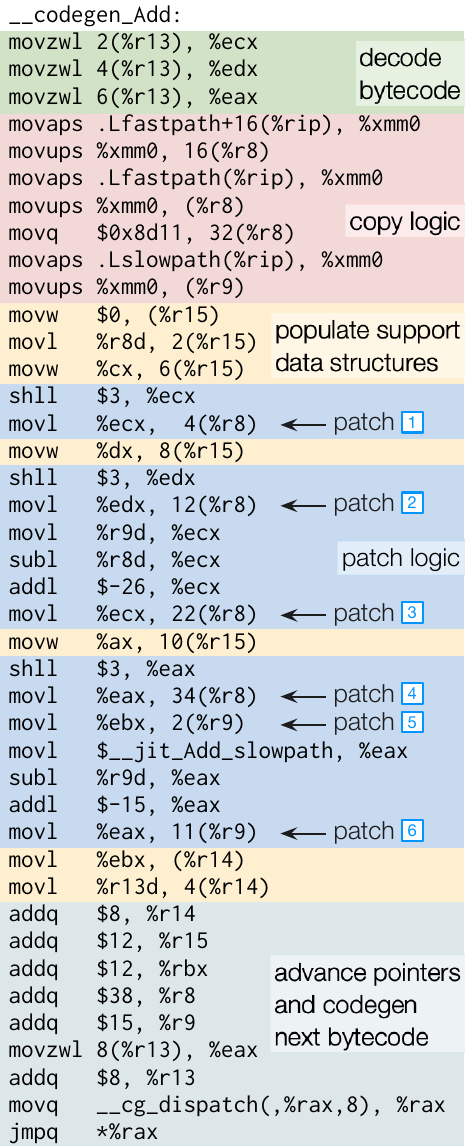}
        \vspace{-2em}
        \caption{
            Real disassembly\\ of the code-generator for\\ \texttt{Add} in LJR's baseline JIT.
            \label{fig:baseline-codegen}
        }
    \end{minipage}
    \hfill
    \begin{minipage}[b]{0.474\textwidth}
        \includegraphics[height=\bigfigheight]{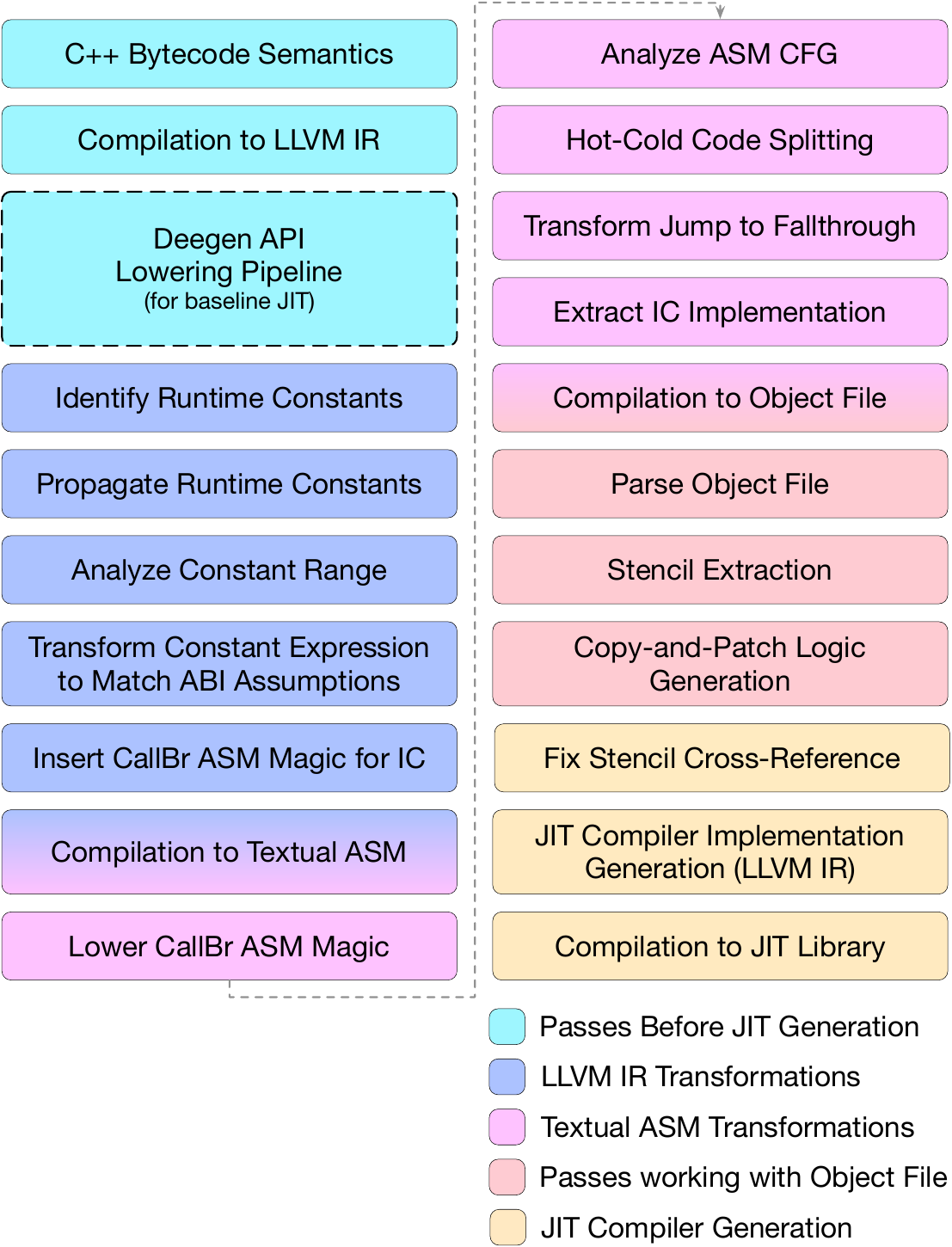}
        \vspace{-2em}
        \caption{
            The build-time pipeline that automatically generates the baseline JIT from bytecode semantics.
            \figref{fig:baseline-jit-compilation-pipeline,fig:baseline-jit-architecture,fig:polymorphic-ic-design} show the generated architecture.%
            \label{fig:baseline-jit-generation-pipeline}
        }
    \end{minipage}
    \hfill
    \begin{minipage}[b]{0.21\textwidth}
        \includegraphics[height=\bigfigheight]{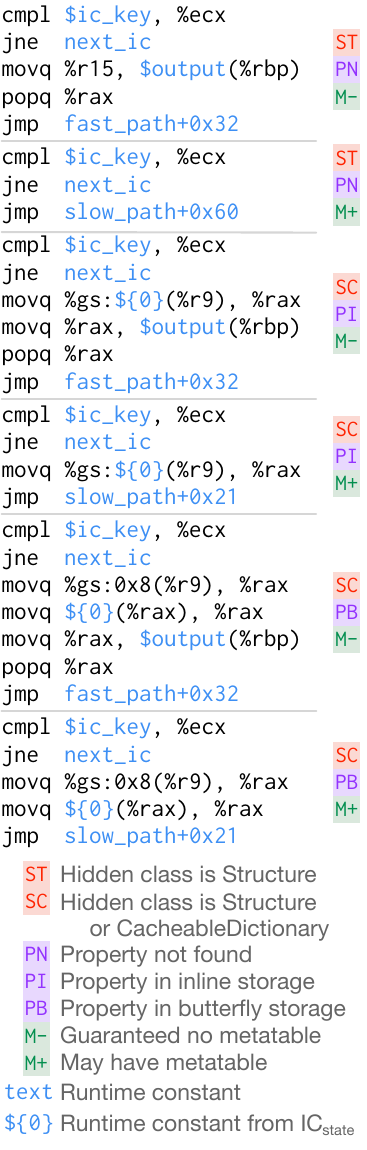}
        \vspace{-1em}
        \caption{
            Real disasm of the six kinds of IC stubs in \texttt{GetById}. 
            \label{fig:ic-entry-baseline-jit-assembly}
        }
    \end{minipage}
\end{figure}

\vspace{2em}

\begin{wrapfigure}{r}{0.40\linewidth}
    \centering
    \vspace{-0.18em}
    \includegraphics[width=\linewidth]{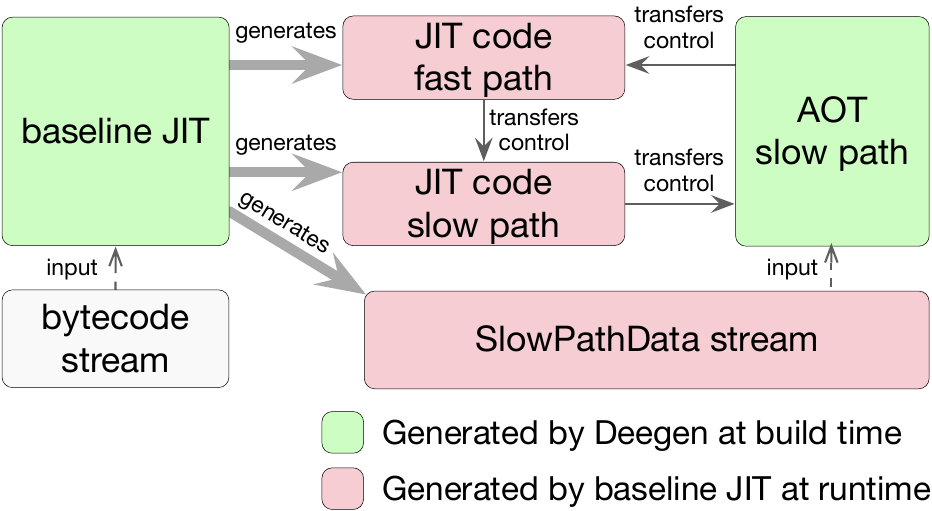}
    \vspace{-2.15em}
    \caption{
        Diagram of the Deegen-generated baseline JIT and its generated JIT code. See also \figref{fig:polymorphic-ic-design} for the polymorphic IC design.%
        \label{fig:baseline-jit-architecture}
    }
    \vspace{-2.2em}
\end{wrapfigure}

\figref{fig:baseline-jit-architecture} illustrates the components of the baseline JIT, the generated JIT code, and their interactions. The rest of this section elaborates on each of the components.

\paragraph{AOT Slow Paths} As explained in \secref{sec:semantic-description-framework}, the AOT slow paths come from a bytecode's \textit{slow path} components and their associated return continuations. Their logic in the baseline JIT tier is very similar to their interpreter counterparts, except that control must be transferred back to the JIT code at the end. This requires information about the JIT code, which is not available in the original bytecode. Deegen solves this problem by letting the baseline JIT generate a \textit{SlowPathData stream} that contains all information about the bytecode and the JIT code needed by the AOT slow paths. 

\paragraph{Baseline JIT Compiler} The baseline JIT compilation happens in four steps (\figref{fig:baseline-jit-compilation-pipeline}): compute total code and data sizes, allocate memory, generate everything, and fix branch targets. In the generation step, we scan through the bytecode once, and generate code and data for each bytecode using the Copy-and-Patch~\cite{copyandpatch} technique. 
\figref{fig:baseline-codegen} shows the code generator generated by Deegen for LJR's \texttt{Add} bytecode. As one can see, Copy-and-Patch allows us to generate code by literally copying and patching, and the logic is always branchless by design, so a modern CPU can leverage its ILP to the utmost. The code also employs continuation-passing and register-pinning: the same techniques that made our interpreter fast. With all of these, we are able to achieve end-to-end compilation throughputs measured in GB/s, which to our knowledge, is unprecedented.


\paragraph{Polymorphic IC} Polymorphic IC is the most important optimization in the baseline JIT. It is also the only high-level optimization,\footnote{This is the typical design choice used by state-of-the-art baseline JITs, e.g., those in JavaScriptCore and SpiderMonkey.} as startup delay is our top priority.  \figref{fig:polymorphic-ic-design} shows its high-level design, and \figref{fig:getbyid-baseline-jit-assembly} shows the actual disassembly of LJR's \texttt{GetById} JIT code. The JIT code contains a \textit{self-modifying code (SMC) region}, which initially only contains a jump to the IC miss slow path (the $\lambda_i$ logic). At execution time, new IC stubs (the $\lambda_e$ logic) are created by the IC miss slow path. Each IC stub is a piece of JIT code (see \figref{fig:ic-entry-baseline-jit-assembly}) that checks whether $\textrm{IC}_{\textrm{key}}$ hits, and branches to the next IC stub on miss. If the IC stub is small enough, and the SMC region is not already holding an IC stub, the IC stub can sit in the SMC region: this is called the \textit{inline slab} optimization. Otherwise, an outlined stub is created to hold the JIT code. The new stub branches to the existing stub chain head on IC miss, and the branch in the SMC region is updated to branch to the new stub. As shown in \figref{fig:ic-entry-baseline-jit-assembly}, IC stubs are not functions: there is no fixed convention to communicate between the main logic and the IC stubs. Instead, the IC stubs work directly on the machine state (e.g., which register holds what value) of the main JIT code, and may directly branch back to different places in the main JIT code. This, together with the inline slab optimization, reduces the overhead of control and data transfer to a minimum. Finally, the VM needs to manage the IC stubs. Thus, as \figref{fig:polymorphic-ic-design} shows, each IC stub is accompanied by a piece of metadata, and the SlowPathData for the bytecode contains an \textit{IC site} that chains all the IC metadata into a linked list. 

\paragraph{OSR-Entry} The JIT code uses the same stack frame layout as the interpreter, and has no inter-bytecode optimizations. OSR-entry (\secref{sec:generated-vm-architecture}) is thus simply a branch to the right JIT code location.

\subsection{Baseline JIT Compiler Generation}
\label{sec:baseline-jit-generation}

\figref{fig:baseline-jit-generation-pipeline} illustrates the build-time pipeline that automatically generates the baseline JIT compiler from the bytecode semantics.
This happens in five stages (illustrated with five colors in \figref{fig:baseline-jit-generation-pipeline}):

\begin{enumerate}
    \item Generate the bytecode implementation function and lower all Deegen API to concrete implementations. This is similar to executing steps 1--3 of the interpreter generation pipeline in \secref{sec:interpreter-generation}, except that many Deegen APIs need to be lowered differently. 
    \item Identify runtime constants and their derived constant expressions in the LLVM IR, and replace them with Copy-and-Patch stencil holes (addresses of external symbols).
    \item Compile the LLVM IR to textual assembly (\texttt{.s} files), parse the assembly file, and run a series of analysis, transformation, and extraction passes on the textual assembly. The passes are mostly architecture-agnostic: we only need to know the control flow destinations of each instruction (i.e., whether an instruction is a branch and/or a barrier), and transformations are limited to instruction reordering: we never modify the instructions themselves. 
    \item Compile the transformed textual assembly to object files and parse them to generate Copy-and-Patch stencils~\cite{copyandpatch} for the main JIT code and the IC stubs. 
    \item Generate the continuation-passing style baseline JIT code generator (see \figref{fig:baseline-codegen}) that puts together all the stencils (a bytecode may have multiple stencils due to return continuations) and populates the support data (SlowPathData and others). If IC is used, we also need to generate the IC code generator and IC management logic for the call IC and the generic IC. 
\end{enumerate}

\vspace{-0.01em}

Due to page limits, we will focus on how the optimizations in \secref{sec:baseline-jit-design} are implemented. See \secref{appendix:life-of-the-add-bytecode} for a step-by-step example of the baseline JIT generation pipeline on the \texttt{Add} bytecode.

\vspace{-0.16em}

\paragraph{Burn in Runtime Constants} For a bytecode, at execution time, everything in the bytecode and in the $\textrm{IC}_{\textrm{state}}$ (once it is created) are runtime constants. By design, Deegen has control over all of them, as Deegen is responsible for decoding the bytecode and feeding $\textrm{IC}_{\textrm{state}}$ into $\lambda_e$. So instead of generating the decoding logic to fetch the value, we simply use a magic function to identify the runtime constant. We can then find all uses of these magic functions to identify all the derived expressions that are runtime constants. For example, \textcolor{BlueBoxColor}{\fboxrule=.1em\fboxsep=.1em\fbox{1}} in \figref{fig:add-baseline-jit-code} (defined as \texttt{lhsSlot}$\times$8) arises from the LLVM \texttt{getelementptr} node that computes the address of local \texttt{lhsSlot}.

There is one complication: in Copy-and-Patch, runtime constants are represented by addresses of external symbols, which in x86-64 are assumed by the ABI to lie in the range $[1, 2^{31}-2^{24})$~\cite{systemvABI}. This assumption must be honored for correctness: for example, the upper bound $2^{31}-2^{24}$ allows LLVM to represent the value with a 32-bit signed literal in the machine instruction (see \figref{fig:add-baseline-jit-code}), so if the expression overflows \texttt{int32}, the instruction would be incorrect. Therefore, we must \textit{statically prove} that all runtime constant expressions will fit in the assumed range when evaluated at codegen time. To do this, we must know the possible range of each runtime constant. 
For $\textrm{IC}_{\textrm{state}}$, information from the high-level design is often needed to narrow down the range (e.g., that \texttt{slot} is a small non-negative integer in \figref{fig:generic-ic-api-example}), so Deegen provides the \deegenKeyword{AnnotateRange} API (\figref{fig:generic-ic-api-list}) for users to annotate the possible ranges. If the proven range of an expression does not fit the assumption, we have two choices: if the range length fits the assumption but the endpoints do not, we can add an adjustment value to the expression to adjust it into the assumed range, and subtract that value in LLVM IR. LLVM is smart enough so doing so will not affect final code quality. Otherwise, we give up and do the evaluation at runtime, but this case rarely happens in practice.
Finally, at code generation time, we replay the expressions using the actual runtime values in the bytecode or $\textrm{IC}_{\textrm{state}}$, and patch the stencil holes accordingly: see \figref{fig:baseline-codegen} for an example. 

\vspace{-0.02em}

\paragraph{Hot-Cold Code Splitting and Jump-to-Fallthrough} We use textual assembly transforms (\figref{fig:baseline-jit-generation-pipeline}) to implement hot-cold splitting and jump-to-fallthrough optimizations.
We parse the assembly text and use debug information as a hack to map assembly lines back to LLVM basic blocks. We then use LLVM's block frequency analysis to identify cold assembly lines, and reorder the assembly to move these cold code to a separate \texttt{text} section. Finally, if the fast path can branch to the next bytecode, we attempt to reorder instructions to move the branch to the last instruction and eliminate it. 

\paragraph{Polymorphic IC} To implement polymorphic IC, we must generate a function that can dynamically patch itself to append a dynamic chain of parametrizable code stubs (\figref{fig:polymorphic-ic-design}). Moreover, the code stubs work directly on the existing machine state, and they may branch back to other places in the function to continue execution (\figref{fig:ic-entry-baseline-jit-assembly}). LLVM is clearly not designed for such use cases.

\begin{wrapfigure}{r}{0.28\linewidth}
    \centering
    \vspace{-1.0em}
    \includegraphics[width=\linewidth]{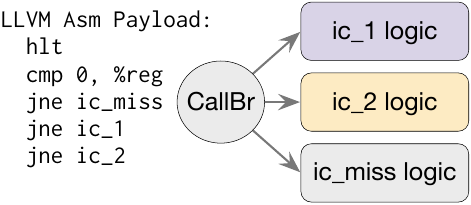}
    \vspace{-2.0em}
    \caption{
        Model IC with CallBr.
        \label{fig:ic-callbr}
    }
    \vspace{-1.2em}
\end{wrapfigure}

The key observation is that the dynamic IC check chain can be viewed as a black box that takes in $\textrm{IC}_{\textrm{key}}$ and outputs a branch target. After this abstraction, the function semantics becomes: we execute the logic before the IC, then execute this black box to select the IC that hits (or the IC miss handler), and finally transfer control to the logic selected by the black box. Interestingly, the GCC ASM-goto extension~\cite{gccAsmGoto} provides \textit{exactly} the semantics we want for the black box: an \textit{opaque} piece of logic that takes in certain inputs and redirects control to somewhere else in the function (opacity is critical since LLVM must not peek into it). In LLVM, ASM-goto is supported by a special \texttt{CallBr} IR node~\cite{llvmCallBrNode}. This allows us to model the IC execution semantics inside LLVM, as illustrated in \figref{fig:ic-callbr}. We use a \texttt{CallBr} node with an assembly payload that takes in the $\textrm{IC}_{\textrm{key}}$ and the LLVM labels for the IC miss logic ($\lambda_i$) and all different IC hit logic ($\lambda_e$). The \texttt{CallBr} has no \texttt{output} or \texttt{clobber}, but branches to one of the given LLVM labels. The content of the assembly payload is by design opaque to LLVM, so we use it to carry information down to the assembly level: as shown in \figref{fig:ic-callbr}, it starts with a \texttt{hlt} so we can identify it in the assembly. Next comes the IC check logic and the list of labels for $\lambda_i$ and each $\lambda_e$, so we know which assembly label implements each IC case. 

At the textual assembly level, we identify the assembly payload by locating the \texttt{hlt}, remove the payload, and change it into a direct branch to the IC miss slow path. We then analyze the control flow graph (CFG) of the assembly. The only hard part is that one cannot determine the possible targets of an indirect branch from the assembly. To work around this, we modified a few lines of the LLVM backend to let it dump indirect branch targets as comments next to the assembly. 

\begin{wrapfigure}{r}{0.37\linewidth}
    \centering
    \vspace{-0.6em}
    \includegraphics[width=\linewidth]{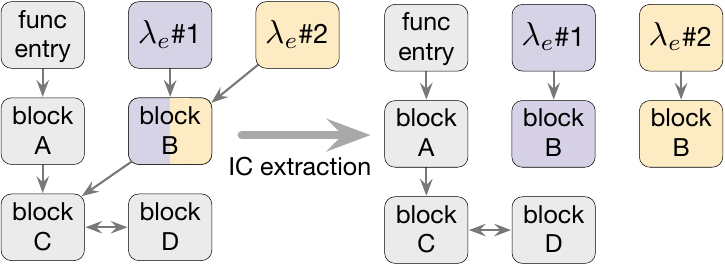}
    \vspace{-2.0em}
    \caption{
        CFG analysis and IC extraction.%
        \label{fig:cfg-analysis}
    }
    \vspace{-1.1em}
\end{wrapfigure}

Now that we have the CFG and the assembly label for the entry point of each $\lambda_e$, we can extract the implementations of the main logic and each IC stub logic (\figref{fig:cfg-analysis}): the main logic should contain all the basic blocks reachable from the function entry, and the logic for an IC stub should contain all the basic blocks reachable from its entry label that do~not belong to the main logic. For example, the IC stub for $\lambda_e$\#1 in \figref{fig:cfg-analysis} should contain its entry block and block B, but not block C or D as they are already available in the main logic. Finally, for each extracted IC stub, we reorder its basic blocks to minimize branches, put them in a separate \texttt{text} section, and compile it into a Copy-and-Patch stencil.

The inline slab optimization is just a bit more engineering. We heuristically select a good SMC region size based on the IC stub sizes. If an IC stub ends with a jump to right after the SMC region, it can be eliminated to a fallthrough, and NOPs must be padded at the end to fill the SMC region. 

Finally, we generate the Copy-and-Patch code generator logic that generates the IC stub, and generate the management logic that implements everything else in the polymorphic IC design. This is not trivial, but mostly engineering work, so we omit it due to space. 

\section{Evaluation}

To evaluate Deegen in practice, we used it to implement LuaJIT Remake (LJR), a standard-compliant VM for Lua 5.1. We targeted Lua 5.1 so we can have a meaningful comparision with LuaJIT~\cite{luajit}, the fastest Lua VM to date (which only supports Lua 5.1).  
LJR reused the parser of LuaJIT, but everything else is written from scratch. 
LJR supports all Lua 5.1 language features, but to focus on our research, we did not implement all Lua 5.1 standard library, and we did not yet implement GC. 

We stress that LJR and LuaJIT have \textit{drastically different} designs. Two main differences are: 
\begin{enumerate}
    \item LuaJIT does not support inline caching. In contrast, IC is a central optimization in LJR.
    \item LuaJIT has an interpreter{ \footnotesize\textrightarrow } tracing JIT architecture. LJR is designed to have an interpreter{ \footnotesize\textrightarrow } baseline method JIT{ \footnotesize\textrightarrow } optimizing JIT architecture, but the optimizing JIT is future work.
\end{enumerate}

Due to the drastic design differences, merely comparing the performance of LJR and LuaJIT is insufficient to derive conclusions about Deegen. For example, LJR outperforms LuaJIT in interpreter performance, but that does \textit{not} necessarily imply that the assembly generated by Deegen has better quality than what \citet{luajit21} hand-wrote, as the high-level designs are different. In fact, the assembly of both LJR and LuaJIT are close-to-optimal under \textit{their respective designs} (\secref{sec:eval-generated-code-quality}), but Deegen supports high-level optimizations (mainly IC) that lead to its better performance.

As such, we evaluate Deegen as follows: \secref{sec:eval-user-engineering-cost} assesses the engineering cost of using Deegen to describe the bytecode semantics in LJR. \secref{sec:eval-generated-code-quality} assesses the quality of the code generated by Deegen. And \secref{sec:eval-performance-evaluation} evaluates the performance of LJR on a variety of benchmarks, to provide direct evidence that Deegen can support real-world languages efficiently, and to corroborate \secref{sec:eval-generated-code-quality} on the capability of Deegen by putting LJR in the context of the state of the art.


\subsection{Engineering Cost for Users}\label{sec:eval-user-engineering-cost}

\begin{wrapfigure}{r}{0.54\linewidth}
\footnotesize
\vspace{-1.2em}
\begin{tabular}{r!{\color{gray}\vrule}c!{\color{gray}\vrule}c!{\color{gray}\vrule}c}
\arrayrulecolor{gray}
                & LuaJIT Remake                                                  & LuaJIT                                                               & PUC Lua \\ \hline
\#LLOC          & 1500                                                           & 4500 (for x64)                                                       & 900     \\ \hline
\#Bytecodes     & 42 def $\rightarrow$ 255                                                            & 86                                                                   & 37      \\ \hline
\#LLOC/Bc & 36/def $\rightarrow$ 6                                                              & 52                                                                   & 24      \\ \hline
Language        & \renewcommand{\arraystretch}{0.8}\begin{tabular}[c]{@{}c@{}}\vspace{-0.8em}\\C++ with \\ Deegen APIs\end{tabular} & \renewcommand{\arraystretch}{0.75}\begin{tabular}[c]{@{}c@{}}\vspace{-0.8em}\\Assembly with\\ DynASM macros\end{tabular} & C       \\ 

\end{tabular}
\vspace{-1.3em}
    \caption{
        Interpreter logical lines of code (LLOC) comparison.
        \label{tab:interpreter-lloc}
    }
    \vspace{-3.1em}
\end{wrapfigure}

We evaluate the engineering cost for users to describe their bytecode semantics using Deegen. \figref{tab:interpreter-lloc} lists the number of logical lines of code (LLOC\footnote{We use the typical LLOC measurement rule for C/C++~\cite{wikipediaLogicalSloc} where each semicolon counts as one logical line.}) used to describe the bytecode semantics in LJR, as well as the \#LLOC used to implement the interpreter in LuaJIT and in the official PUC Lua. Thanks to Deegen's \deegenKeyword{Variant} API, LJR only has 42 bytecode definitions (half of LuaJIT and comparable to PUC Lua), but specializes them into 255 bytecodes for better performance. Moreover, our semantic descriptions are in C++. Compared with LuaJIT's hand-coded x86-64 assembly, our approach has far fewer lines of code, lower engineering barrier, better maintainability, better portability, and the ability to automatically get a JIT for free. PUC Lua's interpreter has 600 fewer LLOC than LJR, but it also has few optimizations, and the line difference is small compared with the size of the whole VM (e.g., the PUC Lua parser has 1300 LLOC). 
As such, we conclude that Deegen allows users to implement a language at a low engineering cost similar to writing a naive interpreter.

\subsection{Qualitative Assessment of Generated Code Quality}\label{sec:eval-generated-code-quality}

As stated earlier, the performance of LJR is not sufficient to justify the quality of the code generated by Deegen. We therefore manually assessed the code quality of the interpreter and baseline JIT compiler generated by Deegen, and the quality of the JIT code generated by the baseline JIT. Many examples are already given in  \figref{fig:interpreter-add-assembly,fig:interpreter-getbyid-assembly,fig:baseline-codegen,fig:add-baseline-jit-code,fig:getbyid-baseline-jit-assembly,fig:ic-entry-baseline-jit-assembly}, but interested readers may also find the full disassembly of all LJR bytecodes in the artifact for their own assessment.

We used our best judgement, reference code from state-of-the-art VMs, and the uiCA latency predictor~\cite{uica} to look for improvements in the Deegen-generated code. For the interpreter, we used LuaJIT and JavaScriptCore's hand-coded-in-assembly interpreters for reference. For the JIT code, we used JavaScriptCore's baseline JIT MacroAssembler logic for reference. 

For the baseline JIT compiler, we are unable to find any improvement. This is not surprising, since the logic is branchless by design and modern compilers like LLVM excel at branchless code.


For the interpreter and the JIT code, we are unable to find improvements in the majority of them. However, for a few complex bytecodes, we are able to identify four areas of minor improvements:
\begin{enumerate}
    \item Stack operations (e.g., \texttt{push}, \texttt{rsp} adjustments) that can be sinked to the slow path. 
    \item Unwanted hoisting of simple instructions (mainly \texttt{mov} between registers) to the fast path.
    \item Sub-optimal tail duplication and tail merge decisions when a lot of tail blocks exist. 
    \item Register shuffling to fulfill C calling convention when making a runtime call.
\end{enumerate}
The former three can be attributed to minor LLVM deficiencies. The last is an inherent limitation of calling conventions, but can be fixed by introducing flexible calling conventions to LLVM. 

Nevertheless, most of the code is optimal to our eyes. Additionally, all four issues above can be characterized as having one simple extra instruction (e.g., \texttt{push}, \texttt{mov}, \texttt{jmp}). Modern CPUs excel at instruction-level parallelism, so these issues are unlikely to cause noticeable slowdown. 

As such, we claim that the code quality of the interpreter and baseline JIT generated by Deegen rivals existing state-of-the-art interpreters and baseline JITs hand-written in assembly by experts.

\subsection{Performance Evaluation}\label{sec:eval-performance-evaluation}

We evaluate LuaJIT Remake, LuaJIT 2.1~\cite{luajit21}, and PUC Lua~\cite{puclua51} on 44 benchmarks from AWFY~\cite{awfy}, CLBG~\cite{clbg}, LuaJIT-Bench~\cite{luajitbench} and LuaBench~\cite{luabench}. Since GC is not implemented in LJR, we turned off GC in LuaJIT and PUC Lua as well. Due to the large design difference between LJR and LuaJIT, we also labeled the benchmarks based on their main workload types, to shed light on where the performance differences come from.

We ran all the experiments on a laptop with an Intel i7-12700H CPU with turbo boost on. The OS is Ubuntu 22.04. All experiments are single-threaded, and task-pinned to a fixed P-core to remove noise due to core discrepancy and OS scheduling. We report the average across three runs. 

\begin{figure}
    \begin{minipage}[b]{\textwidth}
        \includegraphics[width=\textwidth]{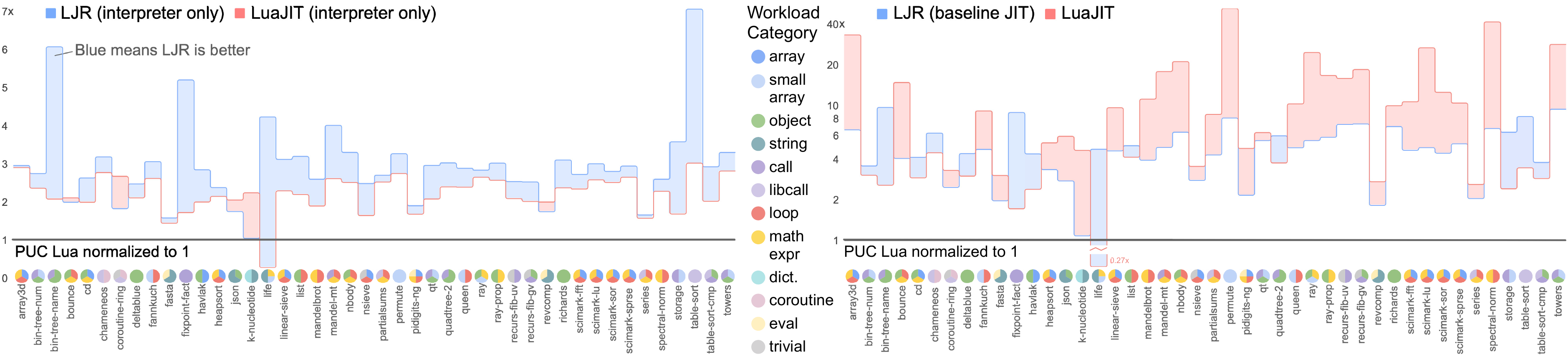}
    \end{minipage}
    
    \vspace{-1em}
    
    \begin{minipage}[b]{\textwidth}
        \begin{minipage}[b]{0.005\textwidth}\end{minipage}\hfill
        \begin{minipage}[b]{0.49\textwidth}
            \caption{Performance comparison between LJR,\\ LuaJIT and PUC Lua (interpreter-only mode).}
            \label{fig:interpreter-perf-comparison}
        \end{minipage}
        \hfill
        \begin{minipage}[b]{0.06\textwidth}\end{minipage}\hfill
        \begin{minipage}[b]{0.45\textwidth}
            \caption{Performance comparison between LJR,\\ LuaJIT and PUC Lua (JIT-enabled mode).}
            \label{fig:jit-perf-comparison}
        \end{minipage}
    \end{minipage}

    \vspace{-0.7em}
    
\end{figure}

\paragraph{Interpreter Performance} \figref{fig:interpreter-perf-comparison} shows the interpreter-only performance of the three systems, with PUC Lua normalized to 1. On average, LJR's interpreter outperforms LuaJIT's interpreter by 31\%, and outperforms PUC Lua by 179\%. LuaJIT has a different string implementation from the rest, leading to 3 cases where it outperforms LJR and one case (life) where its performance plummets.

\paragraph{Baseline JIT Startup Delay and Memory Cost} The top priority of the baseline JIT is to compile fast. Averaged across the 44 benchmarks, the baseline JIT in LJR can compile 19.1 million Lua bytecodes per second, generating machine code at 1.62GiB/s. As such, the startup delay is practically negligible. On average, each bytecode is translated to 91 bytes of machine code. Based on our assessment of the JIT code disassembly (\secref{sec:eval-generated-code-quality}), the room for code size improvement exists but is small.

\paragraph{JIT Performance} Generating an optimizing JIT is future work. Thus, LJR currently only has a baseline JIT, which \textit{by design} prioritizes compilation speed over generated JIT code quality, and is not designed to compete with an optimizing JIT (such as the tracing JIT in LuaJIT) in terms of peak throughput. Despite not being an apple-to-apple comparison, \figref{fig:jit-perf-comparison} shows the performance of LJR's baseline JIT and LuaJIT's optimizing tracing JIT, normalized against the performance of the PUC Lua interpreter. LJR is on average 33\% slower than LuaJIT, and 360\% faster than PUC Lua. As one can see, the majority of the cases where LJR loses to LuaJIT are number-crunching workloads (loops of mathematical expressions), which is expected due to our lack of the optimizing JIT tier. Nevertheless, the only high-level optimization in our baseline JIT, inline caching, is not supported by LuaJIT. As a result, LJR is already faster than LuaJIT on 13 of the 44 benchmarks.

\section{Related Work}

We compare Deegen with work in four areas: interpreter generators, generic JIT compilers via meta-tracing and the first Futamura projection, LLVM-based JIT compilers, and compiler generators.


\subsection{Interpreter Generators}

Interpreter generators can be traced at least back to \citet{10.1145/582153.582178}, who generated a Pascal interpreter from formal denotational language semantics, though it was too slow to use in practice. \texttt{vmgen}~\cite{10.1002/spe.434} and its successor \texttt{Tiger}~\cite{tigerInterpreterGenerator} are possibly the first practical interpreter generators. They take bytecode implementations in a light DSL on top of C and support optimizations such as direct-threading, bytecode specialization, superinstruction, top-of-stack caching, and code replication. TruffleDSL~\cite{truffleDSL} targets dynamic languages by generating a type-based self-optimizing AST interpreter~\cite{truffleSelfOptimizingInterpreter}, similar to dynamic quickening~\cite{10.1145/1899661.1869633}. However, the overhead of their AST interpreter proved to be high, so they are developing a new OperationDSL~\cite{operationdsl, operationdslgithub} to generate bytecode interpreters. \texttt{eJSTK}~\cite{ejstk} and its follow-up ~\cite{10.1145/3555776.3577712} can trim and fine-tune a full JavaScript interpreter to obtain a minimal interpreter tailored to a specific application.

The most relevant work to Deegen is TruffleDSL (and its planned successor OperationDSL), as both aim to generate highly optimized interpreters for dynamic languages. The main difference is real-world performance. Deegen's interpreter significantly outperforms LuaJIT's highly-optimized hand-coded-in-assembly interpreter. In contrast, as reported by \citet{Marr2022GraalWorkshop}, Truffle's interpreters are $\sim$10x slower than even the less optimized interpreters like CPython or CRuby.


\subsection{Generic JIT Compilers}

There are two prior approaches to get a JIT-capable VM without writing a JIT: meta-tracing~\cite{BolzTereick2009TracingTM} and the first Futamura projection~\cite{futamuraPaper}. Both approaches lead to a \textit{generic} JIT compiler that can work on any guest language specification. 

Meta-tracing, pioneered by PyPy~\cite{BolzTereick2009TracingTM} and followed by GVMT~\cite{gvmt} and YK~\cite{ykjit}, traces the execution of an interpreter written in a traceable language (in contrast to traditional tracing that traces the bytecodes in the user program). The meta-tracer is thus a generic tracing-JIT that works for any interpreter written in the traceable language. PyPy traces RPython code, while GVMT and YK trace C code. Meta-tracing has good user ergonomics~\cite{convergemetatracing} and can naturally desugar complex language constructs. Nevertheless, its interpreter suffers from high tracing overhead (100$\times$-200$\times$ when tracing a loop~\cite{ykhalmet}), it lacks a baseline JIT (there are proposals~\cite{dls2020metatracing, izawa2022twolevel}, but their performance and maturity are still not enough for real-world use~\cite{pypypersonalcommunication}), the inherently-large trace (compared with traditional tracing) increases JIT startup delay~\cite{convergemetatracing}, and the tracing approach can hit degenerative cases on some workloads~\cite{mozzilaAbandonTracingJit}.

The first Futamura projection, theorized by \citet{futamuraPaper}, is a generic method-JIT that works by partially evaluating an interpreter on an interpreted program at runtime. Work to practicalize this idea dates back to ~\cite{10.1145/258366.258408, 10.1023/A:1010078412711}, but Truffle~\cite{trufflePaper} is the first to practicalize it for dynamic languages. While Truffle has a high optimizing JIT peak throughput, this comes at the cost of poor interpreter performance, high interpreter memory footprint, lack of a true baseline JIT tier,
and high JIT startup delay (\secref{sec:introduction}). Other projects in building a generic JIT through the first Futamura projection include LMS~\cite{10.1145/1868294.1868314} and HolyJIT~\cite{holyjit}, which work by partial-evaluating the Scala and Rust AST respectively.

Deegen does not use a generic JIT. Instead, Deegen is a compiler generator that statically generates specialized JITs (and interpreters). This allows us to generate the most suitable implementation for each VM tier, thus fully avoiding the drawbacks of the prior approaches above. 


\subsection{LLVM-based JIT Compilers}

LLVM's powerful code-generation infrastructure made it a popular choice to build JIT compilers, with wide success in fields such as databases~\cite{neumann2011, pelotonQueryCompilation, postgresJit, memsql} and scientific computing languages~\cite{numba, julia}. 

There were also many attempts to use LLVM in dynamic language VMs~\cite{unladenswallow,rubinius,llvmlua,sulong,jscftl}, but none have mainstream adoption today. LLVM's high startup delay and lack of support for dynamic-language-specific optimizations (e.g., inline caching) are considered as the main reasons~\cite{unladenswallowretrospective, jscb3backend}. 

Deegen does not use LLVM as a JIT. Instead, Deegen uses LLVM to statically \textit{generate} the JIT. This renders LLVM's high compilation cost irrelevant, and allows us to support domain-specific optimizations via assembly transforms, thus giving us LLVM's power without its drawbacks.






\subsection{Compiler Generators (aka., Meta-Compilers, Compiler-Compilers)}

Most of the prior work on compiler generators, e.g., Yacc~\cite{yacc}, focus on parser generation. Research on generating full compiler backends is surprisingly rare.
The most recent work on full compiler generation is the PQCC project~\cite{pqcc, pqccretrospective} in the early 1980s. PQCC focuses on low-level optimizations of C-like languages, and languages of higher abstraction levels (including dynamic languages) are explicitly not considered. To the best of our knowledge, Deegen is the first work after 40 years of silence in full compiler backend generation research.



\section{Conclusion}

We introduced Deegen, a practical and novel compiler generator for building high-performance dynamic language VMs at low engineering cost. We empirically demonstrated Deegen's capabilities via LuaJIT Remake, a standard-compliant Lua 5.1 VM powered by Deegen. 

There are still numerous opportunities to further improve Deegen. The top priority is clearly the third-tier optimizing JIT, but just to name a few more: optimizations around built-in integer types to efficiently support Lua 5.3 and Python; a dedicated compiler thread to hide JIT startup delay; persistent JIT code caching to completely remove JIT cost across runs; portability to the ARM architecture; optimizations on library functions; a better story for C-bindings and FFI, the GC, multi-threading, perhaps even a fourth-tier heavyweight optimizing JIT in the far future. With all of this future work, we envision Deegen to eventually become the infrastructure for a wide range of dynamic languages, bringing them state-of-the-art performance at low engineering cost.

\section*{Acknowledgements}

We are deeply grateful to Saam Barati for spending countless hours of his after-work time teaching us everything we want to know about JavaScriptCore. Without his help, this project would not have reached its current state. We would also like to thank all individuals who provided helpful discussions, comments, and references throughout the course of the project. This work was supported by the Stanford Agile Hardware Center.

\bibliography{paper}

\newpage

\appendix
\section{Deegen API Reference}\label{appendix:full-deegen-api-reference}

This appendix lists all the Deegen APIs, organized into five categories:
\begin{itemize}\setlength\itemsep{0.1em}
    \item The bytecode specification APIs (see \figref{fig:specialization-language} for example).
    \item The boxing scheme description APIs (see \secref{sec:type-based-optimization}).
    \item The bytecode execution semantics APIs (see \figref{fig:bytecode-add-example} for example).
    \item The bytecode builder and decoder APIs (see \figref{fig:bytecode-builder-api} for example).
    \item The builtin C++ library function definition APIs.
\end{itemize}

\subsection{Bytecode Specification APIs}

\subsubsection{Bytecode Input Operand Specification}

\

\vspace{-0.3em}
\noindent\textcolor[RGB]{220,220,220}{\rule{\linewidth}{0.2pt}}

\begin{itemize}[leftmargin=2.5em]\setlength\itemsep{0.1em}
  \item[\textcolor{Gray}{API.}]  \deegenKwdDark{Local}\texttt{(\CKeyword{const char}* name) \textcolor{CKeywordColor}{->} \CType{Operand}}
  \item[\textcolor{Gray}{Desc.}] Define a bytecode operand that is a local variable (aka., virtual register) in the stack frame. 
  \item[\textcolor{Gray}{Ex.}]  
  \deegenKwdDark{Local}\texttt{("lhs")\ \ \textcolor{Gray}{// An operand named "lhs" that is a local variable}}
  
\end{itemize}

\vspace{-0.5em}
\noindent\textcolor[RGB]{220,220,220}{\rule{\linewidth}{0.2pt}}

\begin{itemize}[leftmargin=2.5em]\setlength\itemsep{0.1em}
  \item[\textcolor{Gray}{API.}]  \deegenKwdDark{Constant}\texttt{(\CKeyword{const char}* name) \textcolor{CKeywordColor}{->} \CType{Operand}}

  \item[\textcolor{Gray}{Desc.}] Define a bytecode operand that is a constant boxed value.
  \item[\textcolor{Gray}{Ex.}]  
  \deegenKwdDark{Constant}\texttt{("lhs")\ \ \textcolor{Gray}{// A operand named "lhs" that is a constant boxed value}}
\end{itemize}

\vspace{-0.5em}
\noindent\textcolor[RGB]{220,220,220}{\rule{\linewidth}{0.2pt}}

\begin{itemize}[leftmargin=2.5em]\setlength\itemsep{0.1em}
  \item[\textcolor{Gray}{API.}]  \deegenKwdDark{Constant}\texttt{<\CType{tTypeMask}>(\CKeyword{const char}* name) \textcolor{CKeywordColor}{->} \CType{Operand}}

  \item[\textcolor{Gray}{Desc.}] Define a bytecode operand that is a constant boxed value, with statically known types.

  The bytecode builder will report a runtime error if one attempts to create a bytecode with this operand specified a value whose type does not belong to \CType{tTypeMask}.
  \item[\textcolor{Gray}{Ex.}]  
  \texttt{\textcolor{Gray}{// A constant boxed value that is statically known to have type tString}}
  
  \deegenKwdDark{Constant}\texttt{<\CType{tString}>("propName")}
\end{itemize}

\vspace{-0.5em}
\noindent\textcolor[RGB]{220,220,220}{\rule{\linewidth}{0.2pt}}

\begin{itemize}[leftmargin=2.5em]\setlength\itemsep{0.1em}
  \item[\textcolor{Gray}{API.}]  \deegenKwdDark{LocalOrConstant}\texttt{(\CKeyword{const char}* name) \textcolor{CKeywordColor}{->} \CType{Operand}}

  \item[\textcolor{Gray}{Desc.}] Define a bytecode operand that may either be a \deegenKwdDark{Local}, or be a \deegenKwdDark{Constant}.

  Each bytecode variant must specify whether this operand is \deegenKwdDark{Local} or \deegenKwdDark{Constant}, since this information is always statically known thus should be specialized for performance.
  \item[\textcolor{Gray}{Ex.}]  
  \deegenKwdDark{LocalOrConstant}\texttt{("lhs")\ \ \textcolor{Gray}{// An operand that is either a Local or a Constant}}
\end{itemize}

\vspace{-0.5em}
\noindent\textcolor[RGB]{220,220,220}{\rule{\linewidth}{0.2pt}}

\begin{itemize}[leftmargin=2.5em]\setlength\itemsep{0.1em}
  \item[\textcolor{Gray}{API.}]  \deegenKwdDark{BytecodeRangeBaseRO}\texttt{(\CKeyword{const char}* name) \textcolor{CKeywordColor}{->} \CType{Operand}}

  \item[\textcolor{Gray}{Desc.}] Define a bytecode operand that consists of a consecutive range of local variables, starting at an ordinal. The execution semantics may read, but not modify local variables in the range. 
  \item[\textcolor{Gray}{Ex.}]  
  \texttt{\textcolor{Gray}{// An operand that represents a range of local variables (read-only)}}
  
  \deegenKwdDark{BytecodeRangeBaseRO}\texttt{("range")}
\end{itemize}

\vspace{-0.5em}
\noindent\textcolor[RGB]{220,220,220}{\rule{\linewidth}{0.2pt}}

\begin{itemize}[leftmargin=2.5em]\setlength\itemsep{0.1em}
  \item[\textcolor{Gray}{API.}]  \deegenKwdDark{BytecodeRangeBaseRW}\texttt{(\CKeyword{const char}* name) \textcolor{CKeywordColor}{->} \CType{Operand}}

  \item[\textcolor{Gray}{Desc.}] Define a bytecode operand that consists of a consecutive range of local variables, starting at an ordinal. The execution semantics may read and write local variables in the range. 
  \item[\textcolor{Gray}{Ex.}]  
  
  \deegenKwdDark{BytecodeRangeBaseRW}\texttt{("range")\ \ \textcolor{Gray}{// An operand that represents a range of locals}}
\end{itemize}

\vspace{-0.5em}
\noindent\textcolor[RGB]{220,220,220}{\rule{\linewidth}{0.2pt}}

\begin{itemize}[leftmargin=2.5em]\setlength\itemsep{0.1em}
  \item[\textcolor{Gray}{API.}]  \deegenKwdDark{Literal}\texttt{<\CType{IntType}>(\CKeyword{const char}* name) \textcolor{CKeywordColor}{->} \CType{Operand}}

  \item[\textcolor{Gray}{Desc.}] Define a bytecode operand that is an integer literal value.
  \item[\textcolor{Gray}{Ex.}]  
  \deegenKwdDark{Literal}\texttt{<\CKeyword{uint16\_t}>("length")\ \ \textcolor{Gray}{// An operand that is a uint16\_t literal value}}
\end{itemize}

\subsubsection{Bytecode Core Information Specification}

\

\vspace{-0.3em}
\noindent\textcolor[RGB]{220,220,220}{\rule{\linewidth}{0.2pt}}

\begin{itemize}[leftmargin=2.5em]\setlength\itemsep{0.1em}
  \item[\textcolor{Gray}{API.}]  \deegenKwdDark{Operands}\texttt{(\CType{Operand}... operands)}

  \item[\textcolor{Gray}{Desc.}] Specify the list of input operands of a bytecode.
  \item[\textcolor{Gray}{Ex.}]  
  \texttt{\textcolor{Gray}{// A bytecode that takes a local variable "lhs" and a local variable "rhs"}}
  
  \texttt{\deegenKwdDark{Operands}(\deegenKwdDark{Local}("lhs"), \deegenKwdDark{Local}("rhs"))}

  \texttt{\textcolor{Gray}{// A bytecode that takes a range of local variables and a uint16\_t literal}}

   \texttt{\deegenKwdDark{Operands}(\deegenKwdDark{BytecodeRangeBaseRW}("range"), \deegenKwdDark{Literal}<\CKeyword{uint16\_t}>("length"))}
\end{itemize}

\vspace{-0.5em}
\noindent\textcolor[RGB]{220,220,220}{\rule{\linewidth}{0.2pt}}

\begin{itemize}[leftmargin=2.5em]\setlength\itemsep{0.1em}
  \item[\textcolor{Gray}{API.}]  \texttt{\deegenKwdDark{Result}()}

  \item[\textcolor{Gray}{Desc.}] Specify that the bytecode does not generate an output value, and will not perform a branch to another bytecode. 
\end{itemize}

\vspace{-0.5em}
\noindent\textcolor[RGB]{220,220,220}{\rule{\linewidth}{0.2pt}}

\begin{itemize}[leftmargin=2.5em]\setlength\itemsep{0.1em}
  \item[\textcolor{Gray}{API.}]  \texttt{\deegenKwdDark{Result}(\deegenKwdDark{BytecodeValue})}

  \item[\textcolor{Gray}{Desc.}] Specify that the bytecode generates an output value (a boxed value) that should be written back to the stack frame, but will not perform a branch to another bytecode. 
\end{itemize}

\vspace{-0.5em}
\noindent\textcolor[RGB]{220,220,220}{\rule{\linewidth}{0.2pt}}

\begin{itemize}[leftmargin=2.5em]\setlength\itemsep{0.1em}
  \item[\textcolor{Gray}{API.}]  \texttt{\deegenKwdDark{Result}(\deegenKwdDark{ConditionalBranch})}

  \item[\textcolor{Gray}{Desc.}] Specify that the bytecode does not generate an output value, but it may conditionally or unconditionally branch to another bytecode. 
\end{itemize}

\vspace{-0.5em}
\noindent\textcolor[RGB]{220,220,220}{\rule{\linewidth}{0.2pt}}

\begin{itemize}[leftmargin=2.5em]\setlength\itemsep{0.1em}
  \item[\textcolor{Gray}{API.}]  \texttt{\deegenKwdDark{Result}(\deegenKwdDark{BytecodeValue}, \deegenKwdDark{ConditionalBranch})}

  \item[\textcolor{Gray}{Desc.}] Specify that the bytecode generates an output value (a boxed value) that should be written back to the stack frame, and may conditionally or unconditionally branch to another bytecode. 
\end{itemize}

\vspace{-0.5em}
\noindent\textcolor[RGB]{220,220,220}{\rule{\linewidth}{0.2pt}}

\begin{itemize}[leftmargin=2.5em]\setlength\itemsep{0.1em}
  \item[\textcolor{Gray}{API.}]  \texttt{\deegenKwdDark{Implementation}(\CType{FnPtr} func)}

  \item[\textcolor{Gray}{Desc.}] Specify the main execution semantics function (see \secref{appendix:execution-semantics-func-proto}) of the bytecode. Reports a build time error if the function prototype of \texttt{func} does not match the expected prototype of the bytecode execution semantics based on the operand description.

  \item[\textcolor{Gray}{Ex.}]  
  \texttt{\CKeyword{void} AddBytecodeImpl(\deegenKwdDark{TValue} lhs, \deegenKwdDark{TValue} rhs) \{ ... \}}

  \deegenKwdDark{DEEGEN\_DEFINE\_BYTECODE}\texttt{(Add) \{ ...; \deegenKwdDark{Implementation}(AddBytecodeImpl); ...; \}}
\end{itemize}

\vspace{-0.5em}
\noindent\textcolor[RGB]{220,220,220}{\rule{\linewidth}{0.2pt}}

\begin{itemize}[leftmargin=2.5em]\setlength\itemsep{0.1em}
  \item[\textcolor{Gray}{API.}]  \texttt{\deegenKwdDark{CheckForInterpreterOsrEntry}()}

  \item[\textcolor{Gray}{Desc.}] Specify that the bytecode is worth checking for interpreter OSR-entry into the baseline JIT. This may be used by any bytecode, but is normally used by bytecodes that represent a control flow back-edge, e.g., loop jump, since the check for OSR-entry always incurs a small overhead. In the interpreter, this bytecode will check if the function has reached the hotness threshold for baseline JIT compilation. If so, we will trigger baseline JIT compilation, and OSR-enter the JIT code from where we are, so the rest of this function (and any future invocation of this function) will be executed in the baseline JIT tier.
\end{itemize}

\subsubsection{Bytecode Variant Specification}

\

\vspace{-0.3em}
\noindent\textcolor[RGB]{220,220,220}{\rule{\linewidth}{0.2pt}}

\begin{itemize}[leftmargin=2.5em]\setlength\itemsep{0.1em}
  \item[\textcolor{Gray}{API.}]  \texttt{\deegenKwdDark{Op}(\CKeyword{const char}* name) \CKeyword{->} \CType{OperandRef}}

  \item[\textcolor{Gray}{Desc.}] Returns a reference to a bytecode operand, so one can further specify specializations. Reports a build error if the operand name does not refer to an operand in the bytecode.

  \item[\textcolor{Gray}{Ex.}]  
  \texttt{\deegenKwdDark{Op}("lhs")\ \ \textcolor{Gray}{// Returns a reference to the operand named "lhs"}}

\end{itemize}

\vspace{-0.5em}
\noindent\textcolor[RGB]{220,220,220}{\rule{\linewidth}{0.2pt}}

\begin{itemize}[leftmargin=2.5em]\setlength\itemsep{0.1em}
  \item[\textcolor{Gray}{API.}]  \texttt{\CType{OperandRef}::\deegenKwdDark{IsConstant}<\CType{tTypeMask}>() \CKeyword{->} \CType{SpecializedOperandRef}}

  \item[\textcolor{Gray}{Desc.}] Specify a specialization that the bytecode operand (which must be a \deegenKwdDark{LocalOrConstant}) is a constant boxed value, optionally with a statically known type mask. 
  Reports a build error if used on an operand that is not a \deegenKwdDark{LocalOrConstant}.

  \item[\textcolor{Gray}{Ex.}]  
  \texttt{\textcolor{Gray}{// Specify that operand "lhs" is a constant boxed value that is a tDouble}}
  
  \texttt{\deegenKwdDark{Op}("lhs").\deegenKwdDark{IsConstant}<\CType{tDouble}>()}

\end{itemize}

\vspace{-0.5em}
\noindent\textcolor[RGB]{220,220,220}{\rule{\linewidth}{0.2pt}}

\begin{itemize}[leftmargin=2.5em]\setlength\itemsep{0.1em}
  \item[\textcolor{Gray}{API.}]  \texttt{\CType{OperandRef}::\deegenKwdDark{IsLocal}() \CKeyword{->} \CType{SpecializedOperandRef}}

  \item[\textcolor{Gray}{Desc.}] Specify a specialization that the bytecode operand (which must be a \deegenKwdDark{LocalOrConstant}) is a local variable.
  Reports a build error if used on an operand that is not a \deegenKwdDark{LocalOrConstant}.

  \item[\textcolor{Gray}{Ex.}]  
  \texttt{\deegenKwdDark{Op}("lhs").\deegenKwdDark{IsLocal}()\ \ \textcolor{Gray}{// Specify that operand "lhs" is a local variable}}

\end{itemize}

\vspace{-0.5em}
\noindent\textcolor[RGB]{220,220,220}{\rule{\linewidth}{0.2pt}}

\begin{itemize}[leftmargin=2.5em]\setlength\itemsep{0.1em}
  \item[\textcolor{Gray}{API.}]  \texttt{\CType{OperandRef}::\deegenKwdDark{HasValue}(\CKeyword{int64\_t} val) \CKeyword{->} \CType{SpecializedOperandRef}}

  \item[\textcolor{Gray}{Desc.}] Specify a specialization that the bytecode operand (which must be a \deegenKwdDark{Literal} operand) has the specified value.
  Reports a build error if used on an operand that is not a \deegenKwdDark{Literal}, or if the specified value does not fit in the integer range for the type of the \deegenKwdDark{Literal} operand.

  \item[\textcolor{Gray}{Ex.}]  
  \texttt{\textcolor{Gray}{// Specify a specialization that operand "length" has value 2}}
  
  \texttt{\deegenKwdDark{Op}("length").\deegenKwdDark{HasValue}(2)}

\end{itemize}

\vspace{-0.5em}
\noindent\textcolor[RGB]{220,220,220}{\rule{\linewidth}{0.2pt}}

\begin{itemize}[leftmargin=2.5em]\setlength\itemsep{0.1em}
  \item[\textcolor{Gray}{API.}]  \texttt{\CType{OperandRef}::\deegenKwdDark{HasType}<\CType{tTypeMask}>() \CKeyword{->} \CType{SpecializedOperandRef}}

  \item[\textcolor{Gray}{Desc.}] Specify that the \deegenKwdDark{Local} bytecode operand should be speculated to have the specified type mask, or the \deegenKwdDark{Constant} bytecode operand is known to have the specified type mask.

  \item[\textcolor{Gray}{Ex.}]  
  \texttt{\textcolor{Gray}{// Operand "lhs" should be speculated to have type tDoubleNotNaN}}
  
  \texttt{\deegenKwdDark{Op}("lhs").\deegenKwdDark{HasType}<\CType{tDoubleNotNaN}>()}

\end{itemize}

\vspace{-0.5em}
\noindent\textcolor[RGB]{220,220,220}{\rule{\linewidth}{0.2pt}}

\begin{itemize}[leftmargin=2.5em]\setlength\itemsep{0.1em}
  \item[\textcolor{Gray}{API.}]  \texttt{\deegenKwdDark{Variant}(\CType{SpecializedOperandRef}... specializations) \CKeyword{->} \CType{VariantRef}}

  \item[\textcolor{Gray}{Desc.}] Generate a specialized bytecode variant using the specializations specified in the argument. 

  \item[\textcolor{Gray}{Ex.}]  
  \texttt{\textcolor{Gray}{// Generate a specialized bytecode variant with "length" known to be 2}}
  
  \texttt{\deegenKwdDark{Variant}(\deegenKwdDark{Op}("length").\deegenKwdDark{HasValue}(2))}

\end{itemize}

\vspace{-0.5em}
\noindent\textcolor[RGB]{220,220,220}{\rule{\linewidth}{0.2pt}}

\begin{itemize}[leftmargin=2.5em]\setlength\itemsep{0.1em}
  \item[\textcolor{Gray}{API.}]  \texttt{\CType{VariantRef}::\deegenKwdDark{EnableTypeBasedCodeSplitting}(}
  
  \texttt{\quad\CType{SpecializedOperandRef}... specializations) \CKeyword{->} \CType{VariantRef}}

  \item[\textcolor{Gray}{Desc.}] Specify that the main execution semantics of the bytecode variant should be automatically splitted into a fast path and a slow path based on the speculation that the specified operands are likely have the specified types. All specializations must be \deegenKwdDark{HasType} specializations.

  \item[\textcolor{Gray}{Ex.}]  
  \texttt{\textcolor{Gray}{// The main execution semantics of the specified variant should be}}
  
  \texttt{\textcolor{Gray}{// splitted into a fast path and a slow path using the speculation}}
  
  \texttt{\textcolor{Gray}{// that operand "lhs" is likely to have type tDoubleNotNaN}}
  
  \texttt{\deegenKwdDark{Variant}().\deegenKwdDark{EnableTypeBasedCodeSplitting}(\deegenKwdDark{Op}("lhs").\deegenKwdDark{HasType}<\CType{tDoubleNotNaN}>())}

\end{itemize}

\subsubsection{Bytecode Declaration}

\

\vspace{-0.3em}
\noindent\textcolor[RGB]{220,220,220}{\rule{\linewidth}{0.2pt}}

\begin{itemize}[leftmargin=2.5em]\setlength\itemsep{0.1em}
  \item[\textcolor{Gray}{API.}]  \deegenKwdDark{DEEGEN\_DEFINE\_BYTECODE}\texttt{(bytecodeName) \{ ... \}}
  \item[\textcolor{Gray}{Desc.}] Define a bytecode named \texttt{bytecodeName}. 
  
  The function body should contain all specifications for the bytecode (by calling relevant Deegen APIs). It may use any C++ logic that satisfies C++20 \texttt{consteval} requirements.
  \item[\textcolor{Gray}{Ex.}] \texttt{\textcolor{Gray}{// Define a bytecode named Add}}

   \deegenKwdDark{DEEGEN\_DEFINE\_BYTECODE}\texttt{(Add) \{ ... \}}
\end{itemize}

\vspace{-0.5em}
\noindent\textcolor[RGB]{220,220,220}{\rule{\linewidth}{0.2pt}}

\begin{itemize}[leftmargin=2.5em]\setlength\itemsep{0.1em}
  \item[\textcolor{Gray}{API.}]  \deegenKwdDark{DEEGEN\_DEFINE\_BYTECODE\_TEMPLATE}\texttt{(bytecodeTplName, tpl-params...) \{ ... \}}

  \item[\textcolor{Gray}{Desc.}] Define a template of bytecodes parametrized by \texttt{tpl-params}.

  One can treat it as if it declares a function with template paramters given by \texttt{tpl-params}.
  \item[\textcolor{Gray}{Ex.}]  
  \texttt{\textcolor{Gray}{// Define a template of bytecodes parametrized by ArithKind kind}}
  
  \texttt{\CKeyword{enum class} ArithKind \{ Add, Sub, Mul, Div \};}
  
  \deegenKwdDark{DEEGEN\_DEFINE\_BYTECODE\_TEMPLATE}\texttt{(ArithOp, ArithKind kind) \{ ... \}}
\end{itemize}

\vspace{-0.5em}
\noindent\textcolor[RGB]{220,220,220}{\rule{\linewidth}{0.2pt}}

\begin{itemize}[leftmargin=2.5em]\setlength\itemsep{0.1em}
  \item[\textcolor{Gray}{API.}]  \deegenKwdDark{DEEGEN\_DEFINE\_BYTECODE\_BY\_TEMPLATE\_INSTANTIATION}\texttt{(bcName, tplName, args..);}

  \item[\textcolor{Gray}{Desc.}] Define a bytecode by instantiating a bytecode template. 

  One can treat it as if it instantiates a templated function with the given template parameters.
  \item[\textcolor{Gray}{Ex.}]  
  \texttt{\textcolor{Gray}{// Define bytecode Add by instantiating ArithOp with kind = ArithKind::Add}}
  
  \deegenKwdDark{DEEGEN\_DEFINE\_BYTECODE\_BY\_TEMPLATE\_INSTANTIATION}\texttt{(}
  
  \texttt{\ \ Add, ArithOp, ArithKind::Add);}
\end{itemize}

\vspace{-0.5em}
\noindent\textcolor[RGB]{220,220,220}{\rule{\linewidth}{0.2pt}}

\begin{itemize}[leftmargin=2.5em]\setlength\itemsep{0.1em}
  \item[\textcolor{Gray}{API.}]  \deegenKwdDark{DEEGEN\_END\_BYTECODE\_DEFINITIONS}

  \item[\textcolor{Gray}{Desc.}] A translation unit (C++ file) may define any number of bytecodes using the above-mentioned APIs. This macro must be put at the end of the file. Internally, this macro uses C++ macro and metaprogramming tricks to automatically collect all definitions in this translation unit into a \texttt{constexpr} array, so Deegen can work through each of the defined bytecodes. 
\end{itemize}

\subsubsection{Bytecode Length Constraints}\label{appendix:bytecode-length-constraints-api}

\

\vspace{-0.3em}
\noindent\textcolor[RGB]{220,220,220}{\rule{\linewidth}{0.2pt}}

\begin{itemize}[leftmargin=2.5em]\setlength\itemsep{0.1em}
  \item[\textcolor{Gray}{API.}]  \texttt{\deegenKwdDark{DEEGEN\_ADD\_BYTECODE\_SAME\_LENGTH\_CONSTRAINT}(bcName1, bcName2);}

  \item[\textcolor{Gray}{Desc.}] Specify that every variant of bytecode \texttt{bcName1} must have equal bytecode length as every variant of bytecode \texttt{bcName2}. This means that padding will be added as necessary so that all affected bytecode variants have equal length. This API allows a user-written parser to substitute a \texttt{bcName1} bytecode with a \texttt{bcName2} bytecode in the middle of a bytecode stream, without affecting the validity of the bytecode stream after the substituted bytecode. Note that different variants of the same bytecode can have different lengths as well, so it is a valid use case to specify \texttt{bcName1 == bcName2} (which makes it legal to substitute a \texttt{bcName1} bytecode with the same bytecode but different operand values).

  \item[\textcolor{Gray}{Ex.}]  
  \texttt{\textcolor{Gray}{// All variants of bytecode Add, Sub, Mul, Div should have equal length}}

  \texttt{\deegenKwdDark{DEEGEN\_ADD\_BYTECODE\_SAME\_LENGTH\_CONSTRAINT}(Add, Sub);}

  \texttt{\deegenKwdDark{DEEGEN\_ADD\_BYTECODE\_SAME\_LENGTH\_CONSTRAINT}(Add, Mul);}

  \texttt{\deegenKwdDark{DEEGEN\_ADD\_BYTECODE\_SAME\_LENGTH\_CONSTRAINT}(Add, Div);}

\end{itemize}

\subsection{Boxing Scheme Description APIs}\label{appendix:boxing-scheme-description-apis}

\

\vspace{-0.3em}
\noindent\textcolor[RGB]{220,220,220}{\rule{\linewidth}{0.2pt}}

\begin{itemize}[leftmargin=2.5em]\setlength\itemsep{0.1em}
  \item[\textcolor{Gray}{API.}]  
  \texttt{\CKeyword{struct} \deegenKwdDark{LeafType};}\hfill(1)
  
  \texttt{\CKeyword{template}<\CKeyword{typename}... \CType{Args}> \CKeyword{struct} \deegenKwdDark{tsm\_or};}\hfill(2)
  
  \texttt{\CKeyword{template}<\CKeyword{typename} \CType{Arg}> \CKeyword{struct} \deegenKwdDark{tsm\_not};}\hfill(3)
  
  \item[\textcolor{Gray}{Desc.}] Metaprogramming combinators to represent the types in the guest languages type hierarchy, and the useful type masks (set of types). (1) represents a base type in the type hierarchy. (2) takes in any number of type masks, and produces the union of these type masks. (3) takes in a type mask, and produces its complement type mask.

  \item[\textcolor{Gray}{Ex.}]  
  \texttt{\textcolor{Gray}{// A type mask that is an empty set (aka. the bottom)}} 

  \texttt{\deegenKwdDark{tsm\_or}<>}

  \texttt{\textcolor{Gray}{// A type mask that contains every possible type (aka. the top)}} 

  \texttt{\deegenKwdDark{tsm\_not}<\deegenKwdDark{tsm\_or}<>{}>}
  
  \texttt{\textcolor{Gray}{// A type mask consisting of tDoubleNotNaN and tDoubleNaN (aka. a double)}} 
  
  \texttt{\deegenKwdDark{tsm\_or}<\CType{tDoubleNotNaN}, \CType{tDoubleNaN}>}

  \texttt{\textcolor{Gray}{// A type mask consisting of all types but tDoubleNotNaN and tDoubleNaN}}

  \texttt{\deegenKwdDark{tsm\_not}<\deegenKwdDark{tsm\_or}<\CType{tDoubleNotNaN}, \CType{tDoubleNaN}>{}>}
\end{itemize}

\vspace{-0.5em}
\noindent\textcolor[RGB]{220,220,220}{\rule{\linewidth}{0.2pt}}

\begin{itemize}[leftmargin=2.5em]\setlength\itemsep{0.1em}
  \item[\textcolor{Gray}{API.}]  \texttt{\CKeyword{struct} \deegenKwdDark{TValue} \{ \CKeyword{uint64\_t} m\_value; \};}

  \item[\textcolor{Gray}{Desc.}] The definition of a boxed value. 
  
  Currently Deegen only supports 8-byte boxing, so the data is hardcoded to 8 bytes. 
\end{itemize}

\vspace{-0.5em}
\noindent\textcolor[RGB]{220,220,220}{\rule{\linewidth}{0.2pt}}

\begin{itemize}[leftmargin=2.5em]\setlength\itemsep{0.1em}
  \item[\textcolor{Gray}{API.}]  
  \texttt{\CKeyword{interface} \CType{tTypeMask} \{}

  \texttt{\ \ \CKeyword{using} \CType{TSMDef} = \deegenKwdDark{LeafType}}\hfill(1)
  
  \texttt{\ \ \ \ \ \ \ \ \ \ \ \ \ \ \ | \deegenKwdDark{tsm\_or}<...>}\hfill(2)

  \texttt{\ \ \ \ \ \ \ \ \ \ \ \ \ \ \ | \deegenKwdDark{tsm\_not}<...>;}\hfill(3)

  \texttt{\ \ \CKeyword{static constexpr size\_t} x\_estimatedCheckCost;}

  \texttt{\ \ \CKeyword{static bool} check(\deegenKwdDark{TValue} v);}

  \texttt{\ \ \CKeyword{static} \deegenKwdDark{TValue} encode(\CType{UnboxedTy} v);\ \ \textcolor{Gray}{// Optional}}

  \texttt{\ \ \CKeyword{static} \CType{UnboxedTy} decode(\deegenKwdDark{TValue} v);\ \ \textcolor{Gray}{// Optional}}

  \texttt{\};}
  
  \item[\textcolor{Gray}{Desc.}] User implements the above interface to define a base type or a type mask in the guest language type hierarchy. (1) specifies that \CType{tTypeMask} is a base type in the type hierarchy (so the type mask is a singleton set), while (2) and (3) specify that \CType{tTypeMask} is a set of types defined as the union or complement of other already-defined type masks. \texttt{x\_estimatedCheckCost} estimates the cost of the \texttt{check} function: a higher value means more expensive. The \texttt{check} function returns whether the type of the given boxed value is within the set of types represented by the type mask. The optional \texttt{decode} function decodes a boxed value which type is within the type mask to an unboxed value, and the \texttt{encode} function does the reverse.

  \item[\textcolor{Gray}{Ex.}]  
  \texttt{\textcolor{Gray}{// For the purpose of exposition, we show the NaN boxing scheme in LJR}} 

  \texttt{\textcolor{Gray}{// Definition for tDoubleNaN (the unboxed value is a NaN double value)}} 
  
  \texttt{\CKeyword{struct} \CType{tDoubleNaN} \{}

  \texttt{\ \ \CKeyword{using} \CType{TSMDef} = \deegenKwdDark{LeafType};}

  \texttt{\ \ \CKeyword{static constexpr size\_t} x\_estimatedCheckCost = 30;}

  \texttt{\textcolor{Gray}{\ \ // Check if a boxed value is a double NaN}}

  \texttt{\textcolor{Gray}{\ \ // Non-double values are boxed as impure NaN, so we must check if}}
  
  \texttt{\textcolor{Gray}{\ \ // the value is a pure NaN: we do this by inspecting its bit pattern}}
  
  \texttt{\ \ \CKeyword{static bool} check(\deegenKwdDark{TValue} v) \{}

  \texttt{\ \ \ \ \CKeyword{return} v.m\_value == 0x7ff8000000000000ULL \textcolor{Gray}{/* IEEE754 pure NaN */};}

  \texttt{\ \ \}}

  \texttt{\textcolor{Gray}{\ \ // There is only one pure NaN bit pattern so encoder takes nothing}}
  
  \texttt{\ \ \CKeyword{static} \deegenKwdDark{TValue} encode() \{}

  \texttt{\ \ \ \ \CKeyword{double} nan = std::numeric\_limits<\CKeyword{double}>::quiet\_NaN();}

  \texttt{\ \ \ \ \CKeyword{return} \{ .m\_value = std::bit\_cast<\CKeyword{uint64\_t}>(nan) \}; }

  \texttt{\ \ \}}

  \texttt{\textcolor{Gray}{\ \ // Decoder may only be called if v is actually a tDoubleNaN}}

  \texttt{\textcolor{Gray}{\ \ // So just bit-cast the value to double (which must yield the pure NaN)}}
  
  \texttt{\ \ \CKeyword{static} \CKeyword{double} decode(\deegenKwdDark{TValue} v) \{}

  \texttt{\ \ \ \ \CKeyword{return} std::bit\_cast<\CKeyword{double}>(v.m\_value); }

  \texttt{\ \ \}}    

  \texttt{\}};

  \item[\textcolor{Gray}{Ex.}]  \texttt{\textcolor{Gray}{// Definition for tDoubleNotNaN (the unboxed value is double but not NaN)}} 
  
  \texttt{\CKeyword{struct} \CType{tDoubleNotNaN} \{}

  \texttt{\ \ \CKeyword{using} \CType{TSMDef} = \deegenKwdDark{LeafType};}

  \texttt{\ \ \CKeyword{static constexpr size\_t} x\_estimatedCheckCost = 10;}

  \texttt{\textcolor{Gray}{\ \ // Non-double values exhibit as NaN when viewed as a double}} 

  \texttt{\textcolor{Gray}{\ \ // So checking for tDoubleNotNaN is simply checking if the value}}  
  
  \texttt{\textcolor{Gray}{\ \ // is not a NaN when viewed as a double, which is very cheap. }}
  
  \texttt{\ \ \CKeyword{static bool} check(\deegenKwdDark{TValue} v) \{} 

  \texttt{\ \ \ \ \CKeyword{return} !isnan(std::bit\_cast<\CKeyword{double}>(v.m\_value));}

  \texttt{\ \ \}}

  \texttt{\textcolor{Gray}{\ \ // Encode/Decode is simply bit-casting}}  
  
  \texttt{\ \ \CKeyword{static} \deegenKwdDark{TValue} encode(\CKeyword{double} v) \{}

  \texttt{\ \ \ \ \CKeyword{return} \{ .m\_value = std::bit\_cast<\CKeyword{uint64\_t}>(v) \}; }

  \texttt{\ \ \}}

  \texttt{\ \ \CKeyword{static} \CKeyword{double} decode(\deegenKwdDark{TValue} v) \{}

  \texttt{\ \ \ \ \CKeyword{return} std::bit\_cast<\CKeyword{double}>(v.m\_value); }

  \texttt{\ \ \}}    
    
  \texttt{\}};

  \item[\textcolor{Gray}{Ex.}]  \texttt{\textcolor{Gray}{// Definition for tDouble (the unboxed value is double)}} 
  
  \texttt{\CKeyword{struct} \CType{tDouble} \{}

  \texttt{\textcolor{Gray}{\ \ // tDouble is a type mask defined as tDoubleNotNaN or tDoubleNaN}}

  \texttt{\ \ \CKeyword{using} \CType{TSMDef} = \deegenKwdDark{tsm\_or}<\CType{tDoubleNaN}, \CType{tDoubleNotNaN}>;}

  \texttt{\ \ \CKeyword{static constexpr size\_t} x\_estimatedCheckCost = 20;}

  \texttt{\ \ \CKeyword{static bool} check(\deegenKwdDark{TValue} v) \{}

  \texttt{\textcolor{Gray}{\ \ \ \ // The bit representation of all IEEE754 double is smaller than the}} 

  \texttt{\textcolor{Gray}{\ \ \ \ // value below. And By design, all non-double values are boxed as}}

  \texttt{\textcolor{Gray}{\ \ \ \ // impure NaN whose bit representation is larger than the value below}}

  \texttt{\textcolor{Gray}{\ \ \ \ // This allows us to check for tDouble by one comparison}}

  \texttt{\ \ \ \ \CKeyword{return}  v.m\_value < 0xFFFBFFFF00000000ULL;}
  
  \texttt{\ \ \}}

  \texttt{\textcolor{Gray}{\ \ // Encode/Decode is simply bit-casting}}  
  
  \texttt{\ \ \CKeyword{static} \deegenKwdDark{TValue} encode(\CKeyword{double} v) \{}

  \texttt{\ \ \ \ \CKeyword{return} \{ .m\_value = std::bit\_cast<\CKeyword{uint64\_t}>(v) \}; }

  \texttt{\ \ \}}

  \texttt{\ \ \CKeyword{static} \CKeyword{double} decode(\deegenKwdDark{TValue} v) \{}

  \texttt{\ \ \ \ \CKeyword{return} std::bit\_cast<\CKeyword{double}>(v.m\_value); }

  \texttt{\ \ \}}    

  \texttt{\}};
\end{itemize}

\vspace{-0.5em}
\noindent\textcolor[RGB]{220,220,220}{\rule{\linewidth}{0.2pt}}

\begin{itemize}[leftmargin=2.5em]\setlength\itemsep{0.1em}
  \item[\textcolor{Gray}{API.}]  \texttt{\deegenKwdDark{DEFINE\_TYPE\_HIERARCHY}(\CType{tTypeMasks}...);}

  \item[\textcolor{Gray}{Desc.}] User defines the guest language type hierarchy by specifying all the base types and useful type masks (which are defined using the API described earlier). 

  \item[\textcolor{Gray}{Ex.}]  
  \texttt{\deegenKwdDark{DEFINE\_TYPE\_HIERARCHY}(\CType{tBottom}, \CType{tDoubleNotNaN}, \CType{tDoubleNaN}, \CType{tDouble},  \CType{tTop});}
  
\end{itemize}

\vspace{-0.5em}
\noindent\textcolor[RGB]{220,220,220}{\rule{\linewidth}{0.2pt}}

\begin{itemize}[leftmargin=2.5em]\setlength\itemsep{0.1em}
  \item[\textcolor{Gray}{API.}]  \texttt{\CKeyword{using }\CType{TypeSpeculationMask} = \CKeyword{uintXX\_t};}

  \item[\textcolor{Gray}{Desc.}] Deegen automatically determines the smallest integer type needed to hold a type mask (that is, its bit width is greater or equal to the number of base types in the guest language). 

\end{itemize}

\vspace{-0.5em}
\noindent\textcolor[RGB]{220,220,220}{\rule{\linewidth}{0.2pt}}

\begin{itemize}[leftmargin=2.5em]\setlength\itemsep{0.1em}
  \item[\textcolor{Gray}{API.}]  \texttt{\CKeyword{template}<\CKeyword{typename }\CType{T}> \CKeyword{constexpr }\CType{TypeSpeculationMask} x\_typeMaskFor;}

  \item[\textcolor{Gray}{Desc.}] Get the bitmask of a type mask from the combinator representation.

  \item[\textcolor{Gray}{Ex.}]  
  \texttt{x\_typeMaskFor<\deegenKwdDark{tsm\_or}<\CType{tDoubleNotNaN}, \CType{tDoubleNaN}>{}>}
\end{itemize}

\vspace{-0.5em}
\noindent\textcolor[RGB]{220,220,220}{\rule{\linewidth}{0.2pt}}

\begin{itemize}[leftmargin=2.5em]\setlength\itemsep{0.1em}
  \item[\textcolor{Gray}{API.}]  \texttt{\deegenKwdDark{DEFINE\_TVALUE\_TYPECHECK\_STRENGTH\_REDUCTION}(P, S, cost, v) \{ ... \}}

  \item[\textcolor{Gray}{Desc.}] Specify a type-check strength reduction rule. \texttt{\CType{TypeSpeculationMask} P} is the precondition type mask the boxed value \texttt{v} is known to have, \texttt{\CType{TypeSpeculationMask} S} is the type mask to check, and \texttt{cost} is the estimated cost of the optimized type check function.

  \item[\textcolor{Gray}{Ex.}]  
  \texttt{\deegenKwdDark{DEFINE\_TVALUE\_TYPECHECK\_STRENGTH\_REDUCTION}(}

  \texttt{\ \ x\_typeMaskFor<\CType{tHeapEntity}>, x\_typeMaskFor<\CType{tTable}>, 60 \textcolor{Gray}{/*cost*/}, v) \{}

  \texttt{\ \ \textcolor{Gray}{// We can directly dereference v since we already know it's a pointer}}

   \texttt{\ \ \CKeyword{return} ((HeapEntityHeader*)v.m\_value)->ty == HeapEntity::Table;}

   \texttt{\}}
\end{itemize}

\subsection{Bytecode Execution Semantics Description APIs}

\subsubsection{Boxed Value Operations}

\

\vspace{-0.3em}
\noindent\textcolor[RGB]{220,220,220}{\rule{\linewidth}{0.2pt}}

\begin{itemize}[leftmargin=2.5em]\setlength\itemsep{0.1em}
  \item[\textcolor{Gray}{API.}]  \texttt{\deegenKwdDark{TValue}::\deegenKwdDark{Is}<\CType{tTypeMask}>() \CKeyword{->} \CKeyword{bool}}

  \item[\textcolor{Gray}{Desc.}] Returns if the type of the given boxed value belongs to the set of types specified by \CType{tTypeMask}. 

  \item[\textcolor{Gray}{Ex.}]  
  \texttt{v.\deegenKwdDark{Is}<\CType{tDouble}>()\ \ \textcolor{Gray}{// Check if v is a tDouble}}
  
\end{itemize}

\vspace{-0.5em}
\noindent\textcolor[RGB]{220,220,220}{\rule{\linewidth}{0.2pt}}

\begin{itemize}[leftmargin=2.5em]\setlength\itemsep{0.1em}
  \item[\textcolor{Gray}{API.}]  \texttt{\deegenKwdDark{TValue}::\deegenKwdDark{As}<\CType{tTypeMask}>() \CKeyword{->} \CType{UnboxedTy}}

  \item[\textcolor{Gray}{Desc.}] Given a boxed value which type belongs to \CType{tTypeMask}, decode the boxed value to an unboxed value. Reports a build time error if the \texttt{decoder} function is not specified for the type mask. Undefined behavior if the type of the boxed value is not in \CType{tTypeMask}.

  \item[\textcolor{Gray}{Ex.}]  
  \texttt{\textcolor{Gray}{// Unbox v (must have type tDouble) to get a double value}}
  
  \texttt{\CKeyword{double }x = v.\deegenKwdDark{As}<\CType{tDouble}>();}
  
\end{itemize}

\vspace{-0.5em}
\noindent\textcolor[RGB]{220,220,220}{\rule{\linewidth}{0.2pt}}

\begin{itemize}[leftmargin=2.5em]\setlength\itemsep{0.1em}
  \item[\textcolor{Gray}{API.}]  \texttt{\CKeyword{static }\deegenKwdDark{TValue}::\deegenKwdDark{Create}<\CType{tTypeMask}>(\CType{UnboxedTy} v) \CKeyword{->} \deegenKwdDark{TValue}}

  \item[\textcolor{Gray}{Desc.}] Create a boxed value from an unboxed value of the specified type. \CType{UnboxedTy} is the unboxed value type corresponding to \CType{tTypeMask}. Reports a build time error if the \texttt{encoder} function is not specified for the type mask.

  \item[\textcolor{Gray}{Ex.}]  
  \texttt{\textcolor{Gray}{// Create a boxed value that is a tDouble with value 1.23}}
  
  \texttt{\deegenKwdDark{TValue} tv = \deegenKwdDark{TValue}::\deegenKwdDark{Create}<\CType{tDouble}>(1.23);}
  
\end{itemize}

\subsubsection{Execution Semantics Function Prototype}\label{appendix:execution-semantics-func-proto}

\

\vspace{-0.3em}
\noindent\textcolor[RGB]{220,220,220}{\rule{\linewidth}{0.2pt}}

\begin{itemize}[leftmargin=2.5em]\setlength\itemsep{0.1em}
  \item[\textcolor{Gray}{API.}]  \texttt{\CKeyword{void} implFunc(\CType{OpTy1} op1, \CType{OpTy2} op2, ...) \{ ... \}}

  \item[\textcolor{Gray}{Desc.}] The execution semantics function must have return type \CKeyword{void}. It must have one argument for each operand, in the same order as the operands are listed in the bytecode specification. The type of the operand determines the type of the argument, as shown below:
  \begin{itemize}
      \item \texttt{\deegenKwdDark{Local} \ \ \ \ \ \ \ \ \ \ \ \ \ \ \CKeyword{->} \deegenKwdDark{TValue}}
      \item \texttt{\deegenKwdDark{Constant} \ \ \ \ \ \ \ \ \ \ \ \CKeyword{->} \deegenKwdDark{TValue}}
      \item \texttt{\deegenKwdDark{LocalOrConstant} \ \ \ \ \CKeyword{->} \deegenKwdDark{TValue}}
      \item \texttt{\deegenKwdDark{BytecodeRangeBaseRO} \CKeyword{->} \CKeyword{const }\deegenKwdDark{TValue}*}
      \item \texttt{\deegenKwdDark{BytecodeRangeBaseRW} \CKeyword{->} \deegenKwdDark{TValue}*}
      \item \texttt{\deegenKwdDark{Literal}<\CType{IntType}>\ \ \ \ \CKeyword{->} \CType{IntType}}
  \end{itemize}
  The \textit{main} bytecode component and the \textit{return continuation} component must not have any additional arguments: doing so will result in a build error.
  The \textit{slow path} component may take additional arguments as user desires, but they must be put after all bytecode operands, and the size of each argument cannot exceed 8 bytes, and they must have trivial copy constructor and no-op destructor. There is also an internal limit on how many additional arguments can be passed in. A build error will be reported if the requirements are not met.
  \item[\textcolor{Gray}{Ex.}]  
  \texttt{\textcolor{Gray}{// Execution semantics of Add bytecode with two LocalOrConstant operands}}
  
  \texttt{\CKeyword{void} addImpl(\deegenKwdDark{TValue} lhs, \deegenKwdDark{TValue} rhs) \{ ... \}}
  
\end{itemize}

\subsubsection{Control Flow Transfer}\label{appendix:bytecode-semantics-control-flow-transfer-apis}

\

\vspace{-0.3em}
\noindent\textcolor[RGB]{220,220,220}{\rule{\linewidth}{0.2pt}}

\begin{itemize}[leftmargin=2.5em]\setlength\itemsep{0.1em}
  \item[\textcolor{Gray}{API.}]  
  \texttt{\deegenKwdDark{Return}() \CKeyword{->} [[no\_return]]}

  \item[\textcolor{Gray}{Desc.}] Dispatch to the next bytecode in the bytecode stream. The current bytecode must have no output operand, otherwise a build error is reported. This API call never returns.
  
\end{itemize}

\vspace{-0.5em}
\noindent\textcolor[RGB]{220,220,220}{\rule{\linewidth}{0.2pt}}

\begin{itemize}[leftmargin=2.5em]\setlength\itemsep{0.1em}
  \item[\textcolor{Gray}{API.}]  
  \texttt{\deegenKwdDark{Return}(\deegenKwdDark{TValue} val) \CKeyword{->} [[no\_return]]}

  \item[\textcolor{Gray}{Desc.}] Stores \texttt{val} to the output local variable of the current bytecode, and dispatch to the next bytecode in the bytecode stream. The current bytecode must have an output operand, otherwise a build error is reported.
  
\end{itemize}

\vspace{-0.5em}
\noindent\textcolor[RGB]{220,220,220}{\rule{\linewidth}{0.2pt}}

\begin{itemize}[leftmargin=2.5em]\setlength\itemsep{0.1em}
  \item[\textcolor{Gray}{API.}]  
  \texttt{\deegenKwdDark{ReturnAndBranch}() \CKeyword{->} [[no\_return]]}

  \item[\textcolor{Gray}{Desc.}] Dispatch to the branch destination of the current bytecode. The current bytecode must have no output operand but may perform branch, otherwise a build error is reported. 
\end{itemize}

\vspace{-0.5em}
\noindent\textcolor[RGB]{220,220,220}{\rule{\linewidth}{0.2pt}}

\begin{itemize}[leftmargin=2.5em]\setlength\itemsep{0.1em}
  \item[\textcolor{Gray}{API.}]  
  \texttt{\deegenKwdDark{ReturnAndBranch}(\deegenKwdDark{TValue} val) \CKeyword{->} [[no\_return]]}

  \item[\textcolor{Gray}{Desc.}] Stores \texttt{val} to the output local variable of the current bytecode, and dispatch to the branch destination of the current bytecode. The current bytecode must have an output operand and may perform branch, otherwise a build error is reported. 
\end{itemize}

\vspace{-0.5em}
\noindent\textcolor[RGB]{220,220,220}{\rule{\linewidth}{0.2pt}}

\begin{itemize}[leftmargin=2.5em]\setlength\itemsep{0.1em}
  \item[\textcolor{Gray}{API.}]  
  \texttt{\deegenKwdDark{ThrowError}(\deegenKwdDark{TValue} val) \CKeyword{->} [[no\_return]]}

  \item[\textcolor{Gray}{Desc.}] Throw out an exception. The exception object is a boxed value specified in the API.

  Currently, exceptions may only be caught by user library functions that are dedicated as exception catchers. This is sufficient to support Lua exception semantics, but supporting more general exception semantics may require introducing new Deegen APIs. 
\end{itemize}

\vspace{-0.5em}
\noindent\textcolor[RGB]{220,220,220}{\rule{\linewidth}{0.2pt}}

\begin{itemize}[leftmargin=2.5em]\setlength\itemsep{0.1em}
  \item[\textcolor{Gray}{API.}]  
  \texttt{\deegenKwdDark{ThrowError}(\CKeyword{const char}* msg) \CKeyword{->} [[no\_return]]}

  \item[\textcolor{Gray}{Desc.}] Throw out an exception. The exception object is a constant C string literal.
\end{itemize}

\vspace{-0.5em}
\noindent\textcolor[RGB]{220,220,220}{\rule{\linewidth}{0.2pt}}

\begin{itemize}[leftmargin=2.5em]\setlength\itemsep{0.1em}
  \item[\textcolor{Gray}{API.}]  
  \texttt{\deegenKwdDark{EnterSlowPath}<\CType{FnPtr} slowPathFunc>(\CType{Args}... args) \CKeyword{->} [[no\_return]]}

  \item[\textcolor{Gray}{Desc.}] Transfer control to a \textit{slow path} component. The types of the extra arguments passed to the slow path must agree with the definition of the extra arguments accepted by the slow path, or a build time error is reported. Deegen automatically detects all the slow path components needed for a bytecode, so there is no need to specially declare the slow path components. 

   \item[\textcolor{Gray}{Ex.}]  
   \texttt{\textcolor{Gray}{// A slow path component for Add, with no additional arguments}}

   \texttt{\CKeyword{void} AddSlowPath(\deegenKwdDark{TValue} lhs, \deegenKwdDark{TValue} rhs) \{ ... \}}

   \texttt{\textcolor{Gray}{// The main component for Add}}

   \texttt{\CKeyword{void} Add(\deegenKwdDark{TValue} lhs, \deegenKwdDark{TValue} rhs) \{ }

   \texttt{\ \ \CKeyword{if} (!lhs.\deegenKwdDark{Is}<\CType{tDouble}>() || !rhs.\deegenKwdDark{Is}<\CType{tDouble}>()) \{ }

   \texttt{\ \ \ \ \deegenKwdDark{EnterSlowPath}<AddSlowPath>();}

   \texttt{\ \ \} else ...}
   
   \texttt{\}}
   
\end{itemize}

\subsubsection{Variadic Arguments and Variadic Results}\label{appendix:var-arg-var-results}

\

\vspace{0.4em}

Guest language functions in Deegen may take variadic arguments and return a variadic number of return values. A bytecode may also generate a variadic number of result values, but such variadic results must be consumed by the immediate next bytecode. This section lists the relevant APIs.

\vspace{-0.3em}
\noindent\textcolor[RGB]{220,220,220}{\rule{\linewidth}{0.2pt}}

\begin{itemize}[leftmargin=2.5em]\setlength\itemsep{0.1em}
  \item[\textcolor{Gray}{API.}]  
  \texttt{\CKeyword{static} \deegenKwdDark{VarArgsAccessor}::\deegenKwdDark{GetPtr}() \CKeyword{->} \CKeyword{const} \deegenKwdDark{TValue}*}

  \item[\textcolor{Gray}{Desc.}] Returns a pointer pointing to the start of the variadic arguments in the current function. 
\end{itemize}

\vspace{-0.5em}
\noindent\textcolor[RGB]{220,220,220}{\rule{\linewidth}{0.2pt}}

\begin{itemize}[leftmargin=2.5em]\setlength\itemsep{0.1em}
  \item[\textcolor{Gray}{API.}]  
  \texttt{\CKeyword{static} \deegenKwdDark{VarArgsAccessor}::\deegenKwdDark{GetNum}() \CKeyword{->} \CKeyword{size\_t}}

  \item[\textcolor{Gray}{Desc.}] Returns the number of variadic arguments in the current function.
\end{itemize}

\vspace{-0.5em}
\noindent\textcolor[RGB]{220,220,220}{\rule{\linewidth}{0.2pt}}

\begin{itemize}[leftmargin=2.5em]\setlength\itemsep{0.1em}
  \item[\textcolor{Gray}{API.}]  
  \texttt{\CKeyword{static} \deegenKwdDark{VarArgsAccessor}::\deegenKwdDark{StoreAllVarArgsAsVariadicResults}()}

  \item[\textcolor{Gray}{Desc.}] Set the variadic results to be all the variadic arguments of the current function. As mentioned earlier, the variadic results must be consumed (or implicitly discarded) by the immediate next bytecode, or it is undefined behavior with no diagnostic reported. 
\end{itemize}

\vspace{-0.5em}
\noindent\textcolor[RGB]{220,220,220}{\rule{\linewidth}{0.2pt}}

\begin{itemize}[leftmargin=2.5em]\setlength\itemsep{0.1em}
  \item[\textcolor{Gray}{API.}]  
  \texttt{\deegenKwdDark{StoreReturnValuesAsVariadicResults}()}

  \item[\textcolor{Gray}{Desc.}] May only be used in the return continuation component, or a build error will be reported. Set up the variadic results to be all the return values returned from the callee.
\end{itemize}

\vspace{-0.5em}
\noindent\textcolor[RGB]{220,220,220}{\rule{\linewidth}{0.2pt}}

\begin{itemize}[leftmargin=2.5em]\setlength\itemsep{0.1em}
  \item[\textcolor{Gray}{API.}]  
  \texttt{\CKeyword{static} \deegenKwdDark{VariadicResultsAccessor}::\deegenKwdDark{GetPtr}() \CKeyword{->} \CKeyword{const} \deegenKwdDark{TValue}*}

  \item[\textcolor{Gray}{Desc.}] Returns the pointer pointing to the variadic results. The most recently executed bytecode must have set up the variadic results, or it is undefined behavior with no diagnostic reported. 
\end{itemize}

\vspace{-0.5em}
\noindent\textcolor[RGB]{220,220,220}{\rule{\linewidth}{0.2pt}}

\begin{itemize}[leftmargin=2.5em]\setlength\itemsep{0.1em}
  \item[\textcolor{Gray}{API.}]  
  \texttt{\CKeyword{static} \deegenKwdDark{VariadicResultsAccessor}::\deegenKwdDark{GetNum}() \CKeyword{->} \CKeyword{size\_t}}

  \item[\textcolor{Gray}{Desc.}] Returns the number of values in the variadic results. The most recently executed bytecode must have set up the variadic results, or it is undefined behavior with no diagnostic reported. 
\end{itemize}

\subsubsection{Guest Language Function Call}\label{appendix:guest-language-function-call-apis}

\

\vspace{-0.3em}
\noindent\textcolor[RGB]{220,220,220}{\rule{\linewidth}{0.2pt}}

\begin{itemize}[leftmargin=2.5em]\setlength\itemsep{0.1em}
  \item[\textcolor{Gray}{API.}]  

  \texttt{\textcolor{Gray}{arg-seq ::= <empty>}}

  \texttt{\textcolor{Gray}{\ \ \ \ \ \ \ \ \ \ | }\deegenKwdDark{TValue} arg \textcolor{Gray}{[, arg-seq]}}

  \texttt{\textcolor{Gray}{\ \ \ \ \ \ \ \ \ \ | }\deegenKwdDark{TValue}* args, \CKeyword{size\_t} numArgs \textcolor{Gray}{[, arg-seq]}}
  
  \item[\textcolor{Gray}{Desc.}] BNF definition of a sequence of arguments passed to the callee. The sequence may be empty, or a combination of singleton arguments (specified individually as \deegenKwdDark{TValue}) and ranges of local variables (specified as the start of the range and the length).

  Currently Deegen allows at most one range of local variables in an argument sequence. This restriction is only for simplicity and may be lifted in the future.
\end{itemize}

\vspace{-0.5em}
\noindent\textcolor[RGB]{220,220,220}{\rule{\linewidth}{0.2pt}}

\begin{itemize}[leftmargin=2.5em]\setlength\itemsep{0.1em}
  \item[\textcolor{Gray}{API.}]  
  \texttt{\deegenKwdDark{MakeCall}(\CType{Function}* callee, \textcolor{Gray}{arg-seq}, \CType{FnPtr} retCont) \CKeyword{->} [[no\_return]]}

  \item[\textcolor{Gray}{Desc.}] Call the guest language function specified by \texttt{callee} using the argument sequence specified by \texttt{\textcolor{Gray}{arg-seq}}. When \texttt{callee} returns, control will be transferred to the return continuation \texttt{retCont}. A build error is reported if the function prototype of \texttt{retCont} does not match the expected execution semantics prototype based on the bytecode operands. If fewer arguments are passed than what is expected by the callee, insufficient arguments will get value of \texttt{Nil}. If more arguments are passed than what is expected by the callee, overflowing arguments will become variadic arguments if the callee accepts variadic arguments, or silently discarded if the callee does not accept variadic arguments. The behavior of insufficient or overflowing argument handling is currently hard-coded (which matches Lua behavior), but it is fairly easy to support other behaviors (e.g., throwing out an exception) in the future.

  \item[\textcolor{Gray}{Ex.}]  
   \texttt{\textcolor{Gray}{// Definition of a return continuation named AddRetCont}}

   \texttt{\CKeyword{void} AddRetCont(\deegenKwdDark{TValue} lhs, \deegenKwdDark{TValue} rhs) \{ ... \}}

   \texttt{\textcolor{Gray}{// Make a guest language function call with no arguments and }}
   
   \texttt{\textcolor{Gray}{// AddRetCont as the return continuation}}

   \texttt{\deegenKwdDark{MakeCall}(callee, AddRetCont);}
\end{itemize}

\vspace{-0.5em}
\noindent\textcolor[RGB]{220,220,220}{\rule{\linewidth}{0.2pt}}

\begin{itemize}[leftmargin=2.5em]\setlength\itemsep{0.1em}
  \item[\textcolor{Gray}{API.}]  
  \texttt{\deegenKwdDark{MakeCallPassingVariadicRes}(\CType{Function}* callee, \textcolor{Gray}{arg-seq}, \CType{FnPtr} retCont)}

  \texttt{\ \ \CKeyword{->} [[no\_return]]}

  \item[\textcolor{Gray}{Desc.}] Same as \deegenKwdDark{MakeCall}, except that the variadic results (set up by the most recently executed bytecode, see \secref{appendix:var-arg-var-results}) are appended to the end of the argument sequence \texttt{\textcolor{Gray}{arg-seq}} and passed to the callee. It is undefined behavior if the most recently executed bytecode did not set up the variadic results, no diagnostic reported.

\end{itemize}

\vspace{-0.5em}
\noindent\textcolor[RGB]{220,220,220}{\rule{\linewidth}{0.2pt}}

\begin{itemize}[leftmargin=2.5em]\setlength\itemsep{0.1em}
  \item[\textcolor{Gray}{API.}]  
  \texttt{\deegenKwdDark{MakeTailCall}(\CType{Function}* callee, \textcolor{Gray}{arg-seq})  \CKeyword{->} [[no\_return]]}

  \item[\textcolor{Gray}{Desc.}] Perform a proper tail call to \texttt{callee} with the argument sequence specified in \texttt{\textcolor{Gray}{arg-seq}}. The results returned from the callee becomes the return values of the current function, and control is transferred to the caller of the current function after the call (so this API does not and cannot take a return continuation). The semantics of proper tail call is honored by Deegen: it is guaranteed that an infinite tail call chain will not result in unbounded stack growth.

\end{itemize}

\vspace{-0.5em}
\noindent\textcolor[RGB]{220,220,220}{\rule{\linewidth}{0.2pt}}

\begin{itemize}[leftmargin=2.5em]\setlength\itemsep{0.1em}
  \item[\textcolor{Gray}{API.}]  
  \texttt{\deegenKwdDark{MakeTailCallPassingVariadicRes}(\CType{Function}* callee, \textcolor{Gray}{arg-seq})}

  \texttt{\ \ \CKeyword{->} [[no\_return]]}

  \item[\textcolor{Gray}{Desc.}] This is the variant of \deegenKwdDark{MakeTailCall} that appends the variadic results to the call arguments.

  See \deegenKwdDark{MakeTailCall} and \deegenKwdDark{\deegenKwdDark{MakeCallPassingVariadicRes}} for detail.

\end{itemize}

\vspace{-0.5em}
\noindent\textcolor[RGB]{220,220,220}{\rule{\linewidth}{0.2pt}}

\begin{itemize}[leftmargin=2.5em]\setlength\itemsep{0.1em}
  \item[\textcolor{Gray}{API.}]  
  \texttt{\deegenKwdDark{MakeInPlaceCall}(\CType{Function}* callee, \deegenKwdDark{TValue}* args, }

  \texttt{\ \ \ \ \ \ \ \ \ \ \ \ \ \ \ \ \CKeyword{size\_t} numArgs, \CType{FnPtr} retCont) \CKeyword{->} [[no\_return]]}

  \item[\textcolor{Gray}{Desc.}] Same as \deegenKwdDark{MakeCall}, except that the call is in-place. This means that the stack frame has been set up in a way so that the callee frame base can directly sit at \texttt{arg}, assuming that the callee does not accept variadic arguments or no argument becomes variadic argument for this call. Specifically, due to the current Deegen stack layout ABI, the user-written parser must reserve scratch space for the stack frame header right before \texttt{args} (which currently takes 32 bytes, or 4 local variables), and all the local variables \texttt{>=} the stack frame header must be assumed to hold garbage value after the call, as they could have been overwritten by callee frames. Failing to do so will result in undefined behavior with no diagnostic reported. The primary motivation of in-place call is performance, since no data movement is needed to set up the callee frame (unless the callee takes variadic arguments and the caller actually passed more arguments than the number of fixed arguments accepted by the callee). This is also why the API only accepts a range of local variables as arguments and nothing else. 

\end{itemize}

\vspace{-0.5em}
\noindent\textcolor[RGB]{220,220,220}{\rule{\linewidth}{0.2pt}}

\begin{itemize}[leftmargin=2.5em]\setlength\itemsep{0.1em}
  \item[\textcolor{Gray}{API.}]  
  \texttt{\deegenKwdDark{MakeInPlaceCallPassingVariadicRes}(\CType{Function}* callee, }

  \texttt{\ \ \deegenKwdDark{TValue}* args, \CKeyword{size\_t} numArgs, \CType{FnPtr} retCont) \CKeyword{->} [[no\_return]]}

  \item[\textcolor{Gray}{Desc.}] This is the in-place variant of \deegenKwdDark{MakeCallPassingVariadicRes}.

\end{itemize}

\vspace{-0.5em}
\noindent\textcolor[RGB]{220,220,220}{\rule{\linewidth}{0.2pt}}

\begin{itemize}[leftmargin=2.5em]\setlength\itemsep{0.1em}
  \item[\textcolor{Gray}{API.}]  
  \texttt{\deegenKwdDark{MakeInPlaceTailCall}(}

  \texttt{\ \ \CType{Function}* callee, \deegenKwdDark{TValue}* args, \CKeyword{size\_t} numArgs) \CKeyword{->} [[no\_return]]}

  \item[\textcolor{Gray}{Desc.}] This is the in-place variant of \deegenKwdDark{MakeTailCall}. Note that despite that the call is in-place, data movement is still needed to maintain the no-unbounded-stack-growth semantics of proper tail calls, which diminishes a lot (but not all) of the performance benefits of in-place calls.

\end{itemize}

\vspace{-0.5em}
\noindent\textcolor[RGB]{220,220,220}{\rule{\linewidth}{0.2pt}}

\begin{itemize}[leftmargin=2.5em]\setlength\itemsep{0.1em}
  \item[\textcolor{Gray}{API.}]  
  \texttt{\deegenKwdDark{MakeInPlaceTailCallPassingVariadicRes}(}

  \texttt{\ \ \CType{Function}* callee, \deegenKwdDark{TValue}* args, \CKeyword{size\_t} numArgs) \CKeyword{->} [[no\_return]]}

  \item[\textcolor{Gray}{Desc.}] This is the in-place variant of \deegenKwdDark{MakeTailCallPassingVariadicRes}, with the same caveats that data movement is needed despite being a in-place call. 

\end{itemize}

\subsubsection{Guest Language Function Return Value Accessors}

\ 

\vspace{-0.3em}
\noindent\textcolor[RGB]{220,220,220}{\rule{\linewidth}{0.2pt}}

\begin{itemize}[leftmargin=2.5em]\setlength\itemsep{0.1em}
  \item[\textcolor{Gray}{API.}]  
  \texttt{\deegenKwdDark{GetReturnValue}(\CKeyword{size\_t} ord) \CKeyword{->} \deegenKwdDark{TValue}}

  \item[\textcolor{Gray}{Desc.}] May only be used in a return continuation component, or a build error is reported. Returns the \texttt{ord}-th (0-based) return value of the call, or \texttt{Nil} if the callee returned fewer return values. The behavior of seeing \texttt{Nil} for insufficient return values is currently hard-coded (which matches Lua behavior), but it is fairly easy to support other behaviors (e.g., throwing out an exception) in the future.

\end{itemize}

\vspace{-0.3em}
\noindent\textcolor[RGB]{220,220,220}{\rule{\linewidth}{0.2pt}}

\begin{itemize}[leftmargin=2.5em]\setlength\itemsep{0.1em}
  \item[\textcolor{Gray}{API.}]  
  \texttt{\deegenKwdDark{GetNumReturnValues}() \CKeyword{->} \CKeyword{size\_t}}

  \item[\textcolor{Gray}{Desc.}] May only be used in a return continuation component, or a build error is reported. Returns the number of return values returned by the callee.

\end{itemize}

\vspace{-0.3em}
\noindent\textcolor[RGB]{220,220,220}{\rule{\linewidth}{0.2pt}}

\begin{itemize}[leftmargin=2.5em]\setlength\itemsep{0.1em}
  \item[\textcolor{Gray}{API.}]  
  \texttt{\deegenKwdDark{StoreReturnValuesTo}(\deegenKwdDark{TValue}* dst, \CKeyword{size\_t} numValuesToStore)}

  \item[\textcolor{Gray}{Desc.}] May only be used in a return continuation component, or a build error is reported. Stores the first \texttt{numValuesToStore} return values of the call to an array starting at \texttt{dst}. Similar to \deegenKwdDark{GetNumReturnValues}, \texttt{Nil} will be appended if insufficient return values are returned from the call, but we can easily support other behaviors in the future.

\end{itemize}

\vspace{-0.5em}
\noindent\textcolor[RGB]{220,220,220}{\rule{\linewidth}{0.2pt}}

\begin{itemize}[leftmargin=2.5em]\setlength\itemsep{0.1em}
  \item[\textcolor{Gray}{API.}]  
  \texttt{\deegenKwdDark{StoreReturnValuesAsVariadicResults}()}

  \item[\textcolor{Gray}{Desc.}] May only be used in the return continuation component, or a build error will be reported. Set up the variadic results (\secref{appendix:var-arg-var-results}) to be all the return values returned from the callee.
\end{itemize}

\subsubsection{Guest Language Function Return}

\ 

\vspace{-0.3em}
\noindent\textcolor[RGB]{220,220,220}{\rule{\linewidth}{0.2pt}}

\begin{itemize}[leftmargin=2.5em]\setlength\itemsep{0.1em}
  \item[\textcolor{Gray}{API.}]  
  \texttt{\deegenKwdDark{GuestLanguageFunctionReturn}() \CKeyword{->} [[no\_return]]}

  \item[\textcolor{Gray}{Desc.}] Return to the caller of the current function, with no return values.
\end{itemize}

\vspace{-0.5em}
\noindent\textcolor[RGB]{220,220,220}{\rule{\linewidth}{0.2pt}}

\begin{itemize}[leftmargin=2.5em]\setlength\itemsep{0.1em}
  \item[\textcolor{Gray}{API.}]  
  \texttt{\deegenKwdDark{GuestLanguageFunctionReturn}(}
  
  \texttt{\ \ \CKeyword{const} \deegenKwdDark{TValue}* retStart, \CKeyword{size\_t} numRets) \CKeyword{->} [[no\_return]]}

  \item[\textcolor{Gray}{Desc.}] Return to the caller of the current function with \texttt{numRets} return values starting at \texttt{retStart}.
\end{itemize}

\vspace{-0.5em}
\noindent\textcolor[RGB]{220,220,220}{\rule{\linewidth}{0.2pt}}

\begin{itemize}[leftmargin=2.5em]\setlength\itemsep{0.1em}
  \item[\textcolor{Gray}{API.}]  
  \texttt{\deegenKwdDark{GuestLanguageFunctionReturnAppendingVariadicResults}(}
  
  \texttt{\ \ \CKeyword{const} \deegenKwdDark{TValue}* retStart, \CKeyword{size\_t} numRets) \CKeyword{->} [[no\_return]]}

  \item[\textcolor{Gray}{Desc.}] Return to the caller of the current function. The return values are the \texttt{numRets} return values starting at \texttt{retStart}, plus all the variadic results (see \secref{appendix:var-arg-var-results}) appended at the end.
\end{itemize}

\subsubsection{Lexical Capture Related APIs}

\ 

\vspace{-0.3em}
\noindent\textcolor[RGB]{220,220,220}{\rule{\linewidth}{0.2pt}}

\begin{itemize}[leftmargin=2.5em]\setlength\itemsep{0.1em}
  \item[\textcolor{Gray}{API.}]  
  \texttt{\deegenKwdDark{CreateNewClosure}(\CType{CodeBlock}* cb, \CKeyword{size\_t} selfReferenceOrdinal) \CKeyword{->} \CType{Function}*}
  
  \item[\textcolor{Gray}{Desc.}] Create a new closure with lexical capture semantics. Deegen uses upvalue mechanism (also known as \textit{flat closure} in the functional language community) to implement lexical capture. Note that despite that the word upvalue is a Lua jargon, the upvalue mechanism is a generic mechanism that can be used to implement lexical capture semantics for any language.
\end{itemize}

\vspace{-0.5em}
\noindent\textcolor[RGB]{220,220,220}{\rule{\linewidth}{0.2pt}}

\begin{itemize}[leftmargin=2.5em]\setlength\itemsep{0.1em}
  \item[\textcolor{Gray}{API.}]  
  \texttt{\deegenKwdDark{GetFEnvGlobalObject}() \CKeyword{->} \deegenKwdDark{TValue}}
  
  \item[\textcolor{Gray}{Desc.}] Return the global environment (root object) captured by the current function. 
\end{itemize}

\vspace{-0.5em}
\noindent\textcolor[RGB]{220,220,220}{\rule{\linewidth}{0.2pt}}

\begin{itemize}[leftmargin=2.5em]\setlength\itemsep{0.1em}
  \item[\textcolor{Gray}{API.}]  
  \texttt{\CKeyword{static} \deegenKwdDark{UpvalueAccessor}::\deegenKwdDark{GetMutable}(\CKeyword{size\_t} ord) \CKeyword{->} \deegenKwdDark{TValue}}
  
  \item[\textcolor{Gray}{Desc.}] Return the value stored in a mutable upvalue. It is undefined behavior with no diagnostic to access an immutable upvalue using this API, or if an invalid \texttt{ord} is provided. 
\end{itemize}

\vspace{-0.5em}
\noindent\textcolor[RGB]{220,220,220}{\rule{\linewidth}{0.2pt}}

\begin{itemize}[leftmargin=2.5em]\setlength\itemsep{0.1em}
  \item[\textcolor{Gray}{API.}]  
  \texttt{\CKeyword{static} \deegenKwdDark{UpvalueAccessor}::\deegenKwdDark{GetImmutable}(\CKeyword{size\_t} ord) \CKeyword{->} \deegenKwdDark{TValue}}
  
  \item[\textcolor{Gray}{Desc.}] Return the value stored in an immutable upvalue. It is undefined behavior with no diagnostic to access a mutable upvalue using this API, or if an invalid \texttt{ord} is provided. 
\end{itemize}

\vspace{-0.5em}
\noindent\textcolor[RGB]{220,220,220}{\rule{\linewidth}{0.2pt}}

\begin{itemize}[leftmargin=2.5em]\setlength\itemsep{0.1em}
  \item[\textcolor{Gray}{API.}]  
  \texttt{\CKeyword{static} \deegenKwdDark{UpvalueAccessor}::\deegenKwdDark{Put}(\CKeyword{size\_t} ord, \deegenKwdDark{TValue} val)}
  
  \item[\textcolor{Gray}{Desc.}] Store \texttt{val} into the mutable upvalue identified by ordinal \texttt{ord}.  It is undefined behavior with no diagnostic if the upvalue is immutable, or if an invalid \texttt{ord} is provided.
\end{itemize}

\vspace{-0.5em}
\noindent\textcolor[RGB]{220,220,220}{\rule{\linewidth}{0.2pt}}

\begin{itemize}[leftmargin=2.5em]\setlength\itemsep{0.1em}
  \item[\textcolor{Gray}{API.}]  
  \texttt{\CKeyword{static} \deegenKwdDark{UpvalueAccessor}::\deegenKwdDark{Close}(\CKeyword{const} \deegenKwdDark{TValue}* start)}
  
  \item[\textcolor{Gray}{Desc.}] Close all upvalues whose stack frame address \texttt{>=start}. 
\end{itemize}

\subsubsection{Generic Inline Caching}

\ 

\vspace{-0.3em}
\noindent\textcolor[RGB]{220,220,220}{\rule{\linewidth}{0.2pt}}

\begin{itemize}[leftmargin=2.5em]\setlength\itemsep{0.1em}
  \item[\textcolor{Gray}{API.}]  
  \texttt{\deegenKwdDark{DeclareInlineCache}() \CKeyword{->} \deegenKwdDark{ICHandler}*}
  
  \item[\textcolor{Gray}{Desc.}] Declare a generic IC. Returns an opaque handler with more APIs to specify the IC semantics. 
\end{itemize}

\vspace{-0.5em}
\noindent\textcolor[RGB]{220,220,220}{\rule{\linewidth}{0.2pt}}

\begin{itemize}[leftmargin=2.5em]\setlength\itemsep{0.1em}
  \item[\textcolor{Gray}{API.}]  
  \texttt{\deegenKwdDark{ICHandler}::\deegenKwdDark{Key}<\CType{IntType}>(\CType{IntType} key)}
  
  \item[\textcolor{Gray}{Desc.}] Specify $\textrm{IC}_{\textrm{key}}$, must be an integer value.
\end{itemize}

\vspace{-0.5em}
\noindent\textcolor[RGB]{220,220,220}{\rule{\linewidth}{0.2pt}}

\begin{itemize}[leftmargin=2.5em]\setlength\itemsep{0.1em}
  \item[\textcolor{Gray}{API.}]  
  \texttt{\deegenKwdDark{ICHandler}::\deegenKwdDark{Body}<\CType{FuncTy}>(\CKeyword{const }\CType{FuncTy}\& lambda)}
  
  \item[\textcolor{Gray}{Desc.}] Specify the idempotent computation $\lambda_i$. The \texttt{lambda} must be a C++ lambda with \texttt{[=]} capture. 
\end{itemize}

\vspace{-0.5em}
\noindent\textcolor[RGB]{220,220,220}{\rule{\linewidth}{0.2pt}}

\begin{itemize}[leftmargin=2.5em]\setlength\itemsep{0.1em}
  \item[\textcolor{Gray}{API.}]  
  \texttt{\deegenKwdDark{ICHandler}::\deegenKwdDark{Effect}<\CType{FuncTy}>(\CKeyword{const }\CType{FuncTy}\& lambda)}
  
  \item[\textcolor{Gray}{Desc.}] Specify the effectful computation $\lambda_e$. This API may only be used inside the \deegenKwdDark{Body} lambda ($\lambda_i$), and must be used in the form of \texttt{\CKeyword{return }ic->\deegenKwdDark{Effect}([=] \{ ... \});} The \texttt{lambda} must be a C++ lambda with \texttt{[=]} capture. The lambda may only capture objects with trivial copy constructor and no-op destructor, or a build error will be reported. Deegen might also have trouble handling capture of complex C++ classes (but doing so is already a bad idea in the first place, since $\lambda_e$ is supposed to be very cheap), and will report build error on such cases. 
\end{itemize}

\vspace{-0.5em}
\noindent\textcolor[RGB]{220,220,220}{\rule{\linewidth}{0.2pt}}

\begin{itemize}[leftmargin=2.5em]\setlength\itemsep{0.1em}
  \item[\textcolor{Gray}{API.}]  
  \texttt{\deegenKwdDark{ICHandler}::\deegenKwdDark{SetUncacheable}()}
  
  \item[\textcolor{Gray}{Desc.}] May only be used in the \deegenKwdDark{Body} lambda ($\lambda_i$). After this API is executed, no IC entry will be created during this execution, even if an \deegenKwdDark{Effect} API is executed. 
\end{itemize}

\vspace{-0.5em}
\noindent\textcolor[RGB]{220,220,220}{\rule{\linewidth}{0.2pt}}

\begin{itemize}[leftmargin=2.5em]\setlength\itemsep{0.1em}
  \item[\textcolor{Gray}{API.}]  
  \texttt{\deegenKwdDark{ICHandler}::\deegenKwdDark{FuseICIntoInterpreterOpcode}()}
  
  \item[\textcolor{Gray}{Desc.}] An optional API that enables interpreter IC dynamic quickening optimization. May only be used if there is only one IC employed in the bytecode, and the IC is not executed in a loop, otherwise it is undefined behavior with no diagnostic reported.
\end{itemize}

\vspace{-0.5em}
\noindent\textcolor[RGB]{220,220,220}{\rule{\linewidth}{0.2pt}}

\begin{itemize}[leftmargin=2.5em]\setlength\itemsep{0.1em}
  \item[\textcolor{Gray}{API.}]  
  \texttt{\deegenKwdDark{ICHandler}::\deegenKwdDark{SpecifyImpossibleValueForKey}<\CType{IntType}>(\CType{IntType} imposVal)}
  
  \item[\textcolor{Gray}{Desc.}] An optional API that specifies an impossible value that will never show up as $\textrm{IC}_{\textrm{key}}$, which improves IC performance in interpreter mode. \CType{IntType} must agree with the $\textrm{IC}_{\textrm{key}}$ type, and \texttt{imposVal} must be a constant value (or can be reduced to a constant value after reasonable LLVM optimizations), or a build error will be reported. It is undefined behavior with no diagnostic if the specified \texttt{imposVal} show up at runtime as $\textrm{IC}_\textrm{key}$.
\end{itemize}

\vspace{-0.5em}
\noindent\textcolor[RGB]{220,220,220}{\rule{\linewidth}{0.2pt}}

\begin{itemize}[leftmargin=2.5em]\setlength\itemsep{0.1em}
  \item[\textcolor{Gray}{API.}]  
  \texttt{\deegenKwdDark{IcAnnotateRange}<\CType{IntType}>(\CType{IntType}\& capture, \CType{IntType} lb, \CType{IntType} ub)}
  
  \item[\textcolor{Gray}{Desc.}] An optional API that specifies the possible range of an $\textrm{IC}_{\textrm{state}}$, which improves IC performance in JIT mode. May only be used inside the \deegenKwdDark{Effect} lambda ($\lambda_e$). Must not be executed conditionally, or undefined behavior with no diagnostic. \texttt{capture} must be a local variable defined in the \deegenKwdDark{Body} lambda ($\lambda_i$). \texttt{lb} and \texttt{ub} must be constant values (or can be reduced to constant values after reasonable LLVM optimizations), or a build error is reported. It is undefined behavior with no diagnostic if at runtime the value of \texttt{capture} is outside the specified range.
\end{itemize}

\vspace{-0.5em}
\noindent\textcolor[RGB]{220,220,220}{\rule{\linewidth}{0.2pt}}

\begin{itemize}[leftmargin=2.5em]\setlength\itemsep{0.1em}
  \item[\textcolor{Gray}{API.}]  
  \texttt{\deegenKwdDark{IcSpecializeVal}<\CType{IntType}>(\CType{IntType}\& capture, \CType{IntType}... vals)}

  \texttt{\deegenKwdDark{IcSpecializeValFull}<\CType{IntType}>(\CType{IntType}\& capture, \CType{IntType}... vals)}
  
  \item[\textcolor{Gray}{Desc.}] Optional APIs that may only be used inside \deegenKwdDark{Effect} lambda ($\lambda_e$). The \texttt{capture} must be an $\textrm{IC}_{\textrm{state}}$, that is, a local variable defined in the \deegenKwdDark{Body} lambda ($\lambda_i$). All \texttt{vals} must be constant values (or can be reduced to constant values after reasonable LLVM optimizations), or a build error is reported. The APIs must not be executed conditionally, or undefined behavior with no diagnostic reported. \deegenKwdDark{IcSpecializeVal} generates specialized $\lambda_e$ implementations for each case where \texttt{capture} equals one of the specified \texttt{vals}, as well as the fallback case where \texttt{capture} does not equal any of the specified \texttt{vals}. \deegenKwdDark{IcSpecializeValFull} does not generate the fallback case, and it is undefined behavior with no diagnostic if at runtime, the value of \texttt{capture} is not within the list of \texttt{vals}. Multiple uses of \deegenKwdDark{IcSpecializeVal} and \deegenKwdDark{IcSpecializeValFull} inside the same \deegenKwdDark{Effect} lambda compose as a Cartesian product.

  \item[\textcolor{Gray}{Ex.}]  
  \texttt{\textcolor{Gray}{// Inside the Body lambda}}

  \texttt{\CKeyword{bool} mayHaveMetatable = ...;}

  \texttt{\CKeyword{bool} isPropertyFound = ...;}

  \texttt{\CKeyword{return} ic->\deegenKwdDark{Effect}([=]() \{}

  \texttt{\textcolor{Gray}{\ \ // This is equivalent to having four hand-written lambdas for}}

  \texttt{\textcolor{Gray}{\ \ // the four combinations of mayHaveMetatable and isPropertyFound}}
  
  \texttt{\ \ \deegenKwdDark{IcSpecializeValFull}(mayHaveMetatable, \CKeyword{false}, \CKeyword{true});}

  \texttt{\ \ \deegenKwdDark{IcSpecializeValFull}(isPropertyFound, \CKeyword{false}, \CKeyword{true});}

  \texttt{\textcolor{Gray}{\ \ // This if-condition-check is always a no-op at runtime}}

  \texttt{\ \ \CKeyword{if} (mayHaveMetatable) \{ ... \} \deegenKwdDark{else} \{ ... \}}

  \texttt{\});}
\end{itemize}

\subsection{Bytecode Builder, Decoder and Editor APIs}\label{appendix:bytecode-builder-apis}

\subsubsection{Bytecode Operand Data Struct}

\ 

\vspace{-0.3em}
\noindent\textcolor[RGB]{220,220,220}{\rule{\linewidth}{0.2pt}}

\begin{itemize}[leftmargin=2.5em]\setlength\itemsep{0.1em}
  \item[\textcolor{Gray}{API.}]  
  \texttt{\CKeyword{struct} \{}

  \texttt{\ \ \CType{OpTy1} OpName1;}

  \texttt{\ \ \CType{OpTy2} OpName2;}

  \texttt{\ \ ...;}

  \texttt{\};}

  \item[\textcolor{Gray}{Desc.}] Deegen automatically generate a data struct definition for each bytecode. The struct contains all operands and the output (if exists) information. Each bytecode operand becomes a member (declared in the same order as the bytecode operands) which name is the name of the operand, and which type is determined based on the bytecode operand type, as follows:
  \begin{itemize}
      \item \texttt{\deegenKwdDark{Local} \ \ \ \ \ \ \ \ \ \ \ \ \ \ \CKeyword{->} \deegenKwdDark{Local}}
      \item \texttt{\deegenKwdDark{Constant} \ \ \ \ \ \ \ \ \ \ \ \CKeyword{->} \deegenKwdDark{CstWrapper}}
      \item \texttt{\deegenKwdDark{LocalOrConstant} \ \ \ \ \CKeyword{->} \deegenKwdDark{LocalOrCstWrapper}}
      \item \texttt{\deegenKwdDark{BytecodeRangeBaseRO} \CKeyword{->} \deegenKwdDark{Local}}
      \item \texttt{\deegenKwdDark{BytecodeRangeBaseRW} \CKeyword{->} \deegenKwdDark{Local}}
      \item \texttt{\deegenKwdDark{Literal}<\CType{IntType}>\ \ \ \ \CKeyword{->} \deegenKwdDark{ForbidUninitialized}<\CType{IntType}>}
  \end{itemize}

  \vspace{0.2em}
  
  Finally, if the bytecode has an output, a member with name \texttt{output} and type \deegenKwdDark{Local} will be declared as the last member. The definition of each type (e.g., \deegenKwdDark{Local}) is as follows:
  \begin{itemize}
      \item \deegenKwdDark{Local} contains the ordinal (\CKeyword{size\_t}) of the local variable. For \deegenKwdDark{BytecodeRangeBaseRO/W}, the ordinal indicates the start of the range. 
      \item \deegenKwdDark{CstWrapper} contains a boxed value (\deegenKwdDark{TValue}), the constant.
      \item \deegenKwdDark{LocalOrCstWrapper} contains a boolean denoting whether it is a \deegenKwdDark{Local} or \deegenKwdDark{CstWrapper}, and the value of the \deegenKwdDark{Local} or \deegenKwdDark{CstWrapper} respectively.
      \item \texttt{\deegenKwdDark{ForbidUninitialized}<\CType{IntType}>} is a wrapper around \CType{IntType} that removes the default constructor, to statically catch the bug if the user forget to specify its value.
  \end{itemize}
  
  \vspace{0.2em}
  
  This data struct is used by both the bytecode builder and the bytecode decoder. User creates a bytecode by specifying each value in the bytecode data struct, and the bytecode decoder generates this data struct so user can easily access the value of each bytecode operand. 
  
  Information of the bytecode branch target (if exists) is not present in the data struct. This is an intentional design choice, and we have dedicated APIs to get and set its value. 

   \item[\textcolor{Gray}{Ex.}]  

   \texttt{\CKeyword{struct} \{ \ \ \textcolor{Gray}{// Definition generated by Deegen for bytecode Add}}

   \texttt{\ \ \deegenKwdDark{LocalOrCstWrapper} lhs;}
     
   \texttt{\ \ \deegenKwdDark{LocalOrCstWrapper} rhs;}

   \texttt{\ \ \deegenKwdDark{Local} output;}

   \texttt{\};}
\end{itemize}

\subsubsection{Bytecode Builder}

\

\vspace{-0.3em}
\noindent\textcolor[RGB]{220,220,220}{\rule{\linewidth}{0.2pt}}

\begin{itemize}[leftmargin=2.5em]\setlength\itemsep{0.1em}
  \item[\textcolor{Gray}{API.}]  
  \texttt{\CKeyword{explicit} \deegenKwdDark{Local}::\deegenKwdDark{Local}(\CKeyword{size\_t} ord)}
  
  \item[\textcolor{Gray}{Desc.}] Constructor for \deegenKwdDark{Local}. If the operand is a \deegenKwdDark{Local} operand, this is just the local ordinal. If the operand is a \deegenKwdDark{BytecodeRangeBaseRO/W} operand, this represents the start ordinal of the range.
\end{itemize}

\vspace{-0.5em}
\noindent\textcolor[RGB]{220,220,220}{\rule{\linewidth}{0.2pt}}

\begin{itemize}[leftmargin=2.5em]\setlength\itemsep{0.1em}
  \item[\textcolor{Gray}{API.}]  
  \texttt{\deegenKwdDark{CstWrapper}::\deegenKwdDark{CstWrapper}(\deegenKwdDark{TValue} val)}
  
  \item[\textcolor{Gray}{Desc.}] Construct a \deegenKwdDark{CstWrapper} by a boxed value.
\end{itemize}

\vspace{-0.5em}
\noindent\textcolor[RGB]{220,220,220}{\rule{\linewidth}{0.2pt}}

\begin{itemize}[leftmargin=2.5em]\setlength\itemsep{0.1em}
  \item[\textcolor{Gray}{API.}]  
  \texttt{\deegenKwdDark{Cst}<\CType{tTypeMask}>(\CType{UnboxedTy} unboxedVal) \CKeyword{->} \deegenKwdDark{CstWrapper}}
  
  \item[\textcolor{Gray}{Desc.}] Construct a \deegenKwdDark{CstWrapper} by an unboxed value and its type in the boxing scheme.

  \item[\textcolor{Gray}{Ex.}]  \texttt{\deegenKwdDark{Cst}<\CType{tDouble}>(123.4)\ \ \textcolor{Gray}{// CstWrapper holding a boxed double value of 123.4}}
\end{itemize}

\vspace{-0.5em}
\noindent\textcolor[RGB]{220,220,220}{\rule{\linewidth}{0.2pt}}

\begin{itemize}[leftmargin=2.5em]\setlength\itemsep{0.1em}
  \item[\textcolor{Gray}{API.}]  
  \texttt{\deegenKwdDark{LocalOrCstWrapper}::\deegenKwdDark{LocalOrCstWrapper}(\deegenKwdDark{Local} val)}

  \texttt{\deegenKwdDark{LocalOrCstWrapper}::\deegenKwdDark{LocalOrCstWrapper}(\deegenKwdDark{CstWrapper} val)}
  
  \item[\textcolor{Gray}{Desc.}] Implicit cast from a \deegenKwdDark{Local} or \deegenKwdDark{CstWrapper} to a \deegenKwdDark{LocalOrCstWrapper}.
\end{itemize}

\vspace{-0.5em}
\noindent\textcolor[RGB]{220,220,220}{\rule{\linewidth}{0.2pt}}

\begin{itemize}[leftmargin=2.5em]\setlength\itemsep{0.1em}
  \item[\textcolor{Gray}{API.}]  
  \texttt{\deegenKwdDark{BytecodeBuilder}::\deegenKwdDark{BytecodeBuilder}()}

  \item[\textcolor{Gray}{Desc.}] Constructor for the \deegenKwdDark{BytecodeBuilder} class that gives access to all the bytecode builder APIs.
\end{itemize}

\vspace{-0.5em}
\noindent\textcolor[RGB]{220,220,220}{\rule{\linewidth}{0.2pt}}

\begin{itemize}[leftmargin=2.5em]\setlength\itemsep{0.1em}
  \item[\textcolor{Gray}{API.}]  
  \texttt{\deegenKwdDark{BytecodeBuilder}::\deegenKwdDark{CreateXXX}(\CType{Operands} operands)}

  \item[\textcolor{Gray}{Desc.}] Appends a bytecode to the end of the bytecode stream. Deegen automatically selects the most specialized bytecode variant based on the specified operand values. If no variant is eligible, an assertion is fired. Since the data struct does not include the branch target, the branch target must be separately specified: we made this design because for a typical parser, the branch target is usually not known at the time the branch bytecode is created.

  \item[\textcolor{Gray}{Ex.}]  
  \texttt{bytecodeBuilder.\deegenKwdDark{CreateAdd}(\{}

  \texttt{\ \ .lhs = \deegenKwdDark{Local}(1),}

  \texttt{\ \ .rhs = \deegenKwdDark{Cst}<\CType{tDouble}>(123.4),}

  \texttt{\ \ .output = \deegenKwdDark{Local}(2)}

  \texttt{\});}
\end{itemize}

\vspace{-0.5em}
\noindent\textcolor[RGB]{220,220,220}{\rule{\linewidth}{0.2pt}}

\begin{itemize}[leftmargin=2.5em]\setlength\itemsep{0.1em}
  \item[\textcolor{Gray}{API.}]  
  \texttt{\deegenKwdDark{BytecodeBuilder}::\deegenKwdDark{GetCurLength}() \CKeyword{->} \CKeyword{size\_t}}

  \item[\textcolor{Gray}{Desc.}] Return the current total length of the bytecode stream.
\end{itemize}

\vspace{-0.5em}
\noindent\textcolor[RGB]{220,220,220}{\rule{\linewidth}{0.2pt}}

\begin{itemize}[leftmargin=2.5em]\setlength\itemsep{0.1em}
  \item[\textcolor{Gray}{API.}]  
  \texttt{\deegenKwdDark{BytecodeBuilder}::\deegenKwdDark{SetBranchTarget}(\CKeyword{size\_t} bcPos, \CKeyword{size\_t} destPos)}

  \item[\textcolor{Gray}{Desc.}] Set the branch target of bytecode at location \texttt{bcPos} to be the bytecode at location \texttt{destPos}. An assertion is fired if the bytecode at \texttt{bcPos} does not have a branch target.
  
\end{itemize}

\vspace{-0.5em}
\noindent\textcolor[RGB]{220,220,220}{\rule{\linewidth}{0.2pt}}

\begin{itemize}[leftmargin=2.5em]\setlength\itemsep{0.1em}
  \item[\textcolor{Gray}{API.}]  
  \texttt{\deegenKwdDark{BytecodeBuilder}::\deegenKwdDark{SetOutputOperand}(\CKeyword{size\_t} bcPos, \deegenKwdDark{Local} local)}

  \item[\textcolor{Gray}{Desc.}] Set the output operand of bytecode at location \texttt{bcPos} to be \texttt{local}. An assertion is fired if the bytecode at \texttt{bcPos} does not have an output operand.
  
\end{itemize}

\vspace{-0.5em}
\noindent\textcolor[RGB]{220,220,220}{\rule{\linewidth}{0.2pt}}

\begin{itemize}[leftmargin=2.5em]\setlength\itemsep{0.1em}
  \item[\textcolor{Gray}{API.}]  
  \texttt{\deegenKwdDark{BytecodeBuilder}::\deegenKwdDark{CheckWellFormedness}() \CKeyword{->} \CKeyword{bool}}

  \item[\textcolor{Gray}{Desc.}] Validate that all the bytecode branch targets are valid (e.g., not branching into the middle of another bytecode, not out of range).
  
\end{itemize}

\vspace{-0.5em}
\noindent\textcolor[RGB]{220,220,220}{\rule{\linewidth}{0.2pt}}

\begin{itemize}[leftmargin=2.5em]\setlength\itemsep{0.1em}
  \item[\textcolor{Gray}{API.}]  
  \texttt{\deegenKwdDark{BytecodeBuilder}::\deegenKwdDark{GetBuiltBytecodeSequence}() \CKeyword{->} std::pair<\CKeyword{uint8\_t}*, \CKeyword{size\_t}>}

  \item[\textcolor{Gray}{Desc.}] Return the built bytecode sequence.
  
\end{itemize}

\vspace{-0.5em}
\noindent\textcolor[RGB]{220,220,220}{\rule{\linewidth}{0.2pt}}

\begin{itemize}[leftmargin=2.5em]\setlength\itemsep{0.1em}
  \item[\textcolor{Gray}{API.}]  
  \texttt{\deegenKwdDark{BytecodeBuilder}::\deegenKwdDark{GetBuiltConstantTable}() \CKeyword{->} std::pair<\deegenKwdDark{TValue}*, \CKeyword{size\_t}>}

  \item[\textcolor{Gray}{Desc.}] Return the built constant table.
  
\end{itemize}

\subsubsection{Bytecode Decoder and Editor}

\

\vspace{-0.3em}
\noindent\textcolor[RGB]{220,220,220}{\rule{\linewidth}{0.2pt}}

\begin{itemize}[leftmargin=2.5em]\setlength\itemsep{0.1em}
  \item[\textcolor{Gray}{API.}]  
  \texttt{\CKeyword{enum class} \deegenKwdDark{BCKind};}

  \item[\textcolor{Gray}{Desc.}] An enum generated by Deegen containing the list of all bytecode names.

  \item[\textcolor{Gray}{Ex.}]  \texttt{\deegenKwdDark{BCKind}::Add\ \ \textcolor{Gray}{// An enum value that refers to bytecode Add}}
  
\end{itemize}

\vspace{-0.5em}
\noindent\textcolor[RGB]{220,220,220}{\rule{\linewidth}{0.2pt}}

\begin{itemize}[leftmargin=2.5em]\setlength\itemsep{0.1em}
  \item[\textcolor{Gray}{API.}]  
  \texttt{\deegenKwdDark{BytecodeBuilder}::\deegenKwdDark{GetBytecodeKind}(\CKeyword{size\_t} bcPos) \CKeyword{->} \deegenKwdDark{BCKind}}

  \item[\textcolor{Gray}{Desc.}] Return the kind of the bytecode at location \texttt{bcPos}.
  
\end{itemize}

\vspace{-0.5em}
\noindent\textcolor[RGB]{220,220,220}{\rule{\linewidth}{0.2pt}}

\begin{itemize}[leftmargin=2.5em]\setlength\itemsep{0.1em}
  \item[\textcolor{Gray}{API.}]  
  \texttt{\deegenKwdDark{BytecodeBuilder}::\deegenKwdDark{DecodeXXX}(\CKeyword{size\_t} bcPos) \CKeyword{->} \CKeyword{auto}}

  \item[\textcolor{Gray}{Desc.}] Decode the bytecode at location \texttt{bcPos}. The bytecode kind specified in the function must match the actual bytecode kind, or an assertion is fired.

  \item[\textcolor{Gray}{Ex.}]  
  \texttt{\textcolor{Gray}{// After exeuction, ops is a struct that looks like }}
  
  \texttt{\textcolor{Gray}{// \{ .lhs = Local(1), .rhs = Cst<tDouble>(123.4), .output = Local(2) \}}}

  \texttt{\CKeyword{auto} ops = bytecodeBuilder.\deegenKwdDark{DecodeAdd}(0);}
\end{itemize}

\vspace{-0.5em}
\noindent\textcolor[RGB]{220,220,220}{\rule{\linewidth}{0.2pt}}

\begin{itemize}[leftmargin=2.5em]\setlength\itemsep{0.1em}
  \item[\textcolor{Gray}{API.}]  
  \texttt{\deegenKwdDark{BytecodeBuilder}::\deegenKwdDark{GetNextBytecodePosition}(\CKeyword{size\_t} bcPos) \CKeyword{->} \CKeyword{size\_t}}

  \item[\textcolor{Gray}{Desc.}] Return the location of the next bytecode following the bytecode at location \texttt{bcPos}.
  
\end{itemize}

\vspace{-0.5em}
\noindent\textcolor[RGB]{220,220,220}{\rule{\linewidth}{0.2pt}}

\begin{itemize}[leftmargin=2.5em]\setlength\itemsep{0.1em}
  \item[\textcolor{Gray}{API.}]  
  \texttt{\deegenKwdDark{BytecodeBuilder}::\deegenKwdDark{BytecodeHasOutputOperand}(\CKeyword{size\_t} bcPos) \CKeyword{->} \CKeyword{bool}}

  \item[\textcolor{Gray}{Desc.}] Return whether the bytecode at location \texttt{bcPos} has an output operand.
  
\end{itemize}

\vspace{-0.5em}
\noindent\textcolor[RGB]{220,220,220}{\rule{\linewidth}{0.2pt}}

\begin{itemize}[leftmargin=2.5em]\setlength\itemsep{0.1em}
  \item[\textcolor{Gray}{API.}]  
  \texttt{\deegenKwdDark{BytecodeBuilder}::\deegenKwdDark{GetOutputOperand}(\CKeyword{size\_t} bcPos) \CKeyword{->} \CKeyword{size\_t}}

  \item[\textcolor{Gray}{Desc.}] Return the local variable ordinal of the output operand of the bytecode at \texttt{bcPos}.
  
\end{itemize}

\vspace{-0.5em}
\noindent\textcolor[RGB]{220,220,220}{\rule{\linewidth}{0.2pt}}

\begin{itemize}[leftmargin=2.5em]\setlength\itemsep{0.1em}
  \item[\textcolor{Gray}{API.}]  
  \texttt{\deegenKwdDark{BytecodeBuilder}::\deegenKwdDark{BytecodeHasBranchTarget}(\CKeyword{size\_t} bcPos) \CKeyword{->} \CKeyword{bool}}

  \item[\textcolor{Gray}{Desc.}] Return whether the bytecode at location \texttt{bcPos} has a branch target.
  
\end{itemize}

\vspace{-0.5em}
\noindent\textcolor[RGB]{220,220,220}{\rule{\linewidth}{0.2pt}}

\begin{itemize}[leftmargin=2.5em]\setlength\itemsep{0.1em}
  \item[\textcolor{Gray}{API.}]  
  \texttt{\deegenKwdDark{BytecodeBuilder}::\deegenKwdDark{GetBranchTarget}(\CKeyword{size\_t} bcPos) \CKeyword{->} \CKeyword{size\_t}}

  \item[\textcolor{Gray}{Desc.}] Return the bytecode location that the bytecode at \texttt{bcPos} may branch to.
  
\end{itemize}

\vspace{-0.5em}
\noindent\textcolor[RGB]{220,220,220}{\rule{\linewidth}{0.2pt}}

\begin{itemize}[leftmargin=2.5em]\setlength\itemsep{0.1em}
  \item[\textcolor{Gray}{API.}]  
  \texttt{\deegenKwdDark{BytecodeBuilder}::\deegenKwdDark{ReplaceBytecode}<\deegenKwdDark{BCKind} k>(\CKeyword{size\_t} bcPos, \CType{Operands} operands)}

  \item[\textcolor{Gray}{Desc.}] Replace the bytecode at \texttt{bcPos} with another bytecode of \texttt{\deegenKwdDark{BCKind} k} and the specified bytecode operands. The old and new bytecode must have equal length, which must be enforced by the bytecode length constraint API (see \secref{appendix:bytecode-length-constraints-api}), or a static assertion is fired. 

  \item[\textcolor{Gray}{Ex.}]  
  \texttt{\textcolor{Gray}{// Replace Add(local, constant) with Sub(local, -constant) }}
  
  \texttt{\CKeyword{auto} ops = bytecodeBuilder.\deegenKwdDark{DecodeAdd}(0);}

  \texttt{bytecodeBuilder.\deegenKwdDark{ReplaceBytecode}<\deegenKwdDark{BCKind}::Sub>(0, \{}
  
  \texttt{\ \ .lhs = ops.lhs,}

  \texttt{\ \ .rhs = \deegenKwdDark{Cst}<\CType{tDouble}>(-ops.rhs.\deegenKwdDark{AsConstant}().\deegenKwdDark{As}<\CType{tDouble}>()),}

  \texttt{\ \ .output = ops.output}

  \texttt{\});}
\end{itemize} 

\subsection{Built-in C++ Library Definition APIs}\label{appendix:cpp-library-definition-apis}

\subsubsection{Execution Semantics APIs}\label{appendix:library-function-execution-semantics-apis}

\ 

\vspace{-0.3em}
\noindent\textcolor[RGB]{220,220,220}{\rule{\linewidth}{0.2pt}}

\begin{itemize}[leftmargin=2.5em]\setlength\itemsep{0.1em}
  \item[\textcolor{Gray}{API.}]  
  \texttt{\deegenKwdDark{GetStackBase}() \CKeyword{->} \deegenKwdDark{TValue}*}

  \item[\textcolor{Gray}{Desc.}] Return the stack base pointer. Everything above it may be used by the library function to store scratch data. At function entry, all the function arguments are stored starting at this pointer, so the function may access the first argument by \texttt{\deegenKwdDark{GetStackBase}()[0]}, etc. 
  
\end{itemize}

\vspace{-0.5em}
\noindent\textcolor[RGB]{220,220,220}{\rule{\linewidth}{0.2pt}}

\begin{itemize}[leftmargin=2.5em]\setlength\itemsep{0.1em}
  \item[\textcolor{Gray}{API.}]  
  \texttt{\deegenKwdDark{GetNumArgs}() \CKeyword{->} \CKeyword{size\_t}}

  \item[\textcolor{Gray}{Desc.}] Return the number of arguments passed to this function.
  
\end{itemize}

\vspace{-0.5em}
\noindent\textcolor[RGB]{220,220,220}{\rule{\linewidth}{0.2pt}}

\begin{itemize}[leftmargin=2.5em]\setlength\itemsep{0.1em}
  \item[\textcolor{Gray}{API.}]  
  \texttt{\deegenKwdDark{GetArg}(\CKeyword{size\_t} ord) \CKeyword{->} \deegenKwdDark{TValue}}

  \item[\textcolor{Gray}{Desc.}] Return the \texttt{ord}-th argument (0-based) passed to this function. Note that this API is simply the short-hand for \texttt{\deegenKwdDark{GetStackBase}()[ord]}, so if you had modified \texttt{\deegenKwdDark{GetStackBase}()[ord]} to another value then you will just see that value.
  
\end{itemize}

\vspace{-0.5em}
\noindent\textcolor[RGB]{220,220,220}{\rule{\linewidth}{0.2pt}}

\begin{itemize}[leftmargin=2.5em]\setlength\itemsep{0.1em}
  \item[\textcolor{Gray}{API.}]  
  \texttt{\deegenKwdDark{GetStackFrameHeader}() \CKeyword{->} \CKeyword{const} \deegenKwdDark{StackFrameHeader}*}

  \item[\textcolor{Gray}{Desc.}] Return a pointer pointing to the stack frame header of the current function stack frame. The \deegenKwdDark{StackFrameHeader} contains information such as the caller frame address, the caller function and the call site where the call is made, which can be used to implement stack walking.
  
\end{itemize}

\vspace{-0.5em}
\noindent\textcolor[RGB]{220,220,220}{\rule{\linewidth}{0.2pt}}

\begin{itemize}[leftmargin=2.5em]\setlength\itemsep{0.1em}
  \item[\textcolor{Gray}{API.}]  
  \texttt{\deegenKwdDark{GetCurrentCoroutine}() \CKeyword{->} \deegenKwdDark{CoroutineContext}*}

  \item[\textcolor{Gray}{Desc.}] Return the current coroutine.
  
\end{itemize}

\vspace{-0.5em}
\noindent\textcolor[RGB]{220,220,220}{\rule{\linewidth}{0.2pt}}

\begin{itemize}[leftmargin=2.5em]\setlength\itemsep{0.1em}
  \item[\textcolor{Gray}{API.}]  
  \texttt{\deegenKwdDark{ThrowError}(\deegenKwdDark{TValue }val) \ \ \ \ \ \ \ \ \ \ \ \ \ \ \ \ \ \ \CKeyword{->} [[no\_return]]}

  \texttt{\deegenKwdDark{ThrowError}(\CKeyword{const char}* msg) \ \ \ \ \ \ \ \ \ \ \ \ \ \CKeyword{->} [[no\_return]]}

  \texttt{\deegenKwdDark{Return}(\deegenKwdDark{TValue}... vals) \ \ \ \ \ \ \ \ \ \ \ \ \ \ \ \ \ \ \CKeyword{->} [[no\_return]]}

  \texttt{\deegenKwdDark{Return}(\deegenKwdDark{TValue}* retStart, \CKeyword{size\_t} numRets) \CKeyword{->} [[no\_return]]}

  \item[\textcolor{Gray}{Desc.}] Same behavior as their counterparts used in the bytecode semantics, see \secref{appendix:bytecode-semantics-control-flow-transfer-apis}.
  
\end{itemize}

\vspace{-0.5em}
\noindent\textcolor[RGB]{220,220,220}{\rule{\linewidth}{0.2pt}}

\begin{itemize}[leftmargin=2.5em]\setlength\itemsep{0.1em}
  \item[\textcolor{Gray}{API.}]  
  \texttt{\deegenKwdDark{MakeInPlaceCall}(\deegenKwdDark{TValue}* args, \CKeyword{size\_t} nargs, \CKeyword{void}* retCont) \CKeyword{->} [[no\_return]]}

  \item[\textcolor{Gray}{Desc.}] Perform a in-place call to a guest language function (see \secref{appendix:guest-language-function-call-apis}). When the call returns, control will be transferred to the return continuation \texttt{retCont}. Note that unlike the \deegenKwdDark{MakeInPlaceCall} API used by the bytecode semantics, the return continuation in this API is a \texttt{\CKeyword{void}*} value, which can be obtained by the \deegenKwdDark{DEEGEN\_LIB\_FUNC\_RETURN\_CONTINUATION} API (to be covered in \secref{appendix:library-function-declaration-apis}). There is no non-in-place variant, since the stack frame is manually maintained by the library function, and Deegen does not know how many slots are currently in use. A \deegenKwdDark{MakeInPlaceTailCall} variant may be added in the future, but currently it is not available due to lack of use cases.
  
\end{itemize}

\vspace{-0.5em}
\noindent\textcolor[RGB]{220,220,220}{\rule{\linewidth}{0.2pt}}

\begin{itemize}[leftmargin=2.5em]\setlength\itemsep{0.1em}
  \item[\textcolor{Gray}{API.}]  
  \texttt{\deegenKwdDark{LongJump}(\deegenKwdDark{StackFrameHeader}* dst, \deegenKwdDark{TValue}* retStart, \CKeyword{size\_t} numRets)}

  \texttt{\ \ \CKeyword{->} [[no\_return]]} 
  
  \item[\textcolor{Gray}{Desc.}] Directly return to an ancestor caller on the stack (the caller of the stack frame identified by \texttt{dst}), similar to C \texttt{longjmp} (however, its internal implementation has nothing to do with C \texttt{longjmp}). This API is designed as a low-level primitive for users to implement exceptions.
  
\end{itemize}

\vspace{-0.5em}
\noindent\textcolor[RGB]{220,220,220}{\rule{\linewidth}{0.2pt}}

\begin{itemize}[leftmargin=2.5em]\setlength\itemsep{0.1em}
  \item[\textcolor{Gray}{API.}]  
  \texttt{\deegenKwdDark{CoroSwitch}(\deegenKwdDark{CoroutineContext}* dst, \CKeyword{uint64\_t} ctxVal) \CKeyword{->} [[no\_return]]} 
  
  \item[\textcolor{Gray}{Desc.}] Transfer control to another coroutine. \texttt{ctxVal} is an opaque value for users to communicate more context (for example, it may be used to indicate how many arguments are passed to the yield or resume, with the actual arguments stored on the stack). This API is designed as a low-level primitive for users to implement coroutine yield and resume semantics.
  
\end{itemize}

\vspace{-0.5em}
\noindent\textcolor[RGB]{220,220,220}{\rule{\linewidth}{0.2pt}}

\begin{itemize}[leftmargin=2.5em]\setlength\itemsep{0.1em}
  \item[\textcolor{Gray}{API.}]  
  \texttt{\deegenKwdDark{GetReturnValues}() \CKeyword{->} \deegenKwdDark{TValue}*}

  \texttt{\deegenKwdDark{GetNumReturnValues}() \CKeyword{->} \CKeyword{size\_t}}
  
  \item[\textcolor{Gray}{Desc.}] May only be used in a library function return continuation. Access the return values returned by the guest language call.
  
\end{itemize}

\subsubsection{Library Function Declaration}\label{appendix:library-function-declaration-apis}

\ 

\vspace{-0.3em}
\noindent\textcolor[RGB]{220,220,220}{\rule{\linewidth}{0.2pt}}

\begin{itemize}[leftmargin=2.5em]\setlength\itemsep{0.1em}
  \item[\textcolor{Gray}{API.}]  
  \texttt{\deegenKwdDark{DEEGEN\_FORWARD\_DECLARE\_LIB\_FUNC}(name);}

  \texttt{\deegenKwdDark{DEEGEN\_FORWARD\_DECLARE\_LIB\_FUNC\_RETURN\_CONTINUATION}(name);}

  \item[\textcolor{Gray}{Desc.}] Forward declare a library function or a return continuation of a library function. This is useful for mutual recursion, and for external code that needs to access the function pointer. 
  
\end{itemize}

\vspace{-0.5em}
\noindent\textcolor[RGB]{220,220,220}{\rule{\linewidth}{0.2pt}}

\begin{itemize}[leftmargin=2.5em]\setlength\itemsep{0.1em}
  \item[\textcolor{Gray}{API.}]  
  \texttt{\deegenKwdDark{DEEGEN\_LIB\_FUNC\_RETURN\_CONTINUATION}(name)}

  \item[\textcolor{Gray}{Desc.}] Returns the entry point of a return continuation that can be used in the \deegenKwdDark{MakeInPlaceCall} API (see \secref{appendix:library-function-execution-semantics-apis}) in the library function execution semantics. 
  
\end{itemize}

\vspace{-0.5em}
\noindent\textcolor[RGB]{220,220,220}{\rule{\linewidth}{0.2pt}}

\begin{itemize}[leftmargin=2.5em]\setlength\itemsep{0.1em}
  \item[\textcolor{Gray}{API.}]  
  \texttt{\deegenKwdDark{DEEGEN\_CODE\_POINTER\_FOR\_LIB\_FUNC}(name)}

  \item[\textcolor{Gray}{Desc.}] Returns the entry point (a \texttt{\CKeyword{void}*} value) of a library function.
  
\end{itemize}

\vspace{-0.5em}
\noindent\textcolor[RGB]{220,220,220}{\rule{\linewidth}{0.2pt}}

\begin{itemize}[leftmargin=2.5em]\setlength\itemsep{0.1em}
  \item[\textcolor{Gray}{API.}]  
  \texttt{\deegenKwdDark{DEEGEN\_DEFINE\_LIB\_FUNC}(name) \{ ... \}}

  \texttt{\deegenKwdDark{DEEGEN\_DEFINE\_LIB\_FUNC\_CONTINUATION}(name) \{ ... \}}

  \item[\textcolor{Gray}{Desc.}] Define the implementation of a library function, or a return continuation of a library function.
  
\end{itemize}

\vspace{-0.5em}
\noindent\textcolor[RGB]{220,220,220}{\rule{\linewidth}{0.2pt}}

\begin{itemize}[leftmargin=2.5em]\setlength\itemsep{0.1em}
  \item[\textcolor{Gray}{API.}]  
  \texttt{\deegenKwdDark{DEEGEN\_END\_LIB\_FUNC\_DEFINITIONS}}

  \item[\textcolor{Gray}{Desc.}]  A translation unit (C++ file) may define any number of library functions using the above-mentioned APIs. This macro must be put at the end of the file. Internally, this macro uses C++ macro and metaprogramming tricks to automatically collect all definitions in this translation unit into a \texttt{constexpr} array, so Deegen can work through each of the defined functions. 
  
\end{itemize}

\newpage

\section{The Life of the Add Bytecode}\label{appendix:life-of-the-add-bytecode}


This appendix elaborates step-by-step how Deegen lowers the C++ execution semantics of LuaJIT Remake's \texttt{Add} bytecode to the baseline JIT code generator illustrated in \figref{fig:baseline-codegen}. 

\figref{fig:baseline-jit-generation-pipeline} illustrated the pipeline that generates the baseline JIT. The rest of the section will use the \texttt{Add} bytecode as example to explain what happens in each pass of the \figref{fig:baseline-jit-generation-pipeline} pipeline.

\subsection{The C++ Bytecode Semantics}

Below is the real C++ implementation of the semantics for arithmetic bytecodes in LJR.

\vspace{0.4em}

\noindent\fboxsep=.5em\fbox{\begin{minipage}{0.97\textwidth}

\texttt{\CComment{// Execution semantics of the arithmetic bytecodes}}

\texttt{\CKeyword{template}<\CType{LuaMetamethodKind} opKind>}

\texttt{\CKeyword{static void} ArithmeticOperationImpl(\deegenKwdDark{TValue} lhs, \deegenKwdDark{TValue} rhs) \{}

\texttt{\ \ \CKeyword{if} (likely(lhs.\deegenKwdDark{Is}<\CType{tDouble}>() \&\& rhs.\deegenKwdDark{Is}<\CType{tDouble}>())) \{}

\texttt{\ \ \ \ \CKeyword{double} ld = lhs.\deegenKwdDark{As}<\CType{tDouble}>();}

\texttt{\ \ \ \ \CKeyword{double} rd = rhs.\deegenKwdDark{As}<\CType{tDouble}>();}

\texttt{\ \ \ \ \CKeyword{double} res;}

\texttt{\ \ \ \ \CKeyword{if constexpr}(opKind == \CType{LuaMetamethodKind}::Add) \{}

\texttt{\ \ \ \ \ \ res = ld + rd;}

\texttt{\ \ \ \ \} \CKeyword{else} \{}

\texttt{\ \ \ \ \ \ \CComment{/* logic for other arithmetic operators redacted */} }

\texttt{\ \ \ \ \}}

\texttt{\ \ \ \ \deegenKwdDark{Return}(\deegenKwdDark{TValue}::\deegenKwdDark{Create}<\CType{tDouble}>(res));}

\texttt{\ \ \} \CKeyword{else} \{ }

\texttt{\ \ \ \ \CComment{/* logic for arithmetic operation behavior on non-numbers redacted */}}

\texttt{\ \ \}}

\texttt{\}}

\texttt{\CComment{// Templated bytecode specifications for all arithmetic bytecodes}}

\texttt{\deegenKwdDark{DEEGEN\_DEFINE\_BYTECODE\_TEMPLATE}(ArithOperation, \CType{LuaMetamethodKind} opKind) \{}

\texttt{\ \ \deegenKwdDark{Operands}(}

\texttt{\ \ \ \ \deegenKwdDark{LocalOrConstant}("lhs"),}

\texttt{\ \ \ \ \deegenKwdDark{LocalOrConstant}("rhs")}

\texttt{\ \ );}

\texttt{\ \ \deegenKwdDark{Result}(\deegenKwdDark{BytecodeValue});}

\texttt{\ \ \deegenKwdDark{Implementation}(ArithmeticOperationImpl<opKind>);}

\texttt{\ \ \deegenKwdDark{Variant}(}

\texttt{\ \ \ \ \deegenKwdDark{Op}("lhs").\deegenKwdDark{IsLocal}(),}

\texttt{\ \ \ \ \deegenKwdDark{Op}("rhs").\deegenKwdDark{IsLocal}()}

\texttt{\ \ ).\deegenKwdDark{EnableTypeBasedCodeSplitting}(}

\texttt{\ \ \ \ \deegenKwdDark{Op}("lhs").\deegenKwdDark{HasType}<\CType{tDoubleNotNaN}>(),}

\texttt{\ \ \ \ \deegenKwdDark{Op}("rhs").\deegenKwdDark{HasType}<\CType{tDoubleNotNaN}>()}

\texttt{\ \ );}

\texttt{\ \ \CComment{/* A few more variants omitted */}}

\texttt{\}}

\texttt{\CComment{// Define the Add bytecode by instantiating the template above}}

\texttt{\deegenKwdDark{DEEGEN\_DEFINE\_BYTECODE\_BY\_TEMPLATE\_INSTANTIATION(}}

\texttt{\ \ Add, ArithOperation, \CType{LuaMetamethodKind}::Add);}

\texttt{\CComment{/* A few more instantiations of other arithemtic bytecodes omitted */}}

\end{minipage}}

\vspace{0.6em}

See \secref{appendix:full-deegen-api-reference} for the documentations of the Deegen APIs (colored in \deegenKwdDark{purple}) used in the code. The code above defines only one variant of the \texttt{Add} bytecode: the one that adds a local variable to another local variable (for brevity, we use \texttt{Add\_0} to refer to this bytecode variant). Throughout this section, we will use \texttt{Add\_0} as our example.

\subsection{Compile to LLVM IR}

The first step of the pipeline is to compile the C++ semantics to LLVM IR. Importantly, all LLVM passes are disabled (via the \texttt{-Xclang -disable-llvm-passes} flag), so the LLVM IR is the raw IR generated by the Clang frontend, and even the functions marked as \texttt{always\_inline} are not inlined. 

At this stage, we add \texttt{no\_inline} to all Deegen APIs that are not magic placeholders (e.g., the \texttt{\deegenKwdDark{TValue}::\deegenKwdDark{Is}<\CType{T}>()} API), so that they will not be inlined by LLVM, and we can always reliably identify them in the IR.

We then run the standard LLVM passes to build up the SSA form, and analyze the IR to figure out all the bytecodes (and their variants) defined in this translation unit. 

For each bytecode variant, we analyze the control flow APIs (guest language function call APIs and \deegenKwdDark{EnterSlowPath}) in the execution semantics to figure out all the components of the bytecode. Specifically, our \texttt{Add\_0} bytecode has a main component and a return continuation component (as the logic that handles the non-number behavior involves calling a Lua function, which is redacted from the code snippet above). 

At this stage, the LLVM IR of the main component of our \texttt{Add\_0} bytecode looks like this:

\vspace{0.3em}

\texttt{\llvmKwd{define dso\_local} \CType{void} @\_\_deegen\_bytecode\_Add\_0\_impl(\CType{i64} \ssa{\%0}, \CType{i64} \ssa{\%1})}

\texttt{\{}

\texttt{\ \ \ssa{\%8} = \llvmKwd{call} \CType{zeroext} \CType{i1} \deegenKwdDark{@\_Z19DeegenImpl\_TValueIsI7tDoubleEb6TValue}(\CType{i64} \ssa{\%0})}

\texttt{\ \ \llvmKwd{br} \CType{i1} \ssa{\%8}, \llvmKwd{label} \ssa{\%.\_crit\_edge}, \llvmKwd{label} \ssa{\%.non\_double\_add}}

\

\texttt{\ssa{.\_crit\_edge:}}

\texttt{\ \ \ssa{\%9} = \llvmKwd{call} \CType{zeroext i1} \deegenKwdDark{@\_Z19DeegenImpl\_TValueIsI7tDoubleEb6TValue}(\CType{i64} \ssa{\%1})}

\texttt{\ \ \llvmKwd{br} \CType{i1} \ssa{\%9}, \llvmKwd{label} \ssa{\%.double\_add}, \llvmKwd{label} \ssa{\%.non\_double\_add}}

\ 

\texttt{\ssa{.double\_add:}}

\texttt{\ \ \ssa{\%11} = \llvmKwd{call} \CType{double} \deegenKwdDark{@\_Z19DeegenImpl\_TValueAsI7tDoubleEDa6TValue}(\CType{i64} \ssa{\%0})}

\texttt{\ \ \ssa{\%12} = \llvmKwd{call} \CType{double} \deegenKwdDark{@\_Z19DeegenImpl\_TValueAsI7tDoubleEDa6TValue}(\CType{i64} \ssa{\%1})}

\texttt{\ \ \ssa{\%13} = \llvmKwd{fadd} \CType{double} \ssa{\%11}, \ssa{\%12}}

\texttt{\ \ \ssa{\%14} = \llvmKwd{bitcast} \CType{double} \ssa{\%13} \llvmKwd{to} \CType{i64}}

\texttt{\ \ \llvmKwd{call} \CType{void} \deegenKwdDark{@DeegenImpl\_ReturnValue}(\CType{i64} \ssa{\%14})}

\texttt{\ \ \llvmKwd{unreachable}}

\ 

\texttt{\ssa{.non\_double\_add:}}     

\texttt{\ \ ; logic implementing semantics for adding non-numbers}

\texttt{\ \ ; $\sim$250 lines of LLVM IR redacted}

\texttt{\}}

\vspace{0.3em}

As one can see, the LLVM IR is a straightforward translation of the C++ semantics. The LLVM IR logic calls the \texttt{\deegenKwdDark{TValue}::\deegenKwdDark{Is}<\CType{tDouble}>()} API (which in LLVM IR, becomes the mangled function name \deegenKwdDark{@\_Z19DeegenImpl\_TValueAsI7tDoubleEDa6TValue}, but Deegen understands this) to check if the two operands are double. If not, it branches to the code path that handles the non-double add (redacted). Otherwise, it calls the \texttt{\deegenKwdDark{TValue}::\deegenKwdDark{As}<\CType{tDouble}>()} API to unbox the values, performs the double addition, call the \texttt{\deegenKwdDark{TValue}::\deegenKwdDark{Create}<\CType{tDouble}>()} API to box the result into a boxed value, and call the \texttt{\deegenKwdDark{Return}} API to dispatch to the next bytecode. The \texttt{\deegenKwdDark{Return}} API (which name in LLVM IR becomes \deegenKwdDark{@DeegenImpl\_ReturnValue}) takes away control, so a \llvmKwd{unreachable} follows.

\subsection{Speculatively Optimize Execution Semantics}

The next step in the pipeline runs the type-based optimizations (\secref{sec:type-based-optimization}). Specifically, as shown in the code, the bytecode specification of \texttt{Add\_0} asked Deegen to \deegenKwdDark{EnableTypeBasedCodeSplitting} based on the speculation that both operands are likely \CType{tDoubleNotNaN}.

As the result of the optimization, the execution semantics is splitted into two functions: 
\begin{itemize}
    \item The fast path assumes (without checking!) that both operands are \CType{tDoubleNotNaN}, and runs the optimized semantics based on such assumptions.
    \item The slow path assumes (without checking!) that at least one of the two operands is not a \CType{tDoubleNotNaN}, and runs the optimized semantics based on such assumptions.
\end{itemize} 

The LLVM IR for the optimized fast path is shown below:

\vspace{0.3em}

\texttt{\llvmKwd{define dso\_local} \CType{void} @\_\_deegen\_bytecode\_Add\_0\_impl(\CType{i64} \ssa{\%0}, \CType{i64} \ssa{\%1})}

\texttt{\{}

\texttt{\ \ \ssa{\%2} = \llvmKwd{call} \CType{double} \deegenKwdDark{@\_Z19DeegenImpl\_TValueAsI7tDoubleEDa6TValue}(\CType{i64} \ssa{\%0})}

\texttt{\ \ \ssa{\%3} = \llvmKwd{call} \CType{double} \deegenKwdDark{@\_Z19DeegenImpl\_TValueAsI7tDoubleEDa6TValue}(\CType{i64} \ssa{\%1})}

\texttt{\ \ \ssa{\%4} = \llvmKwd{fadd} \CType{double} \ssa{\%2}, \ssa{\%3}}

\texttt{\ \ \ssa{\%5} = \llvmKwd{bitcast} \CType{double} \ssa{\%4} \llvmKwd{to} \CType{i64}}

\texttt{\ \ \llvmKwd{call} \CType{void} \deegenKwdDark{@DeegenImpl\_ReturnValue}(\CType{i64} \ssa{\%5})}

\texttt{\ \ \llvmKwd{unreachable}}

\texttt{\}}

\vspace{0.3em}

As one can see, the logic that checks if the operands are \CType{tDouble} is eliminated, since it already \textit{assumes} that they are \CType{tDoubleNotNaN}, which means they must be \CType{tDouble}. 

Of course, this function is only correct if the two operands are actually \CType{tDoubleNotNaN} (which is not necessarily true!), but we will deal with this right afterwards.

As an important side effect of this optimization, the fast path now never makes a guest language function call. So the \texttt{Add\_0} bytecode now consists of a main component, a slow path component (generated from this optimization), and a return continuation component that is only used by the slow path component. Furthermore, since the return continuation is only used by the slow path, there is no reason to compile it into JIT code. So as a result, the JIT code for \texttt{Add\_0} will only consist of the main component, and the AOT code for \texttt{Add\_0} will consist of the generated slow path component and the return continuation component.

Since our goal is to explain the baseline JIT generation process, we will put aside the AOT logic (the generated slow path component, and the return continuation component), and only focus on how the main component (the code above) is further transformed down the pipeline.

\subsection{Generate Bytecode Implementation}

The next step in the pipeline is to generate the concrete implementation of the bytecode. 

As explained in \secref{sec:interpreter-generator}, we use the GHC calling convention to achieve register pinning, and each argument in the function prototype implicitly corresponds to a physical register. So the bytecode implementation is a GHCcc function that implements the full logic of the bytecode.

Inside the function, we need to decode the bytecode operands, and load data from the local variables or the constant table. In a baseline JIT, however, the bytecode operands are runtime constants. So instead of actually decoding the bytecode operands from the bytecode, we simply use a magic placeholder function call to identify them in this step. 

Finally, recall that the fast path function we obtained from the last step does not check whether our type assumptions actually hold. So we must check if the type assumptions hold now. If yes, we can call the optimized fast path. Otherwise, we transfer control to the AOT slow path.

This builds up an LLVM IR function as shown below:

\vspace{0.3em}

\texttt{\llvmKwd{define dso\_local ghccc} \CType{void} @\_\_deegen\_bytecode\_Add\_0(}

\texttt{\ \ \CType{ptr} \ssa{\%coroutineCtx}, \CType{ptr} \ssa{\%stackBase}, \CType{ptr} \ssa{\%0}, \CType{ptr} \ssa{\%codeBlock}, }

\texttt{\ \ \CType{i64} \ssa{\%tagRegister1}, \CType{ptr} \ssa{\%2}, \CType{i64} \ssa{\%3}, \CType{i64} \ssa{\%4}, \CType{i64} \ssa{\%5}, \CType{i64} \ssa{\%tagRegister2}, }

\texttt{\ \ \CType{double} \ssa{\%7}, \CType{double} \ssa{\%8}, \CType{double} \ssa{\%9}, \CType{double} \ssa{\%10}, \CType{double} \ssa{\%11}, \CType{double} \ssa{\%12})}

\texttt{\{}

\texttt{\ \ \ssa{\%lhsSlot} = \llvmKwd{call} \CType{i64} \deegenKwdDark{@\_\_deegen\_constant\_placeholder\_bc\_operand\_0}()}

\texttt{\ \ \ssa{\%rhsSlot} = \llvmKwd{call} \CType{i64} \deegenKwdDark{@\_\_deegen\_constant\_placeholder\_bc\_operand\_1}()}



\texttt{\ \ \ssa{\%16} = \llvmKwd{getelementptr inbounds} \CType{i64}, \CType{ptr} \ssa{\%stackBase}, \CType{i64} \ssa{\%lhsSlot}}

\texttt{\ \ \ssa{\%lhs} = \llvmKwd{load} \CType{i64}, \CType{ptr} \ssa{\%16}, \llvmKwd{align} 8}

\texttt{\ \ \ssa{\%17} = \llvmKwd{getelementptr inbounds} \CType{i64}, \CType{ptr} \ssa{\%stackBase}, \CType{i64} \ssa{\%rhsSlot}}

\texttt{\ \ \ssa{\%rhs} = \llvmKwd{load} \CType{i64}, \CType{ptr} \ssa{\%17}, \llvmKwd{align} 8}

\texttt{\ \ \ssa{\%18} = \llvmKwd{call} \CType{i1} \deegenKwdDark{@\_Z19DeegenImpl\_TValueIsI13tDoubleNotNaNEb6TValue}(\CType{i64} \ssa{\%lhs})}

\texttt{\ \ \llvmKwd{br} \CType{i1} \ssa{\%18}, \llvmKwd{label} \ssa{\%25}, \llvmKwd{label} \ssa{\%slowpath}}

\texttt{\ssa{25:}}

\texttt{\ \ \ssa{\%26} = \llvmKwd{call} \CType{i1} \deegenKwdDark{@\_Z19DeegenImpl\_TValueIsI13tDoubleNotNaNEb6TValue}(\CType{i64} \ssa{\%rhs})}

\texttt{\ \ \llvmKwd{br} \CType{i1} \ssa{\%26}, \llvmKwd{label} \ssa{\%28}, \llvmKwd{label} \ssa{\%slowpath}}

\texttt{\ssa{28:}}

\texttt{\ \ \llvmKwd{call} \CType{void} @\_\_deegen\_bytecode\_Add\_0\_impl(\CType{i64} \ssa{\%lhs}, i64 \ssa{\%rhs})}
  
\texttt{\ \ \llvmKwd{unreachable}}
  
\texttt{\ssa{slowpath:}}

\texttt{\ \ \ssa{\%20} = \llvmKwd{call} \CType{i64} \deegenKwdDark{@\_\_deegen\_constant\_placeholder\_bc\_operand\_103}()}

\texttt{\ \ \ssa{\%21} = \llvmKwd{call} \CType{ptr} \deegenKwdDark{@\_\_DeegenImpl\_GetBaselineJitCBFromCodeBlock}(\CType{ptr} \ssa{\%codeBlock})}

\texttt{\ \ \ssa{\%slowPathDataPtr} = \llvmKwd{getelementptr inbounds} \CType{i8}, \CType{ptr} \ssa{\%21}, \CType{i64} \ssa{\%20}}

\texttt{\ \ \ssa{\%23} = \llvmKwd{bitcast} \CType{i64} \ssa{\%rhs} \llvmKwd{to} \CType{double}}

\texttt{\ \ \ssa{\%24} = \llvmKwd{bitcast} \CType{i64} \ssa{\%lhs} \llvmKwd{to} \CType{double}}

\texttt{\ \ \llvmKwd{musttail call ghccc} \CType{void} @\_\_deegen\_bytecode\_Add\_0\_quickening\_slowpath(}

\texttt{\ \ \ \ \CType{ptr} \ssa{\%coroutineCtx}, \CType{ptr} \ssa{\%stackBase}, \CType{ptr} \ssa{\%slowPathDataPtr}, \CType{ptr} \ssa{\%codeBlock},}

\texttt{\ \ \ \ \CType{i64} \ssa{\%tagRegister1}, \CType{ptr} \llvmKwd{undef}, \CType{i64} \llvmKwd{undef}, \CType{i64} \llvmKwd{undef}, \CType{i64} \llvmKwd{undef},}

\texttt{\ \ \ \ \CType{i64} \ssa{\%tagRegister2}, \CType{double} \ssa{\%24}, \CType{double} \ssa{\%23}, \CType{double} \llvmKwd{undef},}

\texttt{\ \ \ \ \CType{double} \llvmKwd{undef}, \CType{double} \llvmKwd{undef}, \CType{double} \llvmKwd{undef})}

\texttt{\ \ \llvmKwd{ret} \CType{void}}

\texttt{\}}

\vspace{0.3em}

The IR is a bit long, but what it is doing is pretty straightforward. The local variable ordinal of the two operands (\ssa{\%lhsSlot} and \ssa{\%rhsSlot}) are runtime constants, so we obtain their values using two magic function calls. We then load the two operands (\ssa{\%lhs} and \ssa{\%rhs}) from the stack frame.

Next, we check if they satisfy the type assumption, that is, if both operands are \CType{tDoubleNotNaN}. If yes, we can safely call the optimized fast path implementation (\texttt{@\_\_deegen\_bytecode\_Add\_0\_impl}) generated in the previous step. Otherwise, we transfer control to the AOT slow path by a tail call. 

The control transfer to the AOT slow path takes a bit work. The JIT code does not track the bytecode pointer, since everything in the bytecode are runtime constants and directly burnt into the instruction flow. However, the AOT slow path needs the SlowPathData pointer to access the relevant information of the bytecode. Therefore, we first obtain the runtime constant offset value (\ssa{\%20}) of the SlowPathData pointer in the SlowPathData stream (which is a small integer representable by a signed 32-bit integer) using a magic function call, then compute the SlowPathData pointer from there. Finally, note that we also pass the already-loaded value of \ssa{\%lhs} and \ssa{\%rhs} to the slow path so they don't have to be decoded again: this is a very minor optimization, and we do it only because it's a free optimization (LLVM is smart enough to put \ssa{\%lhs} and \ssa{\%rhs} directly in the respective machine registers, so no \texttt{mov} or any extra machine instructions are needed).

\subsection{Lower Deegen APIs for Baseline JIT}

The next step in the pipeline is to lower all Deegen APIs (everything in \deegenKwdDark{purple}) to concrete baseline JIT implementations. For example, we need to lower the \deegenKwdDark{Return} API in \texttt{Add\_0} to the concrete logic that stores the result to the stack frame, and transfer control to the next bytecode.

To do this, we mark the execution semantics (in our case, \texttt{@\_\_deegen\_bytecode\_Add\_0\_impl}) as \texttt{always\_inline}, then let LLVM inline it into our GHCcc bytecode implementation function. The execution semantics then have access to all the VM contexts, so all the Deegen APIs used in the execution semantics can be lowered to concrete implementations accordingly. 

After this pass, the LLVM IR for our GHCcc bytecode implementation function is as below:

\vspace{0.3em}

\texttt{\llvmKwd{define dso\_local ghccc} \CType{void} @\_\_deegen\_bytecode\_Add\_0(}

\texttt{\ \ \CType{ptr} \ssa{\%coroutineCtx}, \CType{ptr} \ssa{\%stackBase}, \CType{ptr} \ssa{\%0}, \CType{ptr} \ssa{\%codeBlock}, }

\texttt{\ \ \CType{i64} \ssa{\%tagRegister1}, \CType{ptr} \ssa{\%2}, \CType{i64} \ssa{\%3}, \CType{i64} \ssa{\%4}, \CType{i64} \ssa{\%5}, \CType{i64} \ssa{\%tagRegister2}, }

\texttt{\ \ \CType{double} \ssa{\%7}, \CType{double} \ssa{\%8}, \CType{double} \ssa{\%9}, \CType{double} \ssa{\%10}, \CType{double} \ssa{\%11}, \CType{double} \ssa{\%12})}

\texttt{\{}

\texttt{\ \ \ssa{\%lhsSlot} = \llvmKwd{call} \CType{i64} \deegenKwdDark{@\_\_deegen\_constant\_placeholder\_bc\_operand\_0}()}

\texttt{\ \ \ssa{\%rhsSlot} = \llvmKwd{call} \CType{i64} \deegenKwdDark{@\_\_deegen\_constant\_placeholder\_bc\_operand\_1}()}

\texttt{\ \ \ssa{\%16} = \llvmKwd{getelementptr inbounds} \CType{i64}, \CType{ptr} \ssa{\%stackBase}, \CType{i64} \ssa{\%lhsSlot}}

\texttt{\ \ \ssa{\%lhs} = \llvmKwd{load} \CType{double}, \CType{ptr} \ssa{\%16}, \llvmKwd{align} 8}

\texttt{\ \ \ssa{\%17} = \llvmKwd{getelementptr inbounds} \CType{i64}, \CType{ptr} \ssa{\%stackBase}, \CType{i64} \ssa{\%rhsSlot}}

\texttt{\ \ \ssa{\%rhs} = \llvmKwd{load} \CType{double}, \CType{ptr} \ssa{\%17}, \llvmKwd{align} 8}

\texttt{\ \ \ssa{\%18} = \llvmKwd{fcmp ord} \CType{double} \ssa{\%lhs}, \ssa{\%rhs}}

\texttt{\ \ \llvmKwd{br} \CType{i1} \ssa{\%18}, \llvmKwd{label} \ssa{\%24}, \llvmKwd{label} \ssa{\%slowpath}}

\texttt{\ssa{24:}}

\texttt{\ \ \ssa{\%fallthruDst} = \llvmKwd{call} \CType{ptr} \deegenKwdDark{@\_\_deegen\_constant\_placeholder\_bc\_operand\_101}()}

\texttt{\ \ \ssa{\%outputSlot} = \llvmKwd{call} \CType{i64} \deegenKwdDark{@\_\_deegen\_constant\_placeholder\_bc\_operand\_100}()}

\texttt{\ \ \ssa{\%25} = \llvmKwd{fadd} \CType{double} \ssa{\%lhs}, \ssa{\%rhs}}

\texttt{\ \ \ssa{\%outputPtr} = \llvmKwd{getelementptr inbounds} \CType{i64}, \CType{ptr} \ssa{\%stackBase}, \CType{i64} \ssa{\%outputSlot}}

\texttt{\ \ \llvmKwd{store} \CType{double} \ssa{\%25}, \CType{ptr} \ssa{\%outputPtr}, \llvmKwd{align} 8}

\texttt{\ \ \llvmKwd{musttail call ghccc} \CType{void} \ssa{\%fallthruDst}(}
  
\texttt{\ \ \ \ \CType{ptr} \ssa{\%coroutineCtx}, \CType{ptr} \ssa{\%stackBase}, \CType{ptr} \llvmKwd{undef}, \CType{ptr} \ssa{\%codeBlock},}

\texttt{\ \ \ \ \CType{i64} \ssa{\%tagRegister1}, \CType{ptr} \llvmKwd{undef}, \CType{i64} \llvmKwd{undef}, \CType{i64} \llvmKwd{undef}, \CType{i64} \llvmKwd{undef},}

\texttt{\ \ \ \ \CType{i64} \ssa{\%tagRegister2}, \CType{double} \llvmKwd{undef}, \CType{double} \llvmKwd{undef}, \CType{double} \llvmKwd{undef},}

\texttt{\ \ \ \ \CType{double} \llvmKwd{undef}, \CType{double} \llvmKwd{undef}, \CType{double} \llvmKwd{undef})}

\texttt{\ \ \llvmKwd{ret} \CType{void}}

\texttt{\ssa{slowpath:}}

\texttt{\ \ \ssa{\%20} = \llvmKwd{call} \CType{i64} \deegenKwdDark{@\_\_deegen\_constant\_placeholder\_bc\_operand\_103}()}

\texttt{\ \ \ssa{\%21} = \llvmKwd{getelementptr inbounds} \CType{\%CodeBlockTy}, \CType{ptr} \ssa{\%codeBlock}, \CType{i64} 0, \CType{i32} 8}

\texttt{\ \ \ssa{\%22} = \llvmKwd{load} \CType{ptr}, \CType{ptr} \ssa{\%21}, \llvmKwd{align} 8}

\texttt{\ \ \ssa{\%slowPathDataPtr} = \llvmKwd{getelementptr inbounds} \CType{i8}, \CType{ptr} \ssa{\%22}, \CType{i64} \ssa{\%20}}

\texttt{\ \ \llvmKwd{musttail call ghccc} \CType{void} @\_\_deegen\_bytecode\_Add\_0\_quickening\_slowpath(}

\texttt{\ \ \ \ \CType{ptr} \ssa{\%coroutineCtx}, \CType{ptr} \ssa{\%stackBase}, \CType{ptr} \ssa{\%slowPathDataPtr}, \CType{ptr} \ssa{\%codeBlock},}

\texttt{\ \ \ \ \CType{i64} \ssa{\%tagRegister1}, \CType{ptr} \llvmKwd{undef}, \CType{i64} \llvmKwd{undef}, \CType{i64} \llvmKwd{undef}, \CType{i64} \llvmKwd{undef},}

\texttt{\ \ \ \ \CType{i64} \ssa{\%tagRegister2}, \CType{double} \ssa{\%lhs}, \CType{double} \ssa{\%rhs}, \CType{double} \llvmKwd{undef},}

\texttt{\ \ \ \ \CType{double} \llvmKwd{undef}, \CType{double} \llvmKwd{undef}, \CType{double} \llvmKwd{undef})}

\texttt{\ \ \llvmKwd{ret} \CType{void}}

\texttt{\}}

\vspace{0.3em}

As one can see, all the Deegen APIs (except the magic functions we internally inserted to identify runtime constants) are lowered to concrete implementations. For example, to lower \deegenKwdDark{Return}, we call magic functions to get the runtime constant of the output slot ordinal (\ssa{\%outputSlot}) and the JIT code address of the next bytecode (\ssa{\%fallthruDst}), store the result to slot \ssa{\%outputSlot} of the stack frame, then transfer control to \ssa{\%fallthruDst} by a tail call.

\subsection{Identify Runtime Constant Expressions}

The next step in the pipeline is to identify all the expressions that can be evaluated at compile time, and statically analyze their ranges (see \secref{sec:baseline-jit-generation} for detail). 

One can easily identify all the constant expressions in the LLVM IR by traversing all uses of the \deegenKwdDark{\_\_deegen\_constant\_placeholder\_bc\_operand\_*} functions. The only tricky thing is that the \llvmKwd{getelementptr} instruction offset may be a runtime constant, even if the base is not (for example, \ssa{\%16} and \ssa{\%17} in the code above).  The range analysis is also a  traditional compiler analysis pass. 

Specifically, we know the possible range of all the ``base'' runtime constants:
\begin{itemize}
    \item A constant that represents a local variable ordinal has range $[0,10^6]$ ($10^6$ is a somewhat arbitrary limitation we set for the maximum number of locals allowed in a function). This applies to \deegenKwdDark{bc\_operand\_0/1} (input operands) and \deegenKwdDark{bc\_operand\_100} (the output slot).
    \item The maximum length of the SlowPathData stream is 256MB (also a somewhat arbitrary limitation we set). This applies to \deegenKwdDark{bc\_operand\_103} 
    (slowPathDataOffset).
    \item A constant that represents a code pointer has range $[1, 2^{31}-2^{24})$ due to x86-64 small code model ABI. This applies to \deegenKwdDark{bc\_operand\_101} (JIT code address for the fallthrough bytecode).
\end{itemize}

So after the analysis, we will detect the following runtime constant expressions as well as their statically-provable ranges:

\vspace{0.3em}

\texttt{\ssa{\%lhsSlot} := \deegenKwdDark{bc\_operand\_0}\ \ \ \ \ \ \ \ \ \ \ \ \ \ \ \ \ \ \ \ }-- proven range $[0, 10^6]$

\texttt{\ssa{\%16}.\llvmKwd{gep-offset} := \deegenKwdDark{bc\_operand\_0} * 8\ \ \ \ \ \ \ \ \ \ }-- proven range [$0, 8\times 10^6$]

\texttt{\ssa{\%rhsSlot} := \deegenKwdDark{bc\_operand\_1}\ \ \ \ \ \ \ \ \ \ \ \ \ \ \ \ \ \ \ \ }-- proven range $[0, 10^6]$

\texttt{\ssa{\%17}.\llvmKwd{gep-offset} := \deegenKwdDark{bc\_operand\_1} * 8\ \ \ \ \ \ \ \ \ \ }-- proven range [$0, 8\times 10^6$]

\texttt{\ssa{\%fallthruDst} := \deegenKwdDark{bc\_operand\_101}\ \ \ \ \ \ \ \ \ \ \ \ \ \ }-- proven range $[1, 2^{31}-2^{24})$

\texttt{\ssa{\%outputSlot} := \deegenKwdDark{bc\_operand\_100}\ \ \ \ \ \ \ \ \ \ \ \ \ \ \ }-- proven range $[0, 10^6]$

\texttt{\ssa{\%outputPtr}.\llvmKwd{gep-offset} := \deegenKwdDark{bc\_operand\_100} * 8\ }-- proven range [$0, 8\times 10^6$]

\texttt{\ssa{\%20} := \deegenKwdDark{bc\_operand\_103}\ \ \ \ \ \ \ \ \ \ \ \ \ \ \ \ \ \ \ \ \ \ \ }-- proven range $[1, 2^{28}]$

\subsection{Insert Copy-and-Patch Stencil Holes}

The next step is to replace all the runtime constant expressions in the LLVM IR with copy-and-patch stencil holes (i.e., addresses of external symbols).

As explained in \secref{sec:baseline-jit-generation}, a constant expression may only be replaced by a stencil hole if its range fits the ABI assumption of $[1, 2^{31}-2^{24})$. If the length of the the proven range fits the assumption, but the range itself does not fit because it covers zero or negative integers, we can add an adjustment value to the expression to make it fit, and subtract the value back in LLVM IR. For example, \texttt{\ssa{\%17}.\llvmKwd{gep-offset}} is the constant expression \texttt{\deegenKwdDark{bc\_operand\_100} * 8}, which has range $[0, 8\times 10^6]$ and does not fit the assumption (since it covers 0). To fix this, we can define an expression \texttt{e := \deegenKwdDark{bc\_operand\_100} * 8 + 1}, and replace \texttt{\ssa{\%17}.\llvmKwd{gep-offset}} with \texttt{e - 1}.

After this transform, we will obtain the following LLVM IR for our \texttt{Add\_0} bytecode:

\vspace{0.3em}

\texttt{; a copy-and-patch stencil hole is an external symbol}

\texttt{\deegenKwdDark{@\_\_deegen\_cp\_stencil\_hole\_0} = \llvmKwd{external dso\_local constant} \CType{i8}, \llvmKwd{align} 1}

\texttt{; declarations for stencil hole 1-4 are identical, omitted}

\ 

\texttt{\llvmKwd{define dso\_local ghccc} \CType{void} @\_\_deegen\_bytecode\_Add\_0(}

\texttt{\ \ \CType{ptr} \ssa{\%coroutineCtx}, \CType{ptr} \ssa{\%stackBase}, \CType{ptr} \ssa{\%0}, \CType{ptr} \ssa{\%codeBlock}, }

\texttt{\ \ \CType{i64} \ssa{\%tagRegister1}, \CType{ptr} \ssa{\%2}, \CType{i64} \ssa{\%3}, \CType{i64} \ssa{\%4}, \CType{i64} \ssa{\%5}, \CType{i64} \ssa{\%tagRegister2}, }

\texttt{\ \ \CType{double} \ssa{\%7}, \CType{double} \ssa{\%8}, \CType{double} \ssa{\%9}, \CType{double} \ssa{\%10}, \CType{double} \ssa{\%11}, \CType{double} \ssa{\%12})}

\texttt{\{}

\texttt{\ \ \ssa{\%14} = \llvmKwd{getelementptr inbounds} \CType{i8}, \CType{ptr} \ssa{\%stackBase},}

\texttt{\ \ \ \ \CType{i64} \llvmKwd{sub} (\CType{i64} \llvmKwd{ptrtoint} (\CType{ptr} \deegenKwdDark{@\_\_deegen\_cp\_stencil\_hole\_0} \llvmKwd{to} \CType{i64}), \CType{i64} 1)}

\texttt{\ \ \ssa{\%lhs} = \llvmKwd{load} \CType{double}, \CType{ptr} \ssa{\%14}, \llvmKwd{align} 8}

\texttt{\ \ \ssa{\%15} = \llvmKwd{getelementptr inbounds} \CType{i8}, \CType{ptr} \ssa{\%stackBase},}

\texttt{\ \ \ \ \CType{i64} \llvmKwd{sub} (\CType{i64} \llvmKwd{ptrtoint} (\CType{ptr} \deegenKwdDark{@\_\_deegen\_cp\_stencil\_hole\_1} \llvmKwd{to} \CType{i64}), \CType{i64} 1)}

\texttt{\ \ \ssa{\%rhs} = \llvmKwd{load} \CType{double}, \CType{ptr} \ssa{\%15}, \llvmKwd{align} 8}

\texttt{\ \ \ssa{\%16} = \llvmKwd{fcmp ord} \CType{double} \ssa{\%lhs}, \ssa{\%rhs}}

\texttt{\ \ \llvmKwd{br} \CType{i1} \ssa{\%16}, \llvmKwd{label} \ssa{\%20}, \llvmKwd{label} \ssa{\%slowpath}}

\texttt{\ssa{20:}}

\texttt{\ \ \ssa{\%21} = \llvmKwd{fadd} \CType{double} \ssa{\%lhs}, \ssa{\%rhs}}

\texttt{\ \ \ssa{\%outputPtr} = \llvmKwd{getelementptr inbounds} \CType{i8}, \CType{ptr} \ssa{\%stackBase},}

\texttt{\ \ \ \ \CType{i64} \llvmKwd{sub} (\CType{i64} \llvmKwd{ptrtoint} (\CType{ptr} \deegenKwdDark{@\_\_deegen\_cp\_stencil\_hole\_3} \llvmKwd{to} \CType{i64}), \CType{i64} 1)}

\texttt{\ \ \llvmKwd{store} \CType{double} \ssa{\%21}, \CType{ptr} \ssa{\%outputPtr}, \llvmKwd{align} 8}

\texttt{\ \ \llvmKwd{musttail call ghccc} \CType{void} \deegenKwdDark{@\_\_deegen\_cp\_stencil\_hole\_4}}

\texttt{\ \ \ \ \CType{ptr} \ssa{\%coroutineCtx}, \CType{ptr} \ssa{\%stackBase}, \CType{ptr} \llvmKwd{undef}, \CType{ptr} \ssa{\%codeBlock},}

\texttt{\ \ \ \ \CType{i64} \ssa{\%tagRegister1}, \CType{ptr} \llvmKwd{undef}, \CType{i64} \llvmKwd{undef}, \CType{i64} \llvmKwd{undef}, \CType{i64} \llvmKwd{undef},}

\texttt{\ \ \ \ \CType{i64} \ssa{\%tagRegister2}, \CType{double} \llvmKwd{undef}, \CType{double} \llvmKwd{undef}, \CType{double} \llvmKwd{undef},}

\texttt{\ \ \ \ \CType{double} \llvmKwd{undef}, \CType{double} \llvmKwd{undef}, \CType{double} \llvmKwd{undef})}

\texttt{\ \ \llvmKwd{ret} \CType{void}}

\texttt{\ssa{slowpath:}}

\texttt{\ \ \ssa{\%17} = \llvmKwd{getelementptr inbounds} \CType{\%CodeBlockTy}, \CType{ptr} \ssa{\%codeBlock}, \CType{i64} 0, \CType{i32} 8}

\texttt{\ \ \ssa{\%18} = \llvmKwd{load} \CType{ptr}, \CType{ptr} \ssa{\%17}, \llvmKwd{align} 8}

\texttt{\ \ \ssa{\%slowPathDataPtr} = \llvmKwd{getelementptr inbounds} \CType{i8}, \CType{ptr} \ssa{\%18},}

\texttt{\ \ \ \ \CType{i64} \llvmKwd{ptrtoint} (\CType{ptr} \deegenKwdDark{@\_\_deegen\_cp\_stencil\_hole\_2} \llvmKwd{to} \CType{i64})}

\texttt{\ \ \llvmKwd{musttail call ghccc} \CType{void} @\_\_deegen\_bytecode\_Add\_0\_quickening\_slowpath(}

\texttt{\ \ \ \ \CType{ptr} \ssa{\%coroutineCtx}, \CType{ptr} \ssa{\%stackBase}, \CType{ptr} \ssa{\%slowPathDataPtr}, \CType{ptr} \ssa{\%codeBlock},}

\texttt{\ \ \ \ \CType{i64} \ssa{\%tagRegister1}, \CType{ptr} \llvmKwd{undef}, \CType{i64} \llvmKwd{undef}, \CType{i64} \llvmKwd{undef}, \CType{i64} \llvmKwd{undef},}

\texttt{\ \ \ \ \CType{i64} \ssa{\%tagRegister2}, \CType{double} \ssa{\%lhs}, \CType{double} \ssa{\%rhs}, \CType{double} \llvmKwd{undef},}

\texttt{\ \ \ \ \CType{double} \llvmKwd{undef}, \CType{double} \llvmKwd{undef}, \CType{double} \llvmKwd{undef})}

\texttt{\ \ \llvmKwd{ret} \CType{void}}

\texttt{\}}

\vspace{0.3em}

And the values of the Copy-and-Patch stencil holes in the code are defined as follow:
\begin{itemize}
    \item \texttt{\deegenKwdDark{cp\_stencil\_hole\_0} := \deegenKwdDark{bc\_operand\_0} * 8 + 1\ \ \ } \textcolor{Gray}{-- \texttt{lhsSlot$\times$8+1}}
    \item \texttt{\deegenKwdDark{cp\_stencil\_hole\_1} := \deegenKwdDark{bc\_operand\_1} * 8 + 1\ \ \ } \textcolor{Gray}{-- \texttt{rhsSlot$\times$8+1}}
    \item \texttt{\deegenKwdDark{cp\_stencil\_hole\_2} := \deegenKwdDark{bc\_operand\_103}\ \ \ \ \ \ \ \ \ } \textcolor{Gray}{-- \texttt{slowPathDataOffset}}
    \item \texttt{\deegenKwdDark{cp\_stencil\_hole\_3} := \deegenKwdDark{bc\_operand\_100} * 8 + 1\ } \textcolor{Gray}{-- \texttt{outputSlot$\times$8+1}}
    \item \texttt{\deegenKwdDark{cp\_stencil\_hole\_4} := \deegenKwdDark{bc\_operand\_101}\ \ \ \ \ \ \ \ \ } \textcolor{Gray}{-- JIT address for the next bytecode}
\end{itemize}

\subsection{Compile to Textual Assembly File}

The next step is to compile the LLVM IR function above to textual assembly file (\texttt{.s} file) using \texttt{-O3}. This results in the following textual assembly:

\vspace{0.3em}

\texttt{\llvmKwd{.text}}

\texttt{\llvmKwd{.globl} \_\_deegen\_bytecode\_Add\_0}    

\texttt{\llvmKwd{.type} \_\_deegen\_bytecode\_Add\_0,\llvmKwd{@function}}

\texttt{\_\_deegen\_bytecode\_Add\_0:}   

\texttt{\ \ \llvmKwd{movsd} \ \ \deegenKwdDark{\_\_deegen\_cp\_stencil\_hole\_0}-1(\%rbp), \%xmm1}

\texttt{\ \ \llvmKwd{movsd} \ \ \deegenKwdDark{\_\_deegen\_cp\_stencil\_hole\_1}-1(\%rbp), \%xmm2} 

\texttt{\ \ \llvmKwd{ucomisd} \%xmm2, \%xmm1}

\texttt{\ \ \llvmKwd{jp} \ \ \ \ .LBB0\_1}

\texttt{\ \ \llvmKwd{addsd} \ \ \%xmm2, \%xmm1}

\texttt{\ \ \llvmKwd{movsd} \ \ \%xmm1, \deegenKwdDark{\_\_deegen\_cp\_stencil\_hole\_3}-1(\%rbp)}

\texttt{\ \ \llvmKwd{jmp} \ \ \ \ \deegenKwdDark{\_\_deegen\_cp\_stencil\_hole\_4}}

\texttt{.LBB0\_1:}

\texttt{\ \ \llvmKwd{movl} \ \ \ \$\deegenKwdDark{\_\_deegen\_cp\_stencil\_hole\_2}, \%r12d}

\texttt{\ \ \llvmKwd{addq} \ \ \ 0x30(\%rbx), \%r12}

\texttt{\ \ \llvmKwd{jmp} \ \ \ \ \_\_deegen\_bytecode\_Add\_0\_quickening\_slowpath}

\vspace{0.3em}

One can see that the assembly makes sense by plugging in the definitions of the stencil holes obtained in the previous step. For example, \deegenKwdDark{\_\_deegen\_cp\_stencil\_hole\_0} is defined as expression \texttt{\deegenKwdDark{bc\_operand\_0} * 8 + 1}, and there is a \texttt{-1} in the assembly that cancels out the \texttt{+1}, so the assembly is effectively loading from \texttt{\%stackBase + lhsSlot * 8} (the register \texttt{\%rbp} stores the stack base of the current function in our register-pinning scheme), which is exactly the local variable \texttt{lhs}.

\subsection{Hot-Cold Code Splitting and Jump-to-Fallthrough}

A few more assembly analysis and transformation passes would be needed if our bytecode had employed inline caching. Fortunately, \texttt{Add\_0} doesn't use IC, so only two assembly transformation passes are needed: hot-cold code splitting and jump-to-fallthrough (see \secref{sec:baseline-jit-generation}).

Hot-cold code splitting works by moving cold assembly code blocks to a separate section, so that the code is splitted into two pieces. In the assembly above, the \texttt{LBB0\_1} block is cold, and everything else is hot. This results in the following transformed assembly (added lines highlighted in red):

\vspace{0.3em}

\texttt{\llvmKwd{.text}}

\texttt{\llvmKwd{.globl} \_\_deegen\_bytecode\_Add\_0}    

\texttt{\llvmKwd{.type} \_\_deegen\_bytecode\_Add\_0,\llvmKwd{@function}}

\texttt{\_\_deegen\_bytecode\_Add\_0:}   

\texttt{\ \ \llvmKwd{movsd} \ \ \deegenKwdDark{\_\_deegen\_cp\_stencil\_hole\_0}-1(\%rbp), \%xmm1}

\texttt{\ \ \llvmKwd{movsd} \ \ \deegenKwdDark{\_\_deegen\_cp\_stencil\_hole\_1}-1(\%rbp), \%xmm2} 

\texttt{\ \ \llvmKwd{ucomisd} \%xmm2, \%xmm1}

\texttt{\ \ \llvmKwd{jp} \ \ \ \ .LBB0\_1}

\texttt{\ \ \llvmKwd{addsd} \ \ \%xmm2, \%xmm1}

\texttt{\ \ \llvmKwd{movsd} \ \ \%xmm1, \deegenKwdDark{\_\_deegen\_cp\_stencil\_hole\_3}-1(\%rbp)}

\texttt{\ \ \llvmKwd{jmp} \ \ \ \ \deegenKwdDark{\_\_deegen\_cp\_stencil\_hole\_4}}

\texttt{\textcolor{red}{.section .text.deegen\_slow,"ax",@progbits}}

\texttt{.LBB0\_1:}

\texttt{\ \ \llvmKwd{movl} \ \ \ \$\deegenKwdDark{\_\_deegen\_cp\_stencil\_hole\_2}, \%r12d}

\texttt{\ \ \llvmKwd{addq} \ \ \ 0x30(\%rbx), \%r12}

\texttt{\ \ \llvmKwd{jmp} \ \ \ \ \_\_deegen\_bytecode\_Add\_0\_quickening\_slowpath}

\vspace{0.3em}

After hot-cold code splitting, one can notice that the hot section (fast path) ends with a jump to \deegenKwdDark{\_\_deegen\_cp\_stencil\_hole\_4}. And the definition of this stencil hole is exactly \deegenKwdDark{bc\_operand\_101}, which is the JIT code address of the immediate next bytecode. Since \texttt{Add\_0} does not have JIT'ed return continuations, the above assembly snippet is all we need for the JIT code of \texttt{Add\_0}. Therefore, the \texttt{\llvmKwd{jmp} \deegenKwdDark{\_\_deegen\_cp\_stencil\_hole\_4}} is the last instruction in the JIT code fast path of \texttt{Add\_0}, and  \deegenKwdDark{\_\_deegen\_cp\_stencil\_hole\_4} is pointing to the immediate next instruction (which is the JIT code entry of the next bytecode). Thus, the \llvmKwd{jmp} is a no-op and can be eliminated. 

After this transform, we obtain the final textual assembly file for \texttt{Add\_0}:

\vspace{0.3em}

\texttt{\llvmKwd{.text}}

\texttt{\llvmKwd{.globl} \_\_deegen\_bytecode\_Add\_0}    

\texttt{\llvmKwd{.type} \_\_deegen\_bytecode\_Add\_0,\llvmKwd{@function}}

\texttt{\_\_deegen\_bytecode\_Add\_0:}   

\texttt{\ \ \llvmKwd{movsd} \ \ \deegenKwdDark{\_\_deegen\_cp\_stencil\_hole\_0}-1(\%rbp), \%xmm1}

\texttt{\ \ \llvmKwd{movsd} \ \ \deegenKwdDark{\_\_deegen\_cp\_stencil\_hole\_1}-1(\%rbp), \%xmm2} 

\texttt{\ \ \llvmKwd{ucomisd} \%xmm2, \%xmm1}

\texttt{\ \ \llvmKwd{jp} \ \ \ \ .LBB0\_1}

\texttt{\ \ \llvmKwd{addsd} \ \ \%xmm2, \%xmm1}

\texttt{\ \ \llvmKwd{movsd} \ \ \%xmm1, \deegenKwdDark{\_\_deegen\_cp\_stencil\_hole\_3}-1(\%rbp)}

\ 

\vspace{-0.8em}

\texttt{\llvmKwd{.section .text.deegen\_slow},"ax",@progbits}

\texttt{.LBB0\_1:}

\texttt{\ \ \llvmKwd{movl} \ \ \ \$\deegenKwdDark{\_\_deegen\_cp\_stencil\_hole\_2}, \%r12d}

\texttt{\ \ \llvmKwd{addq} \ \ \ 0x30(\%rbx), \%r12}

\texttt{\ \ \llvmKwd{jmp} \ \ \ \ \_\_deegen\_bytecode\_Add\_0\_quickening\_slowpath}

\vspace{0.3em}

One can see that the assembly above is exactly \figref{fig:add-baseline-jit-code}. The definition of the holes is also the same if one plugs in the definitions of the stencil holes and simplify the expressions. 

\subsection{Compile to ELF Object File}

Next, we compile the above assembly code to ELF object file. This generates an object file with the following two sections of relocatable machine code and linker relocation records:

\vspace{0.3em}

\noindent\texttt{\llvmKwd{.text}}

\noindent\texttt{\ \ \ \ f2 0f 10 8d {\setlength{\fboxsep}{2pt}\colorbox{BlueBoxColor!30}{00 00 00 00}} f2 0f 10 95 {\setlength{\fboxsep}{2pt}\colorbox{BlueBoxColor!30}{00 00 00 00}} 66 0f 2e ca 0f 8a}

\noindent\texttt{\ \ \ \ {\setlength{\fboxsep}{2pt}\colorbox{BlueBoxColor!30}{00 00 00 00} f2 0f 58 ca f2 0f 11 8d {\setlength{\fboxsep}{2pt}\colorbox{BlueBoxColor!30}{00 00 00 00}}}}

    \begin{itemize}[leftmargin=3.4em]
        \item [\textcolor{CTypeColor}{\raisebox{.5pt}{\textcircled{\raisebox{-.9pt} {1}}}}] \texttt{\CType{R\_X86\_64\_32S} \ \ \deegenKwdDark{\_\_deegen\_cp\_stencil\_hole\_0}-0x1}
        \item [\textcolor{CTypeColor}{\raisebox{.5pt}{\textcircled{\raisebox{-.9pt} {2}}}}] \texttt{\CType{R\_X86\_64\_32S} \ \ \deegenKwdDark{\_\_deegen\_cp\_stencil\_hole\_1}-0x1}
        \item [\textcolor{CTypeColor}{\raisebox{.5pt}{\textcircled{\raisebox{-.9pt} {3}}}}] \texttt{\CType{R\_X86\_64\_PLT32} \llvmKwd{.text.deegen\_slow}-0x4}
        \item [\textcolor{CTypeColor}{\raisebox{.5pt}{\textcircled{\raisebox{-.9pt} {4}}}}] \texttt{\CType{R\_X86\_64\_32S} \ \ \deegenKwdDark{\_\_deegen\_cp\_stencil\_hole\_3}-0x1}
    \end{itemize}

\ 

\vspace{-1em}

\noindent\texttt{\llvmKwd{.text.deegen\_slow}}

\noindent\texttt{\ \ \ \ 41 bc {\setlength{\fboxsep}{2pt}\colorbox{BlueBoxColor!30}{00 00 00 00}} 4c 03 63 30 e9 {\setlength{\fboxsep}{2pt}\colorbox{BlueBoxColor!30}{00 00 00 00}}}

 \begin{itemize}[leftmargin=3.4em]
        \item [\textcolor{CTypeColor}{\raisebox{.5pt}{\textcircled{\raisebox{-.9pt} {1}}}}] \texttt{\CType{R\_X86\_64\_32} \ \ \ \deegenKwdDark{\_\_deegen\_cp\_stencil\_hole\_2}}
        \item [\textcolor{CTypeColor}{\raisebox{.5pt}{\textcircled{\raisebox{-.9pt} {2}}}}] \texttt{\CType{R\_X86\_64\_PLT32} \_\_deegen\_bytecode\_Add\_0\_quickening\_slowpath-0x4}
    \end{itemize}

\subsection{Generate Copy-and-Patch Logic}

Next, we parse the object file to generate the copy-and-patch stencil definition. The \texttt{Add\_0} bytecode is the easy case since the object file doesn't contain a data section, but more complex bytecode may contain both static data (e.g., 64-bit constants) and relocatable data (e.g., jump tables). Great care must be taken to correctly handle all these cases. Extra work is also needed if the bytecode implementation consists of multiple stencils (due to return continuations) or if the bytecode employs inline caching. Fortunately, our \texttt{Add\_0} bytecode does not have any of these complications, so all we need to do is to parse the linker relocation records, and generate the patch logic accordingly. 

The following C logic illustrates the copy-and-patch logic generated from the object file obtained from the last step, that generates the JIT code for the \texttt{Add\_0} bytecode:

\vspace{0.3em}

\texttt{\CKeyword{void} CopyAndPatch\_Add\_0(}

\texttt{\ \ \ \CKeyword{void}* fastPathAddr, \CKeyword{void}* slowPathAddr, \CKeyword{void}* dataSecAddr, \CKeyword{size\_t} lhsSlot,}

\texttt{\ \ \ \CKeyword{size\_t} rhsSlot, \CKeyword{size\_t} outputSlot, \CKeyword{size\_t} slowPathDataOffset)}

\texttt{\{}

\texttt{\ \ \CKeyword{constexpr uint8\_t} fastPathCode[] = \{}

\texttt{\ \ \ \ 0xf2, 0x0f, 0x10, 0x8d, 0x00, 0x00, 0x00, 0x00, 0xf2, 0x0f, 0x10, 0x95,}

\texttt{\ \ \ \ 0x00, 0x00, 0x00, 0x00, 0x66, 0x0f, 0x2e, 0xca, 0x0f, 0x8a, 0x00, 0x00,}

\texttt{\ \ \ \ 0x00, 0x00, 0xf2, 0x0f, 0x58, 0xca, 0xf2, 0x0f, 0x11, 0x8d, 0x00, 0x00,}

\texttt{\ \ \ \ 0x00, 0x00 \};}

\texttt{\ \ memcpy(fastPathAddr, fastPathCode, \CKeyword{sizeof}(fastPathCode));}

\texttt{\ \ *(\CKeyword{uint32\_t}*)(fastPathAddr\CType{ + 4}) = \deegenKwdDark{(lhsSlot * 8 + 1)}\CType{ - 1};}

\texttt{\ \ *(\CKeyword{uint32\_t}*)(fastPathAddr\CType{ + 12}) = \deegenKwdDark{(rhsSlot * 8 + 1)}\CType{ - 1};}

\texttt{\ \ *(\CKeyword{uint32\_t}*)(fastPathAddr\CType{ + 22}) = \llvmKwd{slowPathAddr}\CType{ - (fastPathAddr + 22) - 4};}

\texttt{\ \ *(\CKeyword{uint32\_t}*)(fastPathAddr\CType{ + 34}) = \deegenKwdDark{(outputSlot * 8 + 1)}\CType{ - 1};}

\texttt{\ \ \CKeyword{constexpr uint8\_t} slowPathCode[] = \{}

\texttt{\ \ \ \ 0x41, 0xbc, 0x00, 0x00, 0x00, 0x00, 0x4c, 0x03, 0x63, 0x30, 0xe9, 0x00,}

\texttt{\ \ \ \ 0x00, 0x00, 0x00 \};}

\texttt{\ \ memcpy(slowPathAddr, slowPathCode, \CKeyword{sizeof}(slowPathCode));}

\texttt{\ \ *(\CKeyword{uint32\_t}*)(slowPathAddr\CType{ + 2}) = \deegenKwdDark{slowPathDataOffset};}

\texttt{\ \ \CKeyword{void}* temp = (\CKeyword{void}*)\_\_deegen\_bytecode\_Add\_0\_quickening\_slowpath;}

\texttt{\ \ *(\CKeyword{uint32\_t}*)(slowPathAddr\CType{ + 11}) = temp\CType{ - (slowPathAddr + 11) - 4};}

\texttt{\}}

\vspace{0.3em}

The \deegenKwdDark{\textrm{purple text}} in the code above indicates logic originated from the Copy-and-Patch stencil hole (\deegenKwdDark{\_\_deegen\_cp\_stencil\_hole\_*}) definitions, while the \CType{\textrm{blue text}} in the code indicates logic that originated from the linker relocation record definitions.

As one can see, although this is not the final baseline JIT code generator yet, it is not hard to notice that the logic described in the code above is exactly the copy-and-patch logic in \figref{fig:baseline-codegen}.

\subsection{Generate Baseline JIT Implementation}

As the last step, we generate LLVM IR logic that implements the baseline JIT code generator. This is very straightforward: all we need to do is to decode the bytecode, invoke the copy-and-patch logic described above to generate the JIT code, populate some necessary support data (e.g., the SlowPathData, and the map from bytecode index to JIT code address used for late-patching branch targets), decode the opcode of the next bytecode, and dispatch to the next code generator. 

Finally, we compile our code generator function into object file, which is linked into the final executable. This concludes our long journey that transforms the C++ bytecode semantics of the \texttt{Add\_0} bytecode all the way to the baseline JIT code generator illustrated in \figref{fig:baseline-codegen}.

\end{document}